\documentclass[%
 reprint,
superscriptaddress,
nofootinbib,
 amsmath,amssymb,
 aps,
 aip,
 apl
 prd,
 twocolumn,
]{revtex4-2}

\usepackage{graphicx}
\usepackage{dcolumn}
\usepackage{bm}
\usepackage{color}
\usepackage[cal=boondox]{mathalfa}
\usepackage{ifthen}
\usepackage[normalem]{ulem}

\usepackage{acronym}
\begin{document}

\newboolean{comments}   
\setboolean{comments}{true}
\ifthenelse{\boolean{comments}}{
\newcommand{\steve}[1]{\textcolor{magenta}{[Steve: #1]}}
\newcommand{\charlie}[1]{\textcolor{blue}{[Charlie: #1]}}
\newcommand{\duncan}[1]{\textcolor{green}{[Duncan: #1]}}
\newcommand{\vaibhav}[1]{\textcolor{cyan}{[Vaibhav: #1]}}
}{
\newcommand{\steve}[1]{{}}
\newcommand{\charlie}[1]{{}}
\newcommand{\duncan}[1]{{}}
\newcommand{\vaibhav}[1]{{}}
}

\newcommand{\D}{\mathrm{d}}

\preprint{APS/123-QED}

\title{VARAHA: A Fast Non-Markovian sampler for estimating Gravitational-Wave posteriors} 

\author{Vaibhav Tiwari}
\affiliation{Gravity Exploration Institute, School of Physics and Astronomy, Cardiff University, Queens Buildings, The Parade Cardiff CF24 3AA, UK.}

\author{Charlie Hoy}
\affiliation{University of Portsmouth, Portsmouth, PO1 3FX, United Kingdom}

\author{Stephen Fairhurst}
\affiliation{Gravity Exploration Institute, School of Physics and Astronomy, Cardiff University, Queens Buildings, The Parade Cardiff CF24 3AA, UK.}

\author{Duncan MacLeod}
\affiliation{Gravity Exploration Institute, School of Physics and Astronomy, Cardiff University, Queens Buildings, The Parade Cardiff CF24 3AA, UK.}

\date{\today}

\begin{abstract}
This article introduces VARAHA, an open-source, fast, non-Markovian sampler for estimating gravitational-wave posteriors. VARAHA differs from existing Nested sampling algorithms by gradually discarding regions of low likelihood, rather than gradually sampling regions of high likelihood. This alternative mindset enables VARAHA to freely draw samples from anywhere within the high-likelihood region of the parameter space, allowing for analyses to complete in significantly fewer cycles.
This means that VARAHA can significantly reduce both the wall and CPU time of all analyses. VARAHA offers many benefits, particularly for gravitational-wave astronomy where Bayesian inference can take many days, if not weeks, to complete. For instance, VARAHA can be used to estimate accurate sky locations, astrophysical probabilities and source classifications within minutes, which is particularly useful for multi-messenger follow-up of binary neutron star observations; VARAHA localises GW170817 $\sim 30$ times faster than LALInference. Although only aligned-spin, dominant multipole waveform models can be used for gravitational-wave analyses, it is trivial to extend this algorithm to include additional physics without hindering performance. We envision VARAHA being used for gravitational-wave studies, particularly estimating parameters using expensive waveform models, analysing subthreshold gravitational-wave candidates, generating simulated data for population studies, and rapid posterior estimation for binary neutron star mergers.

\end{abstract}

\maketitle

\acrodef{GW}[GW]{gravitational wave}
\acrodef{PE}{Parameter Estimation}
\acrodef{MC}{Monte Carlo}
\acrodef{MCMC}{Markov Chain Monte Carlo}
\acrodef{PDF}{Probability Density Function}
\acrodef{NS}{Neutron Star}
\acrodef{BH}{Black Hole}
\acrodef{BBH}[BBH]{Binary Black Hole}
\acrodef{BNS}[BNS]{Binary Neutron Star}
\acrodef{CBC}{Compact Binary Coalesence}
\acrodef{EDF}{empirical distribution function}
\acrodef{CDF}{cumulative distribution function}
\acrodef{SNR}{Signal to Noise Ratio}

\maketitle


\section{Introduction}

Compact binary coalescence (CBCs) -- \acp{BBH}, \acp{BNS} and \ac{NS}-\ac{BH} binaries -- are likely the only \acp{GW} sources observed by the network of ground-based \ac{GW} observatories so far~\citep{2019PhRvX...9c1040A, 2021PhRvX..11b1053A, 2021arXiv210801045T, o3b_cat}; although other sources have also been suggested including, for example, cosmic strings~\cite{LIGOScientific:2021nrg} and vector boson-star mergers~\cite{CalderonBustillo:2020fyi,2022arXiv220602551C}. The well-understood \ac{GW} signal morphology produced by CBCs~\cite[see e.g.][and references therein]{Pratten:2020ceb,Estelles:2021gvs,Hamilton:2021pkf,Thompson:2020nei,Ossokine:2020kjp,Matas:2020wab,Varma:2019csw,Riemenschneider:2021ppj,Dietrich:2019kaq} facilitates the estimation of binary parameters through \ac{PE}~\cite[see e.g.][]{first_monday_pe,Christensen:2022bxb}. These parameter estimates are needed to e.g. infer the properties of the source population~\cite[see e.g.][]{2021ApJ...913L...7A, 2021ApJ...913L..19T, o3b_rnp, 2022ApJ...928..155T,Callister:2021fpo,Roulet:2018jbe,Vitale:2022pmu,Edelman:2022ydv}, enhance our understanding of the equation of the state of neutron stars, or probe the cosmological history of the universe~\citep{2019PhRvL.122f1104A,2015PhRvD..92b3012A,LIGOScientific:2021aug}.

These estimated parameters are broadly categorised as a) intrinsic parameters: parameters that are directly responsible for the orbital evolution of the binary, such as masses, spins, tidal parameters, eccentricity, periastron distance, etc., and b) extrinsic parameters: parameters that are observer-dependent, namely, luminosity distance, binary's orientation from the line of sight, sky location, coalescence phase and coalescence time of the \acs{GW} signal. For binaries moving on a quasi-circular orbit with spins aligned with the orbital angular momentum, the extrinsic parameters do not impact the orbital evolution of a binary and consequently can only impart an overall shift to the amplitude or phase evolution of a signal. However, for binaries moving on an eccentric orbit, or with spins misaligned from the orbital angular momentum, the \ac{GW} signal morphology depends on the extrinsic parameters~\citep{1994PhRvD..49.6274A, 2009PhRvD..80h4001Y}. 
Although the number of intrinsic parameters may change depending on the physics of the problem, the number of extrinsic parameters remains fixed at seven~\citep{1996PhRvD..53.6749O}. The basics of \ac{PE} have been thoroughly discussed, and we cite some of the early works~\citep{1994PhRvD..49.2658C,1995PhRvD..52..848P}.

Multiple methods have been developed to perform \ac{PE} on \ac{GW} signals. The minute-scale analysis, BAYESTAR \cite{2016PhRvD..93b4013S}, focuses on the rapid localisation of \acs{GW} signals by estimating only the extrinsic parameters of the signal. It achieves this by keeping the intrinsic parameters fixed to the estimate provided by the \acs{GW} search analysis that first identified the signal. The packages that perform \ac{PE} of both the extrinsic and intrinsic (full) parameters through stochastic sampling methods include LALInference, which was previously the go-to analysis for \ac{GW} PE~\citep{2015PhRvD..91d2003V}, PyCBC Inference~\cite{2019PASP..131b4503B} and Bilby \citep{2019ApJS..241...27A, Romero-Shaw:2020owr, Smith:2019ucc,Ashton:2021anp}, which offer greater flexibility and modularity. These analyses often employ Nested~\cite{10.1214/06-BA127} or \ac{MCMC} sampling~\cite{metropolis1949monte} to obtain estimates for the binaries parameters. The packages RapidPE and RIFT use a non-Markovian approach to create an embarrassingly parallel infrastructure and provide comparatively faster processing times. They can also provide the marginal likelihood for straightforward model selection \citep{2015PhRvD..92b3002P, 2018arXiv180510457L, Wofford:2022ykb} (although see Refs.~\cite{Ashton:2021cub,Hoy:2022tst} for other model selection algorithms). Alternatively, methods to approximately estimate the binaries parameters have also been developed~\citep{2016CQGra..33aLT01T}, including recent advancements with utilising machine-learning techniques~\cite{Gabbard:2019rde,Green:2020dnx,Green:2020hst,Dax:2021tsq,Dax:2022pxd,Shen:2019vep,Williams:2021qyt}.

Typically, stochastic sampling analyses that perform full \ac{PE} can take hundreds or thousands of CPU-hours of processing time \citep{2016ApJ...825..116F, 2015ApJ...804..114B}. This high computational requirement is not sustainable, as the detection rate of \ac{GW} signals is expected to increase~\citep{2020LRR....23....3A} due to the continued improvement in the sensitivity of the \ac{GW} detectors. Fast, minute-scale \ac{PE}, is therefore crucial, especially for low-latency analyses where accurate skymaps and source classification probabilities are needed for timely follow-up by other multi-messenger facilities.

Attempts at improving computation time have primarily focused on speeding up waveform generation and computation of the likelihood function \citep{2014CQGra..31s5010P, 2015PhRvL.114g1104C, 2017CQGra..34k5006V, 2019arXiv190910986S,Hoy:2022tst,Cornish:2021lje}, or by utilising machine learning techniques~\cite{Gabbard:2019rde,Green:2020dnx,Green:2020hst,Dax:2021tsq,Dax:2022pxd,Shen:2019vep,Williams:2021qyt}. However, a significant improvement in computation time can also be achieved by efficiently populating the parameter space. In this paper, we introduce VARAHA, an alternative sampling technique that iteratively discards regions of low likelihood, and converges to the region of the parameter space that contains high posterior probability density (i.e. the posterior mass). We achieve significant gains in speed by introducing a) a non-Markovian method that performs a comparable number of computational operations, resulting in a similar number of effective samples, as Nested sampling but in significantly fewer iterations,
and b) splitting one large-dimensional sampling problem into two small-dimensional problems, where it samples the extrinsic parameters first and uses the acquired information to also sample the intrinsic parameters. These advantages result in significantly reduced processing times arising from greatly improved process parallelisation and array vectorisation in the analysis.

VARAHA can perform \ac{GW} \ac{PE} in a matter of a few minutes. Currently, it is limited to using waveform models that a) assume the spins are aligned with the orbital angular momentum, meaning that the binary does not precess~\cite{1994PhRvD..49.6274A}, and b) restrict attention to only the $\ell = 2$ gravitational-wave multipole, meaning that higher multipoles are neglected.  Nevertheless, VARAHA can meaningfully be used to perform fast \ac{PE} to localise and classify a source for electromagnetic follow-up, estimate parameters for a large number of sub-threshold \ac{GW} candidates, and generate \ac{PE} for simulated populations. We intend to make future extensions that will extend its applicability for the estimation of in-plane spins, eccentricity, and tidal deformability. Future extensions will also include uncertainties arising from the calibration of detector data, and fast methods for waveform generation and matched filtering, which currently consumes a significant portion of the computation.

In section \ref{sec:method}  we describe the basics of the analysis and the factors responsible for the faster processing time. We describe its application to the parameter estimation of \acp{GW} in section~\ref{sec:gwpe}. In section \ref{sec:results}, we present \ac{PE} for the individual observations GW151226~\cite{2016PhRvL.116x1103A} and GW170817~\cite{LIGOScientific:2017vwq}, as well as a population level validation using hundreds of simulated signals. We also discuss the computational requirements of VARAHA as well as its scalability with the number of CPUs.

\section{Method}
\label{sec:method}

\ac{PE} is the process of obtaining the probability distribution of parameters, $\boldsymbol{\theta}$, for a model, $m$, which is believed to describe the observed data, $d$. Given $d$, \ac{PE} generates an estimate for the \emph{posterior} \ac{PDF}, $p(\boldsymbol{\theta} | d, m)$, through Bayes' theorem,

\begin{equation} \label{eq:bayes_theorem}
    p(\boldsymbol{\theta} | d, m) \propto 
    \mathcal{L}(d | \boldsymbol{\theta}, m) \,
    \pi(\boldsymbol{\theta} | m) \, .
\end{equation}
The likelihood, $\mathcal{L}(d | \boldsymbol{\theta}, m)$, is a function of the observed data and model parameters, and $\pi(\boldsymbol{\theta} | m)$ is the prior probability for the model parameters.  The posterior distribution can alternatively be described as a weighted prior, with weights given by

\begin{equation} \label{eq:weights}
    w(\boldsymbol{\theta}) = \frac{\mathcal{L}(d | \boldsymbol{\theta}, m)}{\mathcal{L}_{\mathrm{max}}}
\end{equation}
where $\mathcal{L_{\mathrm{max}}}$ is the maximum likelihood\footnote{Ordinarily the weight is simply $\mathcal{L}(d | \boldsymbol{\theta}, m)$. We use a slightly modified definition to bind the maximum weight to be $w \leq 1$.}.

Typically, obtaining a closed-form expression for the posterior probability across the parameter space $\boldsymbol{\theta}$ is not possible. This means that we are not able to trivially evaluate Eq.~\ref{eq:bayes_theorem}, even if a functional form for the prior distribution is given.  It is therefore common to draw samples from the unknown posterior distribution through stochastic sampling techniques, such as Nested sampling~\cite{10.1214/06-BA127} or Markov-Chain-Monte-Carlo~\cite{metropolis1949monte}. For the case of Nested sampling, a series of contours of increasing likelihood converge, through an iterative process, to the region of high likelihood. Practically, a Nested Sampling routine draws a series of \emph{live points} and, at each cycle, it stores the live point with the lowest likelihood, and replaces it with a new point drawn randomly from the prior; the new point is accepted through the Matropolis-Hasting's algorithm~\cite{10.1093/biomet/57.1.97}. A new contour that encases the current set of live points is generated, and eventually, the contours converge to the regions of the highest likelihood. The stored points, along with their weights, constitute the \emph{samples} drawn from the posterior distribution.

A drawback of Nested sampling is that a few thousand live points are evolved in series, meaning that sampling can take thousands of cycles to complete. In addition, while the recovered posterior distribution becomes more accurate as the number of live points is increased, the number of cycles needed to sample from the unknown posterior distribution also increases (the number of cycles scales linearly with the number of live points~\cite{2004AIPC..735..395S}). Dynamic nested sampling~\cite{higson2019dynamic, Speagle:2019ivv} enables the number of live points to change throughout the analysis. It has been shown that this can be optimised for \ac{PE} analyses, allowing for a reduction in run time by a factor of $\sim 70$ for relatively simple cases. This significant improvement is possible when the region of high probability is contained in a small region of the prior volume (as is typically the case in high-dimensional problems)~\cite{higson2019dynamic}.

\begin{figure*}[t!]
    \centering
    \includegraphics[width=.93\textwidth]{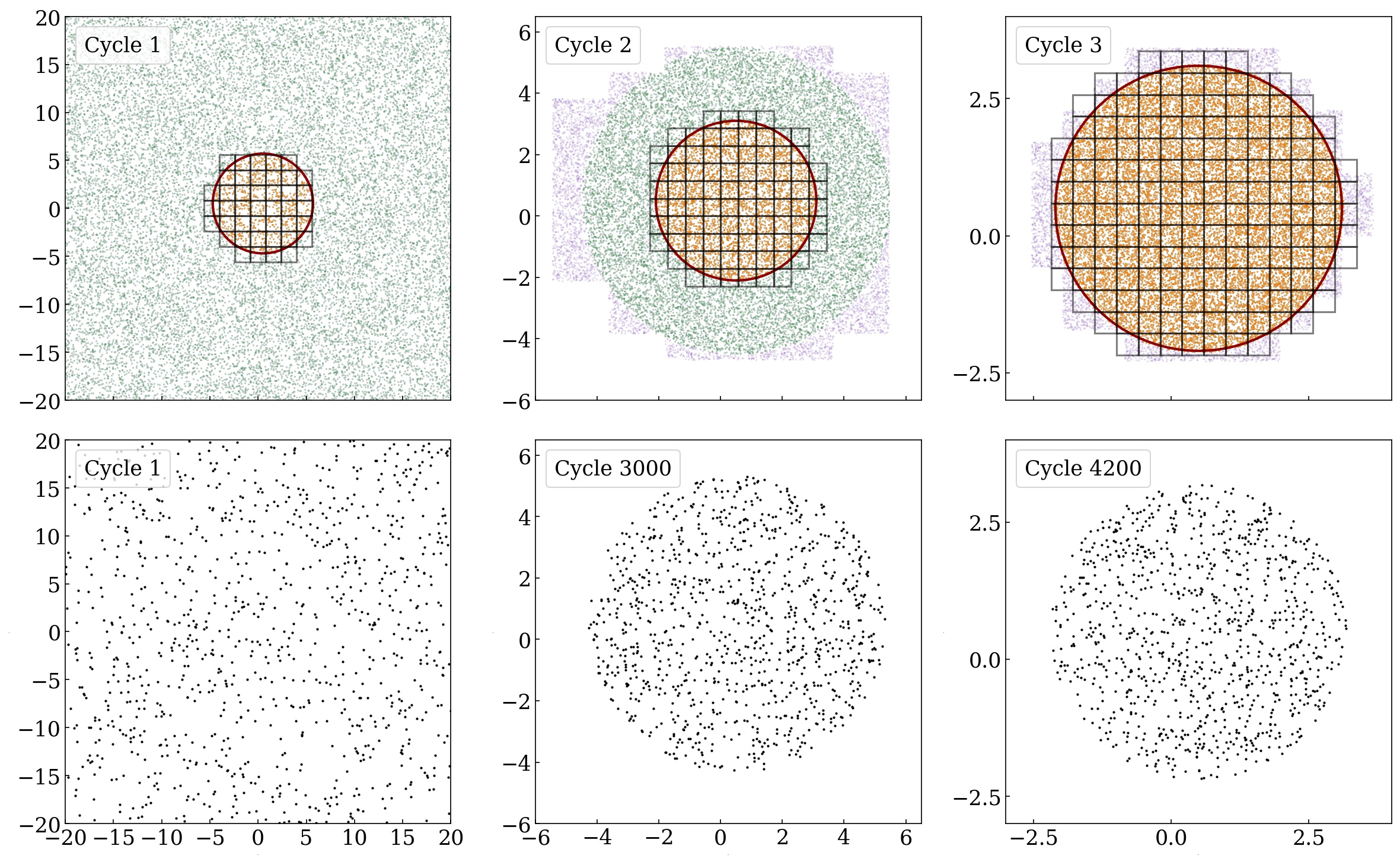}
    \caption{\emph{Top row}: A pictorial representation of VARAHA's sampling algorithm for a two-dimensional Gaussian likelihood distribution with mean [$0.5, 0.5$] and covariance $[0.5, 0.5]$. The left, middle and right panels show cycle 1, cycle 2 and cycle 3 respectively. The purple dots show the points randomly drawn from within the multi-dimensional grid and the green dots show the points that have a likelihood larger than the likelihood threshold from the previous cycle (Step 1). The orange dots show the points with a likelihood larger than the likelihood threshold calculated at the current cycle (Step 2). The red line shows the contour of fixed likelihood equal to the likelihood threshold, representing the live volume, and the black lines show the multi-dimensional grid that surrounds the live volume (Step 5). \emph{Bottom row}: A pictorial representation of a Nested sampling algorithm for the same two-dimensional Gaussian likelihood distribution; for this case, we use the {\sc{Dynesty}}~\cite{Speagle:2019ivv} Nested sampler. The left, middle and right panels show cycle 1, cycle $3,000$ and cycle $4,200$ respectively, and the grey dots show the Nested live points.}
\label{fig:progression_gaussian}
\end{figure*}

In this paper, we introduce an alternate sampling technique that has a major advantage over Nested sampling: a significantly reduced wall and CPU time. Our sampler, VARAHA, achieves this by (a) drawing thousands of points from the \emph{relevant regions in the parameter space that contain the posterior probability mass} and (b) intelligently defining a \emph{likelihood threshold}, below which defines a region of the parameter space that can be safely ignored. This approach means that although VARAHA evaluates the likelihood a comparable number of times as Nested sampling, it computes the likelihood for a large number of points at once, meaning that the computation can be efficiently vectorised and parallelised over multiple CPUs for enhanced performance. This is in contrast to computing the likelihood for a relatively low number of points at once, as is done in Nested sampling. Since evaluating the likelihood is often the most computationally expensive element of \ac{PE}, VARAHA is able to perform \ac{PE} in a fraction of the wall and CPU time compared to other conventional samplers.

VARAHA iteratively discards regions of the parameter space that do not contribute to the posterior distribution and restricts attention to the remaining regions of high likelihood through a series of cycles. Subsequent cycles draw points from within only the regions of high likelihood identified in the previous cycle. Once the final volume has been found, the points contained within the final volume are returned, along with weights given by Eq.~\ref{eq:weights}. We pictorially show VARAHA's algorithm in the top row of panels in Fig.~\ref{fig:progression_gaussian}.

VARAHA defines each volume to contain likelihoods greater than $\mathcal{L}_{\star}$. This volume is defined as,
\begin{equation}\label{eq:vol}
    V(\mathcal{L}_{\star}) = \int \Theta_{} \left[\mathcal{L}(d | \boldsymbol{\theta},m) - \mathcal{L}_{\star}\right] 
    \mathrm{d}\boldsymbol{\theta}
\end{equation}
and the posterior probability mass contained within the volume is 
\begin{equation}\label{eq:probability}
    P(\mathcal{L}_{\star}) = \int \Theta_{} \left[\mathcal{L}(d | \boldsymbol{\theta},m) - \mathcal{L}_{\star}\right] \,p(\boldsymbol{\theta}|d,m)\, 
    \mathrm{d}\boldsymbol{\theta} \, .
\end{equation}
Here $\Theta[x]$ is the Heaviside step function, $\Theta[x] = 1$ for $x\ge0$ and 0 otherwise.  As $\mathcal{L}_{\star} \rightarrow - \infty$, the probability tends to 1 and the volume becomes the full prior volume.  However, by choosing $\mathcal{L}_{\star}$ for which $P(\mathcal{L}_{\star})$ is close to, but slightly smaller than, unity the volume can be significantly smaller than the full prior volume.
VARAHA defines $\mathcal{L}_{\star}$ such that the corresponding volume $V(\mathcal{L}_{\star})$ contains a probability $P_{\mathrm{thr}}$. This likelihood value is hereafter referred to as the likelihood threshold. By identifying this region and sampling only from it, VARAHA can very efficiently generate samples.

The challenge, then, is to efficiently find the volume that contains a probability $P_{\mathrm{thr}}$, referred to as the \emph{live volume}. Since it is generally not possible to evaluate Eq.~\ref{eq:probability} analytically to find the appropriate likelihood threshold, other than for simple toy examples, VARAHA computes it numerically, and iteratively identifies both the appropriate threshold and the corresponding region of parameter space.

At the beginning of the first cycle, shown in the top left panel of Fig.~\ref{fig:progression_gaussian}, the live volume is simply the full prior volume. A large number of live points, $N_{\mathrm{live}} = N_\mathrm{pts}$, are randomly generated within the live volume (shown in green in the figure), and at each point, the likelihood is evaluated.  We can use these points to perform a Monte-Carlo integral of Eq. \ref{eq:probability} to obtain the required likelihood threshold $\mathcal{L}_{\mathrm{thr}}$ that contains the desired $P_{\mathrm{thr}}$.  However, particularly for sharply peaked posteriors in large parameter spaces, it is likely that only a small number of points will contribute significantly to the Monte-Carlo integral.  Therefore, in addition, we also calculate the likelihood threshold $\mathcal{L}_{N_{\mathrm{min}}}$ such that a minimum number of points, $N_{\mathrm{min}}$, lie above the threshold.  If $\mathcal{L}_{\mathrm{thr}} < \mathcal{L}_{N_{\mathrm{min}}}$, we have successfully identified the region which contains $P_{\mathrm{thr}}$. However, if $\mathcal{L}_{N_{\mathrm{min}}} < \mathcal{L}_{\mathrm{thr}}$, we have not sampled the posterior distribution sufficiently densely.  In this case, we exclude the regions of parameter space with likelihood below $\mathcal{L}_{N_{\mathrm{min}}}$ and repeat the process.

Selecting $N_{\mathrm{min}}$ points with the largest likelihoods identifies a region of high likelihood in the parameter space, which is shown by the orange dots in the top row of panels in Fig.~\ref{fig:progression_gaussian}. The volume of this region is given by Eq.~\ref{eq:vol} and is numerically evaluated through the Monte-Carlo integration
\begin{equation}\label{eq:volume}
    V \approx \bar{V} := V_{0} \frac{N_{\mathrm{min}}}{N_{\mathrm{pts}}}
\end{equation}
where $V_{0}$ is the full prior volume.  
The uncertainty in the Monte-Carlo integration is directly related to the Poisson fluctuations in $N_{\mathrm{min}}$: 

\begin{equation}
    \delta V\approx \delta\bar{V} := V_{0} \frac{\sqrt{N_{\mathrm{min}}}}{N_{\mathrm{pts}}} 
    = \frac{\bar{V}}{\sqrt{N_{\mathrm{min}}}} \, .
\end{equation} 
The set of $N_{\mathrm{min}}$ points provide an estimate of the volume and discretely constitute the region of the parameter space enclosed by this volume. With $N_{\mathrm{min}}$ chosen to be a few thousands, the error in the estimated volume is the order of a few per cent. 

The procedure outlined above presents two situations:\\
i) \textbf{The posterior mass enclosed by $\mathcal{L}_{N_{\mathrm{min}}}$ is larger than $P_{\mathrm{thr}}$}: Starting from the full prior volume, the live volume reduces to the volume enclosed by $\mathcal{L}_{N_{\mathrm{min}}}$~(bounded by the red contour in the Fig.~\ref{fig:progression_gaussian}). The fluctuation in $\mathcal{L}_{N_{\mathrm{min}}}$, due to sampling with a finite number of points, will have a negligible impact on the posterior mass contained by it. The largest fluctuation occurs for a uniform distribution and is equal to the \ac{MC} errors of a few per cent in the volume. However, in general, the likelihood distribution is expected to decay from one or several maxima.  \\
ii) \textbf{The posterior mass enclosed by $\mathcal{L}_{N_{\mathrm{min}}}$ is smaller than $P_{\mathrm{thr}}$}: This occurs when there are a large number of points with non-negligible weights, $n > N_{\mathrm{min}}$, and a threshold smaller than $\mathcal{L}_{N_{\mathrm{min}}}$ is required to enclose the posterior mass, $P_{\mathrm{thr}}$. To estimate $n$, VARAHA calculates the \ac{EDF}, $\hat{F}(\mathcal{L}_{\star})$, which simply gives the fraction of points with likelihood less than $\mathcal{L}_{\star}$.\footnote{We really wish to calculate the \ac{CDF}. However, since we only have discrete samples, the true \ac{CDF} is unknown. We, therefore, approximate the \ac{CDF} by computing the EDF and taking a conservative limit to ensure that we enclose \textit{at least} the desired probability.  A bound on the difference between  the \ac{EDF} and \ac{CDF} is given, e.g. by using the Dvoretzky–Kiefer–Wolfowitz–Massart inequality~\citep{10.1214/aoms/1177728174}.} The final likelihood threshold is then calculated as $\mathcal{L}_{\mathrm{thr}} := \hat{F}^{-1}(P_{\mathrm{thr}})$.

VARAHA evolves the first situation (i) using multiple cycles of \ac{MC} integration until it reaches the second situation (ii). Each cycle performs the integration using live volumes, which themselves decrease as cycles progress. A generalised form of Eq.~\ref{eq:vol} for this progression is given by,
\begin{equation}\label{eq:vol_shrinkage}
    V_{i} \approx \bar{V}_{i} :=
    \frac{N_{\mathrm{min}}}{N^{i-1}_{\mathrm{live}}} \,\bar{V}_{i-1}.
\end{equation}

For the second and consecutive cycles, VARAHA samples from the live volume and ignores the remaining parameter space. Since we only store samples contained within the volume and not the volume itself, the structure of the live volume is not known. VARAHA navigates this by a) generating a multi-dimensional grid covering the full parameter space, and b) selecting those hypercubes that contain at least one of the live points from the previous cycle. Once the relevant hypercubes have been identified, the live volume has been reconstructed. $N_\mathrm{pts}$ points are now uniformly scattered within the reconstructed live volume (the purple points in the top middle panel of Fig.~\ref{fig:progression_gaussian}). The likelihood of all the points is calculated and only those with a likelihood above the threshold are kept (the green points in the top middle panel of Fig.~\ref{fig:progression_gaussian}). Since the same number of points are now scattered within a much smaller volume, VARAHA is able to increase the number of live points contained within the live volume and thus setting the stage to perform the next \ac{MC} integration. 

It is important to choose an appropriate spacing for the multi-dimensional grid.  If the grid spacing is too large, the reconstructed live volume is much larger than the value estimated in Eq.~\ref{eq:volume}.  If the grid spacing is too small, the reconstructed live volume will not include relevant regions of the  parameter space that did not get sampled due to random fluctuations in the location of points.  VARAHA constructs the multi-dimensional grid over the full parameter space, requiring that the volume of each hypercube in the grid is equal to the error in the estimated volume $\delta \bar{V}$.  We motivate this choice as follows: the chosen hypercube volume is $\sqrt{N_{\mathrm{min}}}$ times larger than the average volume of $\bar{V}/N_{\mathrm{min}}$ approximately occupied by each live point. This choice ensures that the volume will have a maximum uncertainty of $\delta \bar{V}$ if this uncertainty arises due to Poisson fluctuation near a single live point. In this case, the multi-dimensional grid is expected to enclose most of $V$ even though constructed using information gained from the \ac{MC} volume $\bar{V}$. A uniform grid spacing in all dimensions, therefore, leads to the number of bins per dimension: $N_{\mathrm{bins}} = \left(V_0/\delta\bar{V}\right)^{\left(1 / N_{\mathrm{dim}}\right)}$, where $N_{\mathrm{dim}}$ is the dimensionality of the parameter space.\footnote{In some problems, there is significantly more structure in some dimensions of parameter space than others.  Then, it is desirable to employ a grid with different numbers of bins in each dimension.  The challenge, however, is to derive requirements for the relative number of bins in each dimension.} 

The iterative process described above gradually increases the value of the likelihood threshold and discards the uninteresting regions of the parameter space. However, a very small amount of posterior mass is also lost in the process. A new likelihood threshold approximately increases the discarded posterior mass by an amount,
\begin{equation}
1 - \frac{\sum_j w_j\;\Theta_{} \left[\mathcal{L}_j - \mathcal{L}_{N_{\mathrm{min}}}\right]}{\sum_j w_j},
\end{equation}
where $j$ identifies the samples that are enclosed by the live volume in the current cycle, and $\mathcal{L}_{N_{\mathrm{min}}}$ is the new likelihood threshold calculated for the next cycle.
This corresponds to the posterior mass contained between two concentric circles in Fig.~\ref{fig:progression_gaussian}. In practice, when deciding if $\mathcal{L}_{N_{\mathrm{min}}} > \mathcal{L}_{\mathrm{thr}}$, by calculating the \ac{EDF}, we require the discarded posterior mass not to accumulate to more than 1 - $P_{\mathrm{thr}}$. This implicitly ensures that the posterior mass enclosed by the live volume is at least $P_{\mathrm{thr}}$. VARAHA does not evolve the likelihood threshold any further if the discarded posterior mass becomes very close to 1 - $P_{\mathrm{thr}}$. Even though the likelihood threshold no longer evolves, the number of bins used to create the multi-dimensional grid continues to increase as the number of samples inside the live volume also increases.

Cycles are terminated once the desired accuracy is obtained.  The stopping condition could be determined based on a fixed number of weighted samples or a fixed number of cycles.  At the end of the analysis, VARAHA returns a set of weighted samples, where the weight of each sample is determined by Eq.\ref{eq:weights}.  All the samples have a likelihood value greater than the final value of $\mathcal{L}_{\mathrm{thr}}$.
Samplers that employ MCMC methods typically return a set of unweighted samples, with sample weights equal to $1$.  It is possible to generate these samples from the weighted samples by performing rejection sampling on the weighted samples (see Appendix~\ref{sec:importance_sampling} for details). The final set of unweighted samples has a sample size that is approximately $\sum_{i}w_{i}$~\cite{10.2307/4356322}.  However, since $n_{\mathrm{eff}} \ge \sum_{i}w_{i}$~\cite{vanRavenzwaaij2018}, with equality only if all the weights are equal to 1, the rejection sampling process leads to a reduction in the information contained in the samples. 

\subsection{Implementation} 
\label{sec:implementation}
VARAHA implements this algorithm as follows,

\begin{enumerate}

\item \textbf{Sprinkle points within the multi-dimensional grid}: (a) $N_{\mathrm{pts}}$ points are uniformly drawn from the reconstructed live volume and the likelihood for all points is calculated. (b) Live points with a likelihood larger than the threshold from the previous cycle are then identified. If this is the first cycle, $N_{\mathrm{pts}}$ points are uniformly distributed within the entire prior volume and all points are kept, meaning that $N_{\mathrm{pts}} = N_{\mathrm{live}}$ (this is equivalent to setting the likelihood threshold from the \emph{previous cycle} to negative infinity)

\item \textbf{Calculate the likelihood threshold}: The likelihood that accumulates no more than 1 - $P_{\mathrm{thr}}$ of the truncated posterior mass up to the current cycle, $\mathcal{L}_{\mathrm{thr}}$, is calculated using the $N_{\mathrm{live}}$ live points that are enclosed by the live volume. A second threshold, $\mathcal{L}_{N_{\mathrm{min}}}$, which ensures that $N_{\mathrm{min}}$ points lie above the threshold, is also calculated.  The final likelihood threshold is chosen to be the minimum of these two values. 

\item{\textbf{Calculate the volume of the live volume}: The volume and uncertainty in the live volume is calculated through \ac{MC} integration; see Eq.~\ref{eq:volume}.}

\item \textbf{Calculate effective sample size}: All points from the current or previous cycles that cross the current likelihood threshold are stored as weighted samples. The number of effective samples is calculated using,

\begin{equation} \label{eq:neff}
    n_{\mathrm{eff}} = \frac{(\sum_{i} w_{i})^{2}}{\sum_{i}(w_{i}^{2})},
\end{equation}
where $w_{i}$ is the weight of each sample, defined in Eq.~\ref{eq:weights}~\cite{ESS}.

\item \textbf{Reconstruct the live volume}: A multi-dimensional grid is created that spans the whole parameter space and hypercubes that register at least one live point are kept. The resulting hypercubes reconstruct the live volume.

\item \textbf{Repeat}: Steps 1 to 5 are repeated until a stopping criterion is reached. Example stopping criteria include: terminating once a specific number of effective samples have been obtained or terminating after a specific number of cycles have elapsed. The final output is then a set of weighted samples, where the weights are given by the likelihood evaluated at the sample, normalised by the maximum likelihood; see Eq.~\ref{eq:weights}.

\end{enumerate}

\subsection{Sampler settings}

The number of samples drawn from the uniform distribution $N_\mathrm{pts}$, the posterior mass required to be contained within the live volume, $P_{\mathrm{thr}}$, and the minimum number of points to retain, $N_{\mathrm{min}}$, are the sampler parameters that the user is free to specify. In our testing, we found that the choice of $N_\mathrm{pts}$ has only a small effect on the recovered posterior probability. Although reducing $N_\mathrm{pts}$ decreases the number of computations in each cycle, it increases the number of cycles needed to obtain the required effective number of samples. We find that choosing  $N_\mathrm{pts}$ to be in the range $100,000 \-- 1,000,000$ provides a good compromise between these.  Similarly, $N_{\mathrm{min}}$ sets the minimum fractional error on the estimated \ac{MC} volume. In our testing, we found that as the dimensionality of the problem increases, $N_{\mathrm{min}}$ should increase correspondingly. Depending on the complexity, $N_{\mathrm{min}}$ within the range of $1,000 \-- 10,000$ is adequate for most distributions with dimensionality between 2--8.

As VARAHA only keeps points that cross a specific likelihood threshold, defined through $P_{\mathrm{thr}}$, we deliberately discard part of the parameter space with low likelihood during each cycle. This can have an impact on the recovered posterior distribution if $P_{\mathrm{thr}}$ is too small.  In contrast, if the threshold is chosen to be too large, then we exclude a limited region of parameter space, which leads to a significant increase in analysis time for limited benefit.  In our testing, we found that $P_{\mathrm{thr}} = 0.999$ is sufficient for most cases. 

\subsection{Comparison with Nested Sampling}

In Fig.~\ref{fig:progression_gaussian}, we compare the convergence of VARAHA (top panels) and a Nested sampling algorithm ({\sc{Dynesty}}\footnote{We use {\sc{Dynesty}}=1.0.1, as this is the version in the International Gravitational-Wave Observatory Network (IGWN) Conda environment (\href{https://computing.docs.ligo.org/conda/}{https://computing.docs.ligo.org/conda/}) at the time of writing.}~\cite{Speagle:2019ivv}, bottom panels) for a simple two-dimensional Gaussian likelihood distribution. We used the static version of {\sc{Dynesty}} with default settings, but increased the number of live points to $1,000$, as this is similar to the number of live points used for gravitational-wave analyses~\cite[see e.g.][]{Romero-Shaw:2020owr}. For this example, VARAHA used $N_{\mathrm{pts}} = 20,000$, $N_{\mathrm{min}} = 1,000$ and $P_{\mathrm{thr}} = 0.999$.

As expected, VARAHA rapidly converges to the region of high likelihood (within 3 cycles), while {\sc{Dynesty}} requires significantly more cycles to constrain to a similar region of the parameter space ($\sim 4,200$ cycles). The significant reduction in the number of cycles of VARAHA, compared to {\sc{Dynesty}}, is primarily caused by the likelihood threshold: we see that in VARAHA's first cycle, we constrain the high likelihood region to within a circle of radius 5, while {\sc{Dynesty}} takes $\sim 3,000$ cycles to constrain to a comparable region of the parameter space. VARAHA further constrains the high likelihood region to within a circle of radius 2.5 within 3 cycles, while {\sc{Dynesty}} takes $\sim 4,200$ cycles to obtain a similar constraint. This culminates in a significantly reduced wall and CPU time. VARAHA also obtains a larger number of effective samples, with $n_{\mathrm{eff}} \sim 6,000$ compared to $n_{\mathrm{eff}}\sim 4,000$ for {\sc{Dynesty}}.

\begin{figure*}
    \centering
    \includegraphics[width=.92\textwidth]{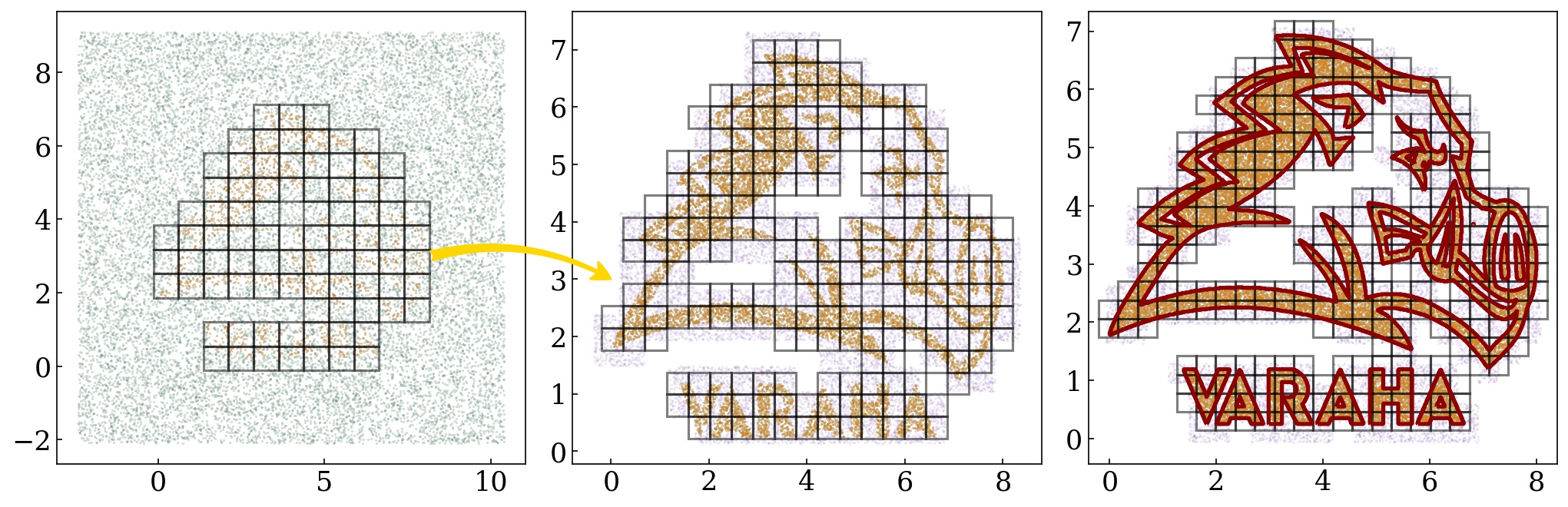}
    \caption{Plot showing the evolution of the multi-dimensional grid that surrounds the live volume, when VARAHA samples from a complex two-dimensional distribution with a constant density. The purple dots show the points randomly drawn from within the multi-dimensional grid and the green dots show the points that have a likelihood larger than the likelihood threshold from the previous cycle (Step 1). The orange dots show the points with a likelihood larger than the likelihood threshold calculated at the current cycle (Step 2) and the black lines show the multi-dimensional grid that surrounds the live volume (Step 5). The left, middle and right panels show to the $1^{\mathrm{st}}$, $3^{\mathrm{rd}}$ and $5^{\mathrm{th}}$ cycles respectively. In the right panel, the red line shows the contour of fixed likelihood equal to the likelihood threshold, representing the final live volume.
}
\label{fig:progression}
\end{figure*}

\subsection{The evolution of the multi-dimensional grid}

One of the key features of VARAHA is that it is able to draw points from within the live volume without having to know the structure. It achieves this by generating a multi-dimensional grid that covers the full parameter space and registering hypercubes that contain at least one point with likelihood above the previous likelihood threshold. To demonstrate this in practice, we analyse a complex two-dimensional distribution and explicitly show how the multi-dimensional grid is constructed and how it converges to the region of high likelihood. The chosen distribution has multiple disconnected regions of equal likelihood. While this example is only two-dimensional, the disconnected peaks in likelihood can prove challenging to identify. Since this distribution is analytically known, it provides a good illustration of VARAHA. For this example, we used $N_{\mathrm{pts}} = 20,000$, $N_{\mathrm{min}}=1,000$ and we terminated VARAHA once 5 cycles have elapsed. 

In Fig.~\ref{fig:progression} we show the evolution of the multi-dimensional grid. We see that in the first cycle, VARAHA is able to identify the rough location of the high-likelihood region and construct a coarse multi-dimensional grid that surrounds the entire volume. As VARAHA progresses, we see that the number of hypercubes increases and the multi-dimensional grid converges to the complex two-dimensional distribution with gaps appearing between adjacent hypercubes where there is little to no probability support. Unlike in Fig.~\ref{fig:progression_gaussian}, we do not see any green dots beyond the first cycle. This is because VARAHA rapidly identifies the high-likelihood region (owing to a constant density throughout) in the first cycle and maintains a constant likelihood threshold for all subsequent cycles. This implies that all points that cross the previously defined likelihood threshold (green dots) are used as live points for the current cycle (orange dots).

\subsection{Example: Bimodal multi-variate Gaussian distribution}
\label{sec:gaussian}

Next, we showcase the full VARAHA sampling algorithm. We chose to analyse a six-dimensional bimodal multi-variate Gaussian distribution and compare results with existing Nested and Markov-Chain-Monte-Carlo samplers. We also explicitly show how the likelihood threshold evolves as the number of cycles increases. For this example, we terminated VARAHA once $20,000$ effective samples were collected. During our sampling we used $P_{\mathrm{thr}} = 0.999$, $N_{\mathrm{pts}} = 400,000$ and $N_{\mathrm{min}} = 1,000$.

We analysed an asymmetric multivariate Gaussian distribution with each mode's mean and covariance randomly chosen: the mean of the first and second modes are randomly chosen between $[-1, 1]$ and $[-5, -3]$ respectively and we use a covariance matrix that is obtained by applying an inverse Wishart distribution~\cite{nydick2012wishart} to a diagonal matrix with elements randomly chosen between 1 and 2.

\begin{table}[t!]
\begin{ruledtabular}
\begin{tabular}{c c c c c}
   Cycle & $N_{\mathrm{bins}}$ & log Likelihood threshold & $n_{\mathrm{eff}}$\\
   \hline
1 & 4 & -170.7 & 3 \\
3 & 12 & -26.9 & 6 \\
4 & 19 & -13.6 & 21 \\
8 & 28 & -13.6 & 2896 \\
13 & 32 & -13.6 & 22323 \\
\end{tabular}
\end{ruledtabular}
\caption{Output from VARAHA showing the evolution of the number of bins in each dimension $N_{\mathrm{bins}}$, the likelihood threshold in each cycle, and the number of effective samples $n_{\mathrm{eff}}$ from the multi-variate Gaussian example. The likelihood threshold in each cycle is set by either $\mathcal{L}_{N_{\mathrm{min}}}$ or $\mathcal{L}_{\mathrm{thr}}$ depending on the situation (see text for details); for this case, the first 3 cycles are set by $\mathcal{L}_{N_{\mathrm{min}}}$ and subsequent cycles are set by $\mathcal{L}_{\mathrm{thr}}$. The sampled distribution is a bimodal multivariate Gaussian and VARAHA collected more than 20,000 effective samples across 13 cycles in 30 seconds on a single CPU thread.
}
\label{table:gaussian_summary}
\end{table}

\begin{figure}[t!]
    \centering
    \includegraphics[width=.48\textwidth]{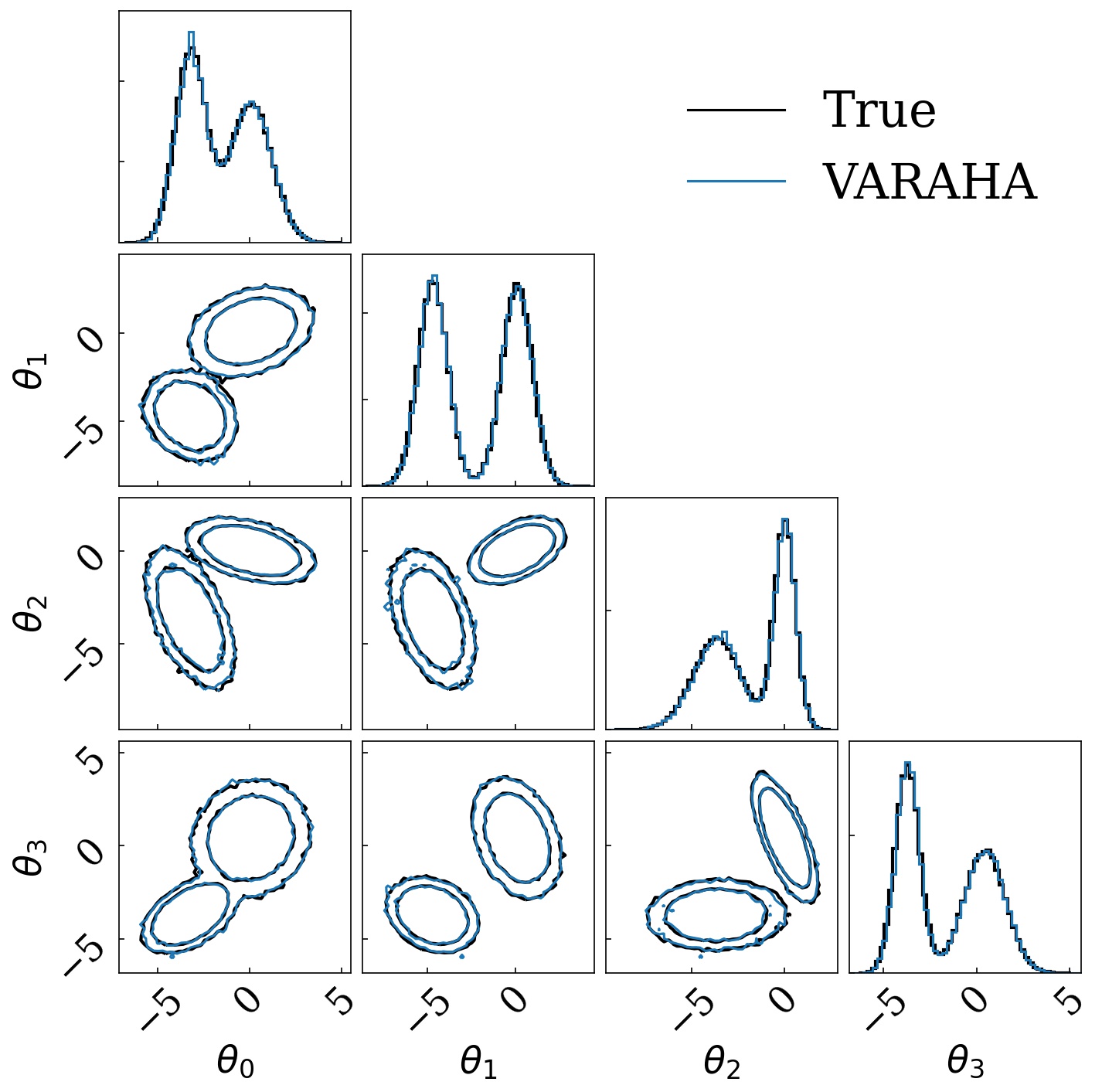}
    \caption{Corner plot~\citep{corner} showing the posterior PDF obtained by VARAHA and the true bimodal multi-variate Gaussian distribution across 4 of the six dimensions. The 2-dimensional contours show the 1.5$\sigma$ and 2.5$\sigma$ confidence levels and the individual histograms on the leading diagonal show the data marginalised to a particular dimension. VARAHA completes sampling in 30 seconds on a single CPU thread.
    }
    \label{fig:sampled_gaussian}
\end{figure}

The output of VARAHA is shown in Table~\ref{table:gaussian_summary}. The prior volume spans from -20 to 20 in each dimension. As VARAHA converges to the region of high likelihood, the number of  bins steadily increases over the cycles, reflecting the decreased uncertainty in the recovered volume. As expected, the likelihood threshold monotonically increases from $-170$ to $-13.6$ as the number of cycles increases. The likelihood threshold for the first three cycles is set by $\mathcal{L}_{N_{\mathrm{min}}}$ with subsequent cycles using $\mathcal{L}_{\mathrm{thr}}$. For the fourth cycle and beyond, the likelihood threshold remains fixed, which reflects the fact that the volume enclosing $P_{\mathrm{thr}}$ of the posterior mass has been found. The number of effective samples is relatively low in the first few cycles since the majority have $\mathcal{L}_{i} \ll \mathcal{L}_{\mathrm{max}}$, but increases steadily as the likelihood threshold increases. VARAHA obtains just over 23,000 effective samples from a total of approximately 800,000 weighted samples. Of course, rejection sampling could be used to obtain a set of unweighted samples, however, this would result in a smaller final sample size. We see that a larger number of cycles is needed than in the previous example, which is a result of the higher dimensionality of the likelihood surface.

In Figure~\ref{fig:sampled_gaussian}, we plot the posterior distribution obtained with VARAHA and the known analytic distribution. We see that VARAHA recovers the true distribution to high accuracy, with the mean and widths of each mode correctly identified. For comparison, we also analysed the same analytic distribution with two external samplers: {\sc{dynesty}}~\cite{Speagle:2019ivv}, a Nested sampler, and {\sc{Bilby MCMC}}~\cite{Ashton:2021anp}, a Markov-Chain-Monte-Carlo sampler, both operated through the {\sc{bilby}} infrastructure~\cite{2019ApJS..241...27A}\footnote{For both {\sc{dynesty}} and {\sc{Bilby MCMC}}, we used the default settings provided by {\sc{bilby}} with some modifications to ensure reasonable convergence. Both samplers used a single CPU. For {\sc{dynesty}}, we used 1000 live points and a nact (the number of autocorrelation times before accepting a point) of 5. For {\sc{Bilby MCMC}}, we used 4 temperature chains and we specified that we would like to obtain 10,000 independent samples.}. In our testing, we found that VARAHA finished sampling in 30 seconds on a single CPU: at least $60\times$ faster than either {\sc{dynesty}} or {\sc{Bilby MCMC}}. Although the runtimes for both {\sc{dynesty}} and {\sc{Bilby MCMC}} can vary significantly depending on the chosen settings, it is unlikely that either sampler can obtain a comparable number of posterior samples as VARAHA in 30 seconds that accurately recovers the mean and widths of each mode, when using a single CPU.

\section{Application to Parameter Estimation of Gravitational Waves}
\label{sec:gwpe}

In this section, we demonstrate the application of VARAHA to the \ac{PE} of gravitational wave signals. We focus only on gravitational waves originating from compact binary mergers and compare results to those obtained with {\sc{LALInference}}, the software that has regularly been used since the first gravitational-wave detection in 2015~\cite{first_monday_pe} (see also~\cite{2015PhRvD..92b3002P, 2019ApJS..241...27A, Romero-Shaw:2020owr, Smith:2019ucc, 2019PASP..131b4503B, 2018arXiv180510457L, 2015PhRvD..91d2003V, Williams:2021qyt, Gabbard:2019rde,Green:2020dnx,2018arXiv180608792Z, Leslie:2021ssu}). For this article, we restrict attention to quasi-circular binaries with spins aligned with the orbital angular momentum, referred to as an aligned-spin binary, meaning that the binary does not precess \cite{1994PhRvD..49.6274A}. In addition, we focus only on the leading (2,2) harmonic of the waveform, which is typically the most significant contribution to the observed \ac{GW} signal \cite{2021PhRvD.103b4042M}. In this case, the binary parameters can be cleanly decomposed into intrinsic parameters, which determine the properties of the binary, and extrinsic parameters, which determine the location and orientation of the binary relative to the earth. Throughout this work, we use the {\sc{IMRPhenomD}}~\cite{2016PhRvD..93d4006H,phenomd2} gravitational-wave model to evaluate the likelihood since it is optimised for aligned-spin binary systems that contain only the leading (2,2) harmonic (see also~\cite{Taracchini:2012ig,Taracchini:2013rva,Bohe:2016gbl,Pratten:2020fqn}).

For an aligned-spin model, the intrinsic parameters primarily affect the amplitude and phase evolution of the gravitational wave, and the extrinsic parameters only affect the overall amplitude, phase and time of arrival of the signal at each of the detectors. Table~\ref{tab:gw_params} lists the extrinsic and intrinsic parameters of the system estimated by VARAHA.  VARAHA allows separable sampling of extrinsic and intrinsic parameters and breaks one large dimensional problem into two small ones.  We note that other analyses, e.g. rapid PE/RIFT \citep{2015PhRvD..92b3002P, 2018arXiv180510457L},  employ a similar methodology as used by VARAHA. This framework is extensible to include higher harmonics in the gravitational-wave signal and additional signal parameters, such as in-plane spins, which lead to additional physical effects in the waveform, and can reduce biases in the recovered \ac{PE}~\cite{Kalaghatgi:2019log,Shaik:2019dym,LIGOScientific:2020ufj,LIGOScientific:2020stg,LIGOScientific:2020zkf,o3b_cat,2022Natur.610..652H,Krishnendu:2021cyi}. It will require accounting for the morphological dependence of a signal on the extrinsic parameters.  This extension will be investigated in future work.

\begin{table}[t]
\begin{tabular}{| c | l |}
\toprule
Label & Description \\
\hline
$\alpha$ & Right ascension of the source \\
$\delta$ & Declination of the source \\
$d_L$ & Luminosity distance of the source \\
$\iota$ & Inclination angle \\
$\psi$ & Polarisation angle \\
$\phi_{c}$ & Coalescence phase \\
$t_c$ & Coalescence time in the reference detector \\
\hline
$\mathcal{M}$ & Detector frame chirp mass  \\
$q$ & Mass ratio defined to be less than 1 \\
$\chi_1$ & First aligned spin component \\
$\chi_2$ & Second aligned spin component \\
\toprule
\end{tabular} 
\caption{\ac{GW} signal parameters sampled by VARAHA.
We group extrinsic parameters, except coalescence time, into one variable $\boldsymbol{\Omega}$ (top section), and all the intrinsic parameters into $\boldsymbol{\theta}$ (bottom section).  The component masses $m_1$ and $m_2$ are characterised by the mass ratio $q=m_2/m_1$ and chirp mass $\mathcal{M} = (m_1m_2)^{(3/5)} / (m_1 + m_2)^{(1/5)}$. The masses are measured in the frame of the detector from the signal that has already suffered cosmological redshift.
}
\label{tab:gw_params}
\end{table}

\subsection{Factorising the likelihood}
\label{sec:gw_like}

For a network of gravitational-wave detectors (e.g.~\cite{LIGOScientific:2014pky,VIRGO:2014yos,KAGRA:2020agh}), the probability that the observed detector data contains a gravitational wave signal from a coalescing binary and with parameters, $(t_c, \boldsymbol{\Omega}, \boldsymbol{\theta})$, is given by Bayes formula as
\begin{equation}
    p(t_c, \boldsymbol{\Omega}, \boldsymbol{\theta}|\vec{d}) = \frac{\mathcal{L}(\vec{d} | t_c, \boldsymbol{\Omega}, \boldsymbol{\theta})\,\pi(t_c, \boldsymbol{\Omega}, \boldsymbol{\theta})}{\{\vec{d}\}},
    \label{eq:bayes}
\end{equation}
where $\vec{d} \equiv \{d_1, d_2, \cdots\}$ represents the strain data observed in each gravitational-wave detector, $t_c$ is the coalescence time in the reference detector, $\boldsymbol{\Omega}$ denotes the remaining extrinsic parameters, and $\boldsymbol{\theta}$ are the intrinsic parameters.
Under the assumption that the data in each detector is independent, stationary, Gaussian and containing a gravitational wave signal, a Gaussian likelihood, $\mathcal{L}(\vec{d} \, | t_c, \boldsymbol{\Omega}, \boldsymbol{\theta})$, is constructed as
\begin{multline}
\log\left(\mathcal{L}\left(\vec{d} \, | t_c, \boldsymbol{\Omega}, \boldsymbol{\theta}\right) \right) =
c\sum_{i \in \mathrm{dets}}\,\langle d^i - h^i| d^i - h^i \rangle\\
= c\sum_{i \in \mathrm{dets}}\,\left( \langle d^i| d^i\rangle - \langle d^i|h^i \rangle - \langle h^i| d^i\rangle + \langle h^i| h^i \rangle\right) 
\label{eq:gauss_lkl}
\end{multline}
where $c = (2\pi)^{-k/2}$ for a $k$ dimensional distribution and $h^i\equiv h^i(t_c, \boldsymbol{\Omega}, \boldsymbol{\theta})$ is the expected \ac{GW} signal in the $i^{\mathrm{th}}$ detector. The term  $\{\vec{d}\}$ is the marginal likelihood (or evidence),
\begin{equation}
\{\vec{d}\} = \int \mathcal{L}(\vec{d} | t_c, \boldsymbol{\Omega}, \boldsymbol{\theta})\,\pi(t_c, \boldsymbol{\Omega}, \boldsymbol{\theta})\, \D \boldsymbol{\theta} \,\D\boldsymbol{\Omega}\,\D t_c.
\end{equation}
In Eq.~\ref{eq:gauss_lkl}, the \textit{complex} noise weighted inner product between two time-domain functions $ \langle a(t)\, |\, b(t) \rangle$ is defined in the frequency domain as,
\begin{equation}
\langle a(t)\,|\, b(t)\rangle  = 4\int_{f_{\mathrm{min}}}^{f_{\mathrm{max}}} \frac{\tilde{a}(f)\tilde{b}(f)^{\star}}{S(f)} \; \D f,
\label{eq:mf}
\end{equation}
where $S(f)$ is the power spectrum of the detector noise \citep{1994PhRvD..49.2658C}, $\tilde{a}(f)$ and $\tilde{b}(f)$ are the Fourier transforms of $a(t)$ and $b(t)$ respectively and star represents the complex conjugate. Eq.~\ref{eq:mf} can be evaluated at an arbitrary time shift by using the convolution theorem and taking the inverse Fourier transform~\citep{2013PhRvD..87b4033B},
\begin{equation}
    \langle a(t)\,|\, b(t + \Delta t)\rangle = 4\int_{f_{\mathrm{min}}}^{f_{\mathrm{max}}} \frac{\tilde{a}(f)\tilde{b}(f)^{\star}}{S(f)} \mathrm{e}^{2\pi i \Delta t}\; \D f.
    \label{eq:invfourier}
\end{equation}
Consequently, the likelihood can be evaluated for a fixed set of intrinsic parameters, from a single inner product calculation. 

Returning to Eq.~\ref{eq:gauss_lkl}, the first term, $\langle d_{i} | d_{i} \rangle$, is independent of the parameters $\boldsymbol{\Omega}$ and $\boldsymbol{\theta}$ and is therefore constant. Since it will be absorbed in the normalisation for the Eq.~\ref{eq:bayes} we neglect it in what follows. The final term is the inner product of the expected signal in the $i^{\mathrm{th}}$ detector with itself,
\begin{equation}\label{eq:deff}
    \langle h^i(t_c, \boldsymbol{\Omega}, \boldsymbol{\theta}) | h^i(t_c, \boldsymbol{\Omega}, \boldsymbol{\theta}) \rangle =: \varrho^i(\boldsymbol{\Omega}, \boldsymbol{\theta}) ^ 2.
\end{equation}
If $h(\boldsymbol{\theta})$ is the signal arriving from a binary with intrinsic parameters $\boldsymbol{\theta}$ at a unit distance, from overhead the detector and with the orbital plane facing the observer, the relative amplitude of a signal from a given distance $d_{L}$ sky location and orientation is determined by the effective distance, given by,
\begin{equation}
    D^{i}_{\mathrm{eff}} = d_L/\sqrt{\left[ F^{i}_+\mathrm{}^2\left( \frac{1 + \cos^2 \iota}{2}\right) + F_i^\times\mathrm{}^2 \cos^2 \iota \right]},
    \label{eq:sky_scale}
\end{equation}
where the detector response functions $F^{i}_{+, \times}$ depend upon the sky location, polarization and time of arrival of the source~\cite{2011CQGra..28l5023S}.  Thus, once 
\begin{equation}
    \varrho_{o}^{i}(\boldsymbol{\theta})^{2}  = \langle h(\boldsymbol{\theta})|h(\boldsymbol{\theta}) \rangle
\end{equation}
has been calculated, it is straightforward to evaluate $\varrho^{i}$ as
\begin{equation}
    \varrho^i(\boldsymbol{\Omega}, \boldsymbol{\theta}) 
    = \frac{\varrho_{o}^i(\boldsymbol{\theta})}{D^{i}_{\mathrm{eff}}} \, .
\end{equation}

Finally, we turn to the two middle terms in Eq.~\ref{eq:gauss_lkl} which constitute the inner product of the data $d^{i}$ with the expected signal.  We have already seen that the variation of the signal amplitude can be encoded in the effective distance $D^{i}_{\mathrm{eff}}$.  Similarly, the phase of the signal observed in detector $i$ is given by \cite{2012PhRvD..85l2006A}
\begin{equation}
    \phi_{\boldsymbol{\Omega}}^i = \phi^i - 2\,\phi_c,
    \quad \mathrm{where} \quad
    \phi^i = \tan^{-1}\left( \frac{F_\times^i}{F_+^i}\frac{2 \cos \iota}{1 + \cos^2 \iota}\right) \, .
\end{equation}
and the time of arrival is given by
\begin{equation}
    t^{i} = t_{c} + \Delta t^{i}(\boldsymbol{\Omega}, t_{c})
\end{equation}
where $\Delta t^{i}$ depends upon the location of the source relative to the detectors.

Since the amplitude and phase evolution of the signal is unchanged by the extrinsic parameters, we can calculate the inner product for any set of extrinsic parameters by rescaling and time-shifting the inner product time series for the reference waveform $h(\boldsymbol{\theta})$.
Thus, the inner product of the data $d^i$ with the expected signal is given by
\begin{align}
    \langle d^i | h^i(t_c, \boldsymbol{\Omega}, \boldsymbol{\theta})\rangle &= 
    \langle d^i | \frac{h^i(\boldsymbol{\theta}, t^{i}) \exp(i \phi_{\boldsymbol{\Omega}}^{i})}{D^{i}_{\mathrm{eff}}} \rangle \nonumber \\
    &= \frac{\varrho_{o}^i(\boldsymbol{\theta})
    \rho^{i}(\boldsymbol{\theta}, t^{i}) \exp(i \phi_{\boldsymbol{\Omega}}^{i})}{D^{i}_{\mathrm{eff}}}
    \label{eq:rho_i}
\end{align}
where we have defined the \ac{SNR} for the template with intrinsic parameters $\boldsymbol{\theta}$ as 
\begin{equation}
    \rho^i(\boldsymbol{\theta}, t_c) = \frac{\langle d_i | h(\boldsymbol{\theta}, t^{i})\rangle}{\sqrt{\langle  h(\boldsymbol{\theta})|h(\boldsymbol{\theta})\rangle}} \, .
    \label{eq:mf_phase}
\end{equation}
The third term in Eq.~\ref{eq:gauss_lkl} is simply the complex conjugate of \ref{eq:rho_i}.

Combining these expressions, the resulting log-likelihood from 
Equation~\ref{eq:gauss_lkl} assumes the form
\begin{align}\label{eq:loglkl}
    \log \left(\mathcal{L}\left(\vec{d} | t_c, \boldsymbol{\Omega}, \boldsymbol{\theta}\right) \right) &= 
    \\
    \sum_{i \in \mathrm{dets}} \Big[
    -\frac{1}{2} \frac{\varrho_{o}^i(\boldsymbol{\theta})^2}{(D^{i}_{\mathrm{eff}})^{2}} 
    &+ \frac{\varrho_{o}^i(\boldsymbol{\theta})
    |\rho^i(\boldsymbol{\theta}, t^{i})|}
    {D^{i}_{\mathrm{eff}}} 
    \cos\left(\Delta \phi^{i}  \right)\Big], \nonumber
\end{align}
where $\Delta \phi^{i}$ is the difference between the measured phase in detector $i$ and the expected phase, given the parameters of the signal:
\begin{equation}
\Delta \phi^{i} = \mathrm{arg}(\rho^i(\boldsymbol{\theta}, t^{i})) - \phi_{\boldsymbol{\Omega}}^i
\end{equation}

The log-likelihood is maximised for a given set of intrinsic parameters when $D^{i}_{\mathrm{eff}} = \varrho^i(\boldsymbol{\theta}) / \rho^i(\boldsymbol{\theta}, t_c^i)$, and the cosine term in the last equation is unity. However, as the cosine term is solely dependent on the phase acquired due to the extrinsic parameters and $\rho_i(t_c^i)$ is dependent on the arrival times in different network detectors, the maximisation puts a time-phase constraint \citep{2018CQGra..35j5002F}.

The expression in Eq.~\ref{eq:loglkl} clearly separates the likelihood dependence on the intrinsic parameters $\boldsymbol{\theta}$ from the extrinsic parameters $\boldsymbol{\Omega}$ and the coalescence time $t_{c}$.  In particular, the effective distance, $D^{i}_{\mathrm{eff}}$, phase, $\phi^{i}_{\boldsymbol{\Omega}}$ and time of arrival $t^{i}$ in each detector depends only upon the extrinsic parameters $\boldsymbol{\Omega}$ and the time of arrival $t_{c}$.  The \ac{SNR} time-series associated to the fiducial waveform $h(\boldsymbol{\theta})$, and its overall normalisation, is dependent only on the intrinsic parameters $\boldsymbol{\theta}$.  However, the specific time at which to evaluate the \ac{SNR} does depend upon the intrinsic parameters through $\Delta t^{i}$.  In the following sections, we make repeated use of this splitting of the likelihood to independently estimate the intrinsic and extrinsic parameters.  In particular, the most time-consuming step is the generation of simulated waveform and evaluation of the \ac{SNR} time-series.  Thus by writing the likelihood in the form of Eq.~\ref{eq:loglkl}, it becomes clear that the entire extrinsic parameter space, for a fixed set of intrinsic parameters, can be explored with a single evaluation of the \ac{SNR} time-series.

\subsection{Extrinsic Parameters}
\label{sec:extrinsic}

VARAHA starts by first fixing the intrinsic parameters $\boldsymbol{\theta}$ to a reference waveform and samples the posterior distribution for the seven extrinsic parameters, $\boldsymbol{\Omega}$ and $t_{c}$. BAYESTAR~\cite{2016PhRvD..93b4013S}, a rapid, non-Markovian sky localisation algorithm commonly used by the LIGO, Virgo and KAGRA collaborations (see e.g.~\cite{LIGOScientific:2017vwq}), also samples the extrinsic parameters by fixing the intrinsic parameters to a reference waveform. In this subsection, we explain how VARAHA varies from BAYESTAR and demonstrate that it is able to compete with BAYESTAR's performance.

In the initial evaluation, it is natural to use the values for the intrinsic parameters, $\boldsymbol{\theta}_o$, which are reported by the search analysis that identified the signal \cite{pycbc, gstlal, 2016CQGra..33q5012A, 2022PhRvD.105b4023C}. The detector which has the largest value of $\rho^i(\boldsymbol{\theta}_{o})$ in the network is chosen as the reference detector. We then perform \ac{PE} over the seven-dimensional extrinsic parameter space using the method described in Section \ref{sec:method}. The first step involves scattering millions of points across the extrinsic parameter space. Since 5 dimensions are angles, it is natural to cover the full range of possible values:
\begin{eqnarray}
\alpha &=& [0, 2\pi] \nonumber \\
\sin(\delta) &=& [-1, 1] \nonumber \\
\cos(\iota) &=& [-1, 1] \nonumber \\
\phi_c &=& [0, 2\pi] \nonumber \\
\psi &=& [0, 2\pi].
\label{extr_ranges}
\end{eqnarray}
\begin{figure}
    \begin{center}
    \includegraphics[width=0.47\textwidth]{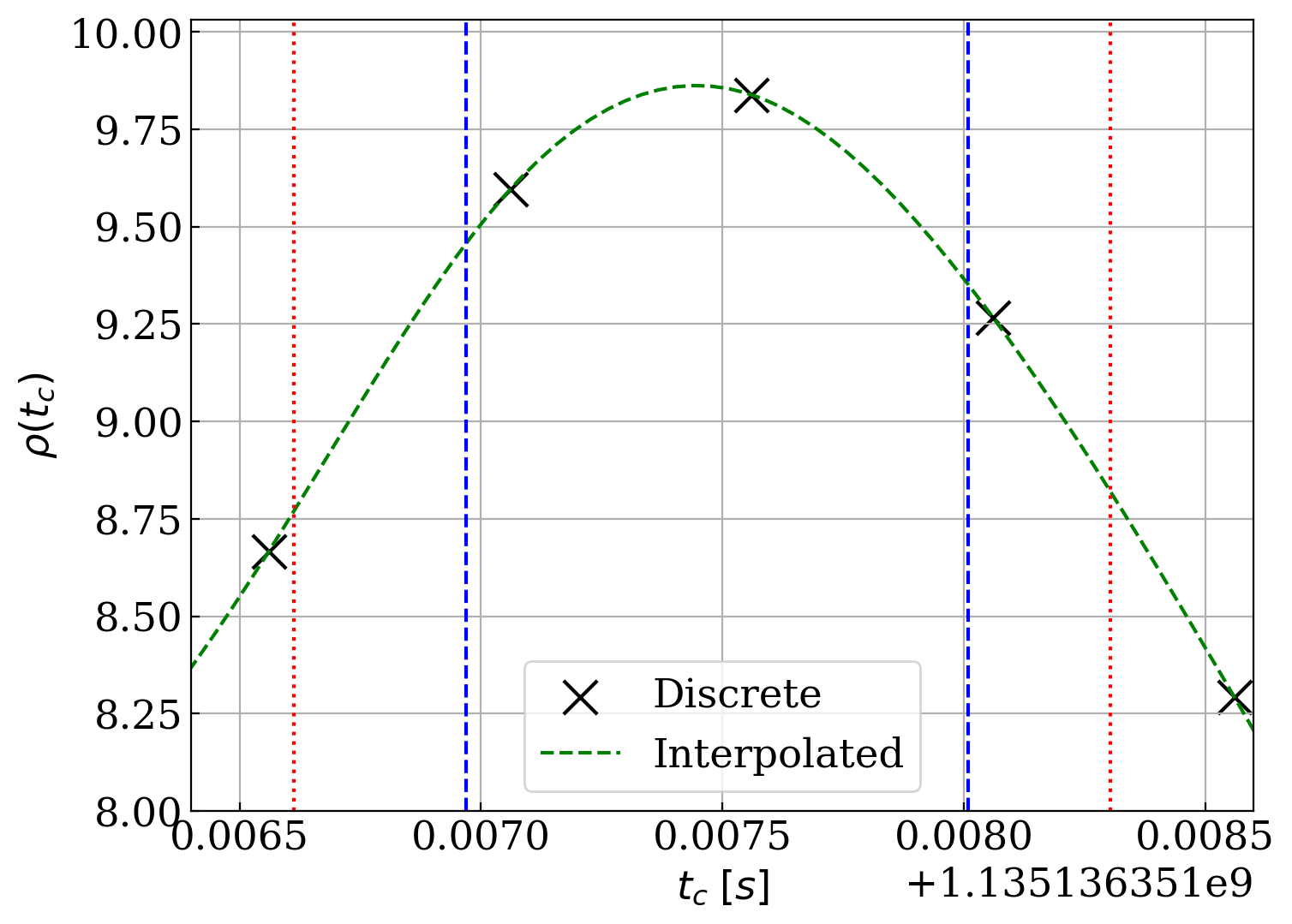}
    \caption{
    The figure plots $\rho(t_c)$ for values of $t_c$ around the observed GPS of the signal GW151226 \cite{2016PhRvL.116x1103A} for the LIGO--Hanford detector. The dotted red lines set the bound on $t_c$ as described in Eq.~\ref{eq:tc_bound}. The dashed blue lines show the 99.9\% credible interval of the posterior distribution on $t_c$ obtained from performing parameter estimation on the full set of extrinsic parameters.
    }
    \label{fig:tcbounds}
    \end{center}
\end{figure}
The initial choice of bounds for the coalescence time and luminosity distance requires more care.  We wish to determine the narrowest ranges of $t_{c}$ and $d_{L}$ that will ensure we chose a range that encloses the posterior mass.  If the initial choice is too narrow, then we risk missing part of the relevant parameter space and if it is too broad this will lead to unnecessary exploration of uninteresting regions of the parameter space which will increase the analysis time.

To fix the range of coalescence time, we restrict attention to the reference detector (the one with the largest \ac{SNR}).  We assume that the extrinsic parameters are chosen to maximise the likelihood contribution from the reference detector, i.e. that $D^{i}_{\mathrm{eff}} = \varrho^i(\boldsymbol{\theta_{o}}) / |\rho^i(\boldsymbol{\theta_{o}}, t_c)|$ and that the phase $\Delta \phi_{i} = 0$.  In that case, the likelihood contribution from the reference detector is $\tfrac{1}{2} |\rho^i(\boldsymbol{\theta_{o}}, t_c)|^{2}$. This is normally distributed in $t_c$. Furthermore, from the \ac{GW} search result, we know the time $t_o$ which gives the maximum \ac{SNR}, $\rho^i(\boldsymbol{\theta_{o}}, t_o)$. As we vary the coalescence time $t_{c}^{i}$ in the reference detector, the observed \ac{SNR} will be reduced and, consequently, the maximum contribution of the reference detector to the likelihood will be reduced.  The initial range of coalescence times is chosen so that the boundary is at least 4-$\sigma$, i.e. we require
\begin{equation}
    \tfrac{1}{2} \left[|\rho^i(\boldsymbol{\theta_{o}}, t_o)|^{2} 
    - |\rho^i(\boldsymbol{\theta_{o}}, t_c)|^{2} \right]
    < 4^2/2.
    \label{eq:tc_bound}
\end{equation}

So far we have restricted attention to a reference detector.  Let us assume that there exists a set of extrinsic parameters which is a good fit to the data in all detectors,  i.e. $D^{i}_{\mathrm{eff}} \approx \varrho^i(\boldsymbol{\theta_{o}}) / |\rho^i(\boldsymbol{\theta_{o}}, t_o)|$ and $\Delta \phi^{i} \approx 0$.  Then, as we vary the coalescence time $t_{o}$, \textit{at best} we will find a set of extrinsic parameters that matches the data in all detectors other than the reference, while in the reference detector, the time is offset from the observed peak.  Thus, the loss in likelihood in the reference detector gives an (approximate) lower limit on the loss in the network likelihood.

We restrict the initial range of allowed coalescence times that satisfies Eq.~\ref{eq:tc_bound}. Figure \ref{fig:tcbounds} plots an example for the bounds on $\rho(t_c)$ estimated for the data from the LIGO Hanford detector corresponding to the signal GW151226 \cite{2016PhRvL.116x1103A}. The discrete values of $\rho(t_c)$ are obtained by taking discrete inverse Fourier transform of the frequency domain inner-product between data and waveform, sampled at $2048$Hz. A smooth function is obtained by interpolating using a cubic spline. 

The initial range of distances is chosen to ensure that the chosen range encloses the posterior mass. We also require the bounds to be as small as possible to prevent VARAHA from sampling in regions of parameter space with low likelihood. Returning to Eq.~\ref{eq:deff}, we see that the effective distance is always equal to or greater than the luminosity distance, with equality only for face on systems ($\cos \iota = \pm 1)$, lying either directly above or below the detector ($\sqrt{{F^{i}_{+}}^{2} + {F^{i}_{\times}}^{2}} = 1$).  We choose the maximum distance to be three times the effective distance in the reference detector:
\begin{equation}
    d_{L}^{\mathrm{max}} = 3 D^{i}_{\mathrm{eff}} = 3 \varrho^i(\boldsymbol{\theta_{o}}) / |\rho^i(\boldsymbol{\theta_{o}}, t_o)|
\end{equation}
as well as choosing a minimum distance of 0. At a distance, $d_{L}^{\mathrm{max}}$, the maximum possible likelihood in the reference detector occurs for a face-on, overhead system.  In that case, the log-likelihood is reduced by an amount $\tfrac{2}{9} |\rho^i(\boldsymbol{\theta_{o}}, t_o)|^{2}$.  At an \ac{SNR} of 7 in the reference detector, which corresponds to a relatively weak signal, this leads to a reduction in the log-likelihood of 11.  However, as with the discussion of the coalescence time, it is likely that there will also be a reduction in the likelihood in the other detectors, meaning this is a lower limit on the loss in the network likelihood. As often is the case, instead of the distance, we use a uniform prior on the volume,
\begin{equation}
    p(d_L) \propto d_{L}^2.
\end{equation}

We have verified the efficacy of our choices on the range of coalescence time and distance by performing parameter estimation runs on hundreds of simulated signals. 

\begin{figure*}
    \centering
    \includegraphics[width=.85\textwidth]{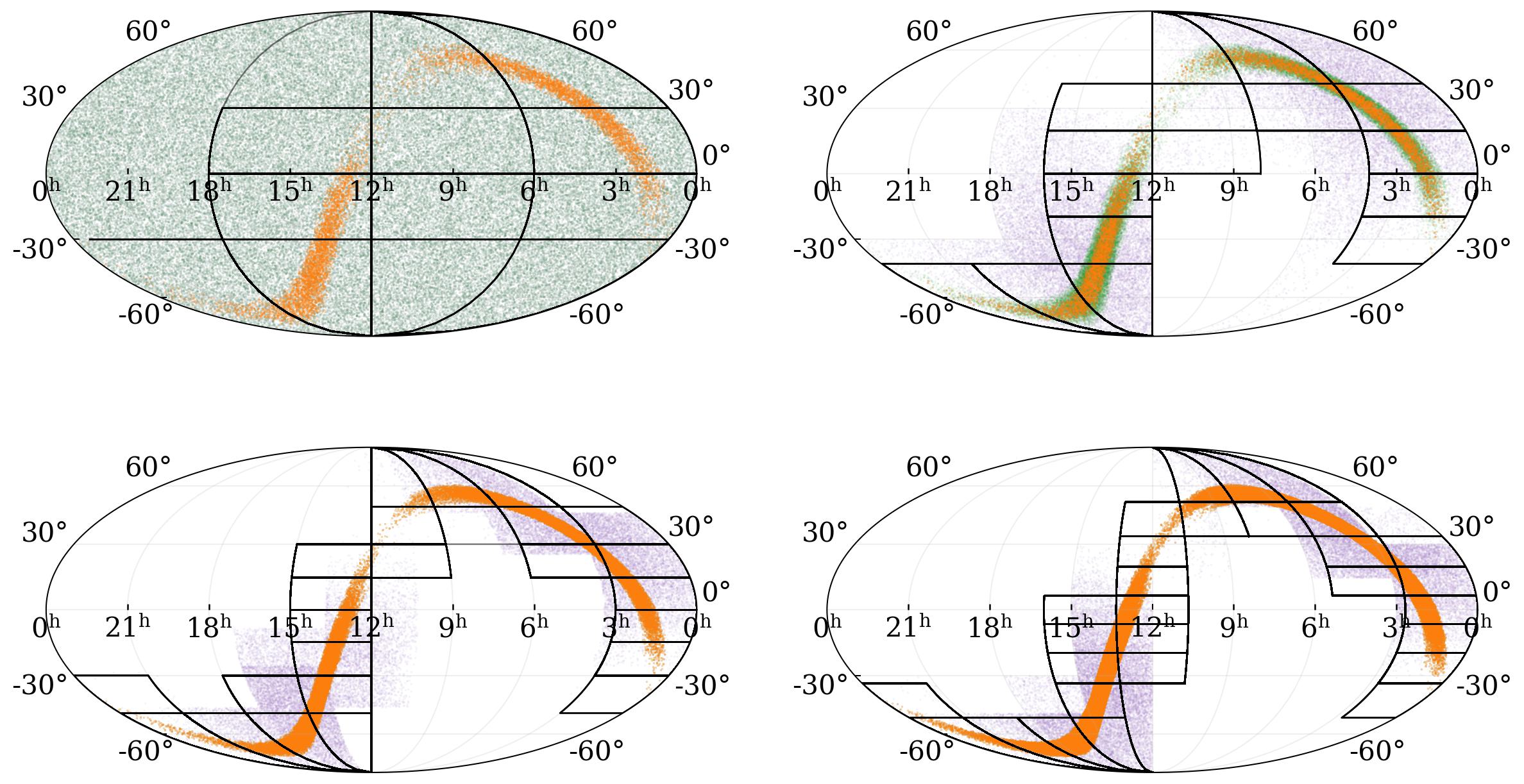}
    \caption{Points obtained when sampling over the extrinsic parameters for the observation GW151226. The top left panel shows the $1^\mathrm{st}$ cycle, the top right shows the $2^\mathrm{nd}$ cycle, the bottom left shows the $4^\mathrm{th}$ cycle and the bottom right shows the $8^\mathrm{th}$ cycle. The purple dots show the points randomly drawn from within the multidimensional grid, the green dots are all of the points that have a likelihood larger than the likelihood threshold from the previous cycle and the orange dots are the points with likelihoods larger than the likelihood threshold for the current cycle. The black lines show the multi-dimensional grid that surrounds the live volume. Eight cycles were completed in less than 45s using one CPU thread and accumulated around 10,000 effective samples.}
    \label{fig:extrinsic_gw151226_binevolution}
\end{figure*}

As an example, we estimate the extrinsic parameters for the signal GW151226 \cite{2016PhRvL.116x1103A}. We scatter $N_{\mathrm{pts}}=1,000,000$ points within the multi-dimensional grid for each cycle (Step 1 in Section~\ref{sec:method}), set $P_{\mathrm{thr}} = 0.9999$ when evaluating $\mathcal{L}_{\mathrm{thr}}$, and keep a minimum of $N_{\mathrm{min}}=8,000$ points at each cycle to evaluate $\mathcal{L}_{N_{\mathrm{min}}}$ (Step 2 in Section~\ref{sec:method}). We terminate sampling once 8 cycles have elapsed. Table~\ref{table:extrinsic_gw151226} shows the evolution of various quantities. The process collects $\sim 10,000$ effective samples from a total of $\sim 160,000$ weighted samples in 8 cycles.  As expected, for the first two cycles, the likelihood threshold is set by $\mathcal{L}_{N_{\mathrm{min}}}$ while for later cycles, where there is a higher probability of randomly scattering points within the high likelihood region, the threshold is set by $\mathcal{L}_{\mathrm{thr}}$. Unlike in previous examples, the likelihood threshold is positive and increases monotonically. This is because we have neglected the $\langle d | d\rangle$ term in Eq.~\ref{eq:gauss_lkl}.
Figure~\ref{fig:extrinsic_gw151226_binevolution} pictorially shows VARAHA converging to the most probable sky location of GW151226. We see that as the number of cycles increases, the live volume shrinks to the region of high likelihood. As a result, the number of bins in a multi-dimensional grid that encloses the live volume steadily increases with each cycle. This indicates that the error on the \ac{MC} volume is steadily decreasing over time. For this example, VARAHA localises GW151226 within 45 seconds on a single CPU thread. Assuming an approximately linear scaling with the number of CPU threads, we expect to localise most gravitational-wave signals in less than 5 seconds when parallelising over 10 CPU threads (see Section~\ref{sec:poptests} for a discussion about CPU scaling). The exact runtime of BAYESTAR is unknown for this case, but we expect that BAYESTAR completed in $\sim 2$ minutes when running on a single CPU thread (based on Fig.12 in~\cite{2016PhRvD..93b4013S}).

\begin{table}[t!]
\begin{ruledtabular}
\begin{tabular}{cccccc}
   Cycle & $N_b$ & log Likelihood threshold & $n_{\mathrm{eff}}$\\
   \hline
1 & 4 & 56.8 & 42 \\
2 & 6 & 69.4 & 257 \\
4 & 8 & 72.1 & 2671 \\
8 & 8 & 72.1 & 9751 \\
\end{tabular}
\end{ruledtabular}
\caption{Output from VARAHA showing the evolution of the number of bins in each dimension $N_{\mathrm{bins}}$, the likelihood threshold in each cycle (either $\mathcal{L}_{N_{\mathrm{min}}}$ or $\mathcal{L}_{\mathrm{thr}}$ depending on the situation, see text for details), and the number of effective samples $n_{\mathrm{eff}}$ when estimating the extrinsic parameters for GW151226.}
\label{table:extrinsic_gw151226}
\end{table}

\subsection{Intrinsic Parameters}
\label{subsec:intrinsic}

In order to obtain samples from the full posterior distribution, we need to vary both the intrinsic and extrinsic parameters in Equation~\ref{eq:loglkl}. However, as changing the extrinsic parameters only changes the overall amplitude, phase and time of arrival of the gravitational wave signal (since we are restricting to the leading multipole of a non-precessing system), samples can be obtained by combining independent and separated sampling over the extrinsic and intrinsic parameters. In order to achieve this, we require two likelihoods: one which is conditioned only on the extrinsic parameters (introduced in Section.~\ref{sec:extrinsic}), and another, marginalised likelihood, that is dependent only on the intrinsic parameters of the source. The marginalised likelihood that is dependent only on the intrinsic parameters is simply~\cite{2015PhRvD..92b3002P},
\begin{equation}
    \mathcal{L}_{\mathrm{intr}}(\vec{d} |\boldsymbol{\theta}) 
    = \int \mathrm{d}t_c\, \mathrm{d}\boldsymbol{\Omega} \, \mathcal{L}(\vec{d} | t_c, \boldsymbol{\Omega}, \boldsymbol{\theta})\, .
\label{eq:reduced_likelihood}
\end{equation}
This procedure breaks one high-dimensional problem into two smaller-dimensional problems and has two significant benefits. First, the computational requirement of sampling decreases with decreased dimensionality~\cite{10.1214/06-BA127} which is expected to reduce the overall cost. Second, an analysis sampling the full parameter space needs to generate a simulated gravitational waveform for each likelihood calculation, and compute the inner product of the waveform with the detector data. Often the computational cost to generate and filter these waveforms is high. By splitting the sampling, for each waveform generation and matched filtering operation, the likelihood over arbitrary values of extrinsic parameters can be calculated using inexpensive operations that change the overall amplitude, phase and time of the signal. 

To evaluate the marginalised likelihood in Equation~\ref{eq:reduced_likelihood}, we integrate over the extrinsic parameter space for each draw of the intrinsic parameters $\boldsymbol{\theta}$. We identify the high-likelihood region of the intrinsic parameter space by following the method outlined in Section~\ref{sec:method}, albeit with the likelihood calculation replaced by the more complex calculation of $\mathcal{L}_{\mathrm{intr}}$.  To begin, we define the region of interest in the intrinsic parameter space.  Here, our four parameters are the chirp mass $\mathcal{M}$, mass ratio $q$, and the spins of each black hole aligned (or anti-aligned) with the orbital angular momentum $\chi_1$ and $\chi_2$.
We set the initial range of the intrinsic parameter space to be
\begin{eqnarray}
\mathcal{M} &=& [\mathcal{M}_0 - \Delta \mathcal{M} \nonumber, \mathcal{M}_0 + \Delta \mathcal{M} \nonumber] \\
q &=& [0.05, 1.0] \nonumber \\
\chi_1 &=& [-0.9, 0.9] \nonumber \\
\chi_2 &=& [-0.9, 0.9] \, .
\label{intr_ranges}
\end{eqnarray}
The ranges for mass ratio and spins are primarily driven by the region of validity of the IMRPhenomD waveform model~\cite{2016PhRvD..93d4006H,phenomd2}, as well as the physical restriction that $q \le 1$, $| \chi | \le 1$.  The central value of the chirp mass range, $\mathcal{M}_o$, is the chirp mass of the \ac{GW} search template that identified the signal.  The width $\Delta\mathcal{M}$ is chosen as,
\begin{equation}\label{eq:chirp_mass_width}
    \Delta \mathcal{M} = \mathrm{min}(1.2 \times 10 ^ {-3} \left(\frac{10}{\rho_0}\right)\mathcal{M}_0^{8/3},\; \mathcal{M}_o^{1.1}/20),
\end{equation}
where $\rho_0$ is the reported \ac{SNR}. The  first term in Eq.~\ref{eq:chirp_mass_width} is motivated by the expected accuracy of measurements of the chirp mass for low-mass signals where the inspiral is the dominant part of the signal observed in the detectors\citep{1994PhRvD..49.2658C}.  The second term in Eq.~\ref{eq:chirp_mass_width} is taken from empirical uncertainties of chirp mass measurements from GWTC-3 \citep{o3b_cat} and is conservatively broad to ensure that the range is broader than the observed distribution.  We again note that the mass values are in the frame of the detector, thus $\mathcal{M} = (1 + z) \mathcal{M}_{\mathrm{source}}$.

For each draw of intrinsic parameters $\boldsymbol{\theta}_{j}$, we marginalise the likelihood by integrating it over a fiducial parameter space for the extrinsic parameters, $\boldsymbol{\Omega}_{o}$. To generate $\boldsymbol{\Omega}_{o}$, we make use of the samples generated in the extrinsic parameter space associated with the intrinsic parameters identified by the search, $\boldsymbol{\theta}_{o}$.  In general, we expect there to be a minimal correlation between the masses and spins and several of the extrinsic parameters.  As discussed in \cite{Fairhurst:2009tc, 2016PhRvD..93b4013S}, while changing the masses and spins will impact the measured coalescence time in each detector, the relative time delay will be only weakly impacted and consequently, the inferred sky location will be largely independent of the intrinsic parameters.  Similarly, the orientation of the binary, encoded in the inclination $\iota$ primarily depends upon the observed ratio of power in the two gravitational wave polarizations and this is unlikely to change significantly with mass or spin.  Finally, although the intrinsic amplitude of a gravitational wave signal does vary linearly with mass, the chirp mass width is constrained to be at most a few per cent of the measured value resulting in the inferred distance varying a few per cent with changes in mass. Thus, the overall change in the volume that contains the high likelihood region in the extrinsic parameter space only varies a few per cent with any change in the intrinsic parameters. To accommodate these fractional changes, we use $\boldsymbol{\theta}_o$ and specify a lower $P_{\mathrm{thr}}$ for the intrinsic parameter space than what is desired for the extrinsic parameters. This means that VARAHA defines each volume in the intrinsic parameter space to accommodate a slightly smaller probability than what is desired for the extrinsic parameters. For instance, if $P_{\mathrm{thr}} = 0.999$ is specified for the extrinsic parameters, VARAHA uses a $P_{\mathrm{thr}} = 0.995$ for the intrinsic parameters. This ad-hoc choice results in the recovery of sane posteriors even at a population level. However, we will ascribe a more rigorous treatment of this problem in a future presentation.

We generate $\boldsymbol{\Omega}_o$ by retaining samples of $(\alpha, \delta, \iota, d_L)$ that cross the likelihood threshold from the extrinsic-only analysis and augment it with samples from the remaining three parameters: $(\psi, \phi_c, t_{c})$ as defined later. For each draw in the intrinsic parameter space, we evaluate Eq.~\ref{eq:reduced_likelihood} by numerically integrating over $\boldsymbol{\Omega}_o$.%
\footnote{Note that the prior remains uniform in this procedure. The samples corresponding to $\boldsymbol{\Omega}_o$ are uniformly distributed.}
The observed gravitational wave phase and coalescence time will vary significantly with the masses and spins. Thus, we draw $(\phi_c, \psi)$ from the full range $(0, 2 \pi)$. In addition, due to the degeneracy between the coalescence phase and the polarisation angle, retaining samples of the latter from the extrinsic-only analysis or regenerating them does not impact the posterior on the intrinsic parameters.  The appropriate range for the coalescence time $t_{c}$ can be derived using the same method as for the initial point $\boldsymbol{\theta}_{o}$, as described around Eq.~\ref{eq:tc_bound}, although using the peak \ac{SNR} in the reference detector for the intrinsic parameters $\boldsymbol{\theta}_{j}$. For each sample in $\boldsymbol{\Omega}_o^i$ we pre-calculate $D^{i}_{\mathrm{eff}}$ and $\phi_{\boldsymbol{\Omega}}^i$. Each draw of $\boldsymbol{\theta}_j$ requires waveform generation, matched filtering of the data and calculation of likelihood using these pre-calculated values. Finally, the (marginalised) likelihood is obtained by approximating the integral in Eq.~\ref{eq:reduced_likelihood} by writing,

\begin{equation}
    \mathcal{L}_{\mathrm{intr}}(\vec{d} |\boldsymbol{\theta}_j) \approx \sum_i \sum_{k \in \mathrm{dets}}\mathcal{L}\left(d_k | t_c, \boldsymbol{\Omega}_o^i, \boldsymbol{\theta}_j\right).
    \label{eq:reduced_likelihood_markov}
\end{equation}
We continue sampling using different draws of $\boldsymbol{\theta}_j$ and follow our sampling procedure guided by their marginalised likelihood values.

Calculating the marginal likelihood is computationally expensive. An approximate but optimistic value can be calculated by maximising the likelihood independently over the time of arrival and phase in each detector. This is done by ignoring the phase term in Eq.~\ref{eq:reduced_likelihood} and using the maximum \ac{SNR} in each detector independently.  By ignoring both of these factors, we obtain a likelihood which will always be equal to or greater than the true likelihood. 
\begin{widetext}
    \begin{align}
    \label{eq:loglkl_optimistic}
        \mathcal{L}_{\mathrm{intr}}(\vec{d} |\boldsymbol{\theta}_j)\nonumber
        & \leq \sum_i\exp\Big(\sum_{k \in \mathrm{dets}} \Big[
        -\frac{1}{2} \frac{\varrho_{o}^k(\boldsymbol{\theta}_{j})^2}{(D^{i}_{\mathrm{eff}})^{2}} 
        + \frac{\varrho_{o}^k(\boldsymbol{\theta}_{j})}
        {D^{i}_{\mathrm{eff}}}
        \mathrm{max}_{t}    |\rho^k(\boldsymbol{\theta}_j, t)|
        \Big]\Big) \, \nonumber \\
        & \leq \exp\left(\sum_{k \in \mathrm{dets}} \frac{1}{2} \left(\mathrm{max}_{t}    |\rho^k(\boldsymbol{\theta}_j, t)|\right)^2\right).
    \end{align}
\end{widetext}
The last line in Eq.~\ref{eq:loglkl_optimistic} further maximises the likelihood on all the extrinsic parameters. The benefit is that the term in the second line can be calculated after matched filtering the data, in addition, for the term in the first line we only need to generate $D^i_{\mathrm{eff}}$ for each of the intrinsic samples. Both these calculations require significantly less computation than the full likelihood calculation. We thus marginalise only if both of these values are larger than the likelihood threshold.

The marginalised likelihood $\mathcal{L}_{\mathrm{intr}}(\vec{d} |\boldsymbol{\theta}_{j})$ assign the weight for each point in the intrinsic parameter space is
\begin{equation}
    w_{j} = \frac{\mathcal{L}_{\mathrm{intr}}(\vec{d} |\boldsymbol{\theta}_{j})}{\mathcal{L}_{\mathrm{max}}(j)}
\end{equation}
where $\mathcal{L}_{\mathrm{max}}(j)$ is the maximum marginalised likelihood value among all the samples.
In addition, we obtain a weight $w_j^{i}$ for each sample $\boldsymbol{\Omega}^{i}_o$ corresponding to the intrinsic parameters $\boldsymbol{\theta}_{j}$, which is simply
\begin{equation}
w_{i}^{j} = \frac{\mathcal{L}\left(\vec{d}\, | \boldsymbol{\theta}_i, \,t_c {}_i^j, \boldsymbol{\Omega}_i^j\right)}
{\mathcal{L}_{\mathrm{max}}(i,j)} \,,
    \label{eq:rejsamp_extr}
\end{equation}
where $\mathcal{L}_{\mathrm{max}}(i,j)$ is the maximum likelihood of all the samples in the fiducial volume estimated for each sample of the intrinsic parameters.

As we are interested in generating samples which correspond to the full eleven-dimensional parameter space of intrinsic and extrinsic parameters combined.  This presents a challenge in generating and storing the samples.
We circumvent this problem by randomly keeping only one set of extrinsic parameters for each set of intrinsic parameters and discarding the rest. This choice is made by performing rejection sampling on the weights $w_{i}^{j}$ to select a single $\boldsymbol{\Omega}_o^{i}$ and $t_{j}^{i}$ for each $\boldsymbol{\theta}_j$.  Thus for each point in the intrinsic parameter space that we sample, we store a single sample $(\boldsymbol{\theta}_{j}, \boldsymbol{\Omega}^{i}_o, t^{i}_{j})$ and the weight,$w_{j}$, associated with the intrinsic parameter, $\boldsymbol{\theta}_j$. 

We are now left with the sampling of intrinsic parameters and the associated marginalised likelihoods, in the successive cycles we identify live volumes across 4 dimensions $(\mathcal{M}, q, \chi_1, \chi_2)$. For each cycle, we evaluate the marginalised likelihood values and continue the cycles until we obtain the desired number of effective samples 
\begin{equation}
    N_{\textrm{int}}^{\textrm{eff}} = \frac{(\sum{w_j})^2}{\sum w_j^2}.
\end{equation}

\begin{figure*}
    \centering
    \includegraphics[width=.95\textwidth]{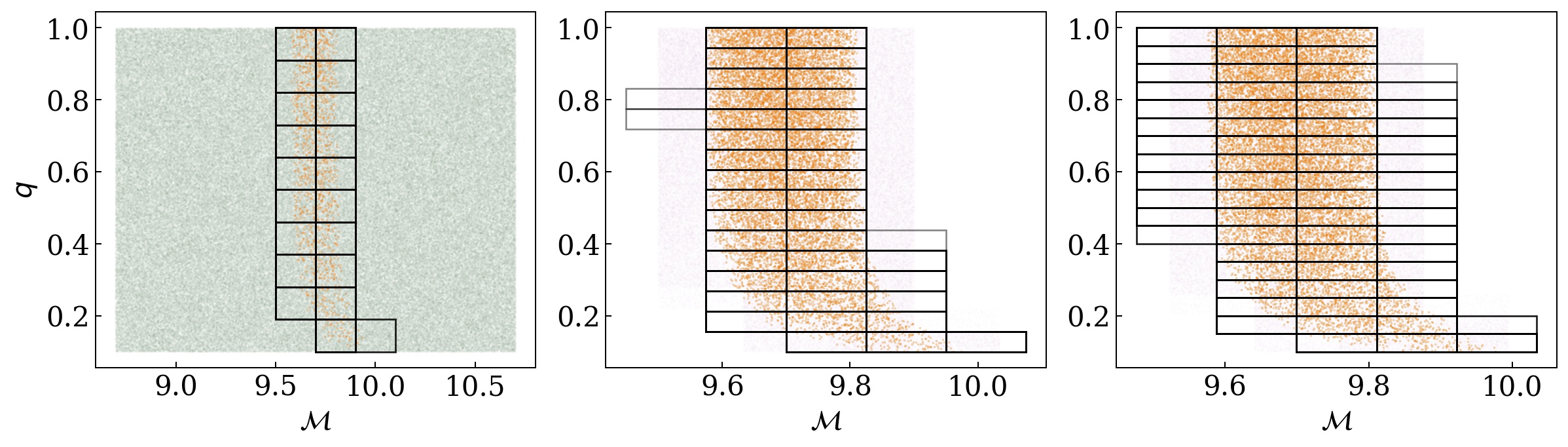}
        \caption{The 2D projection of the evolution in the $\mathcal{M}-q$ plane for the analysis sampling the intrinsic parameters for the observation
GW151226. The left panel shows the $1^\mathrm{st}$ cycle, the middle panel shows the $4^\mathrm{th}$ cycle and the right panel shows the $8^\mathrm{th}$ cycle. The purple dots show the points randomly drawn from within the multi-dimensional grid, the green dots are all of the points that have a likelihood larger than the likelihood threshold from the previous cycle and the orange dots are the points with likelihoods larger than the current likelihood threshold. The black lines show the multi-dimensional grid that surrounds the live volume with likelihood equal to the likelihood threshold.}
    \label{fig:bin_evoltion_intrinsic_gw151226}
\end{figure*}

We continue sampling intrinsic parameters for GW151226. We scatter points within the multi-dimensional grid for each cycle (Step 1 in Section~\ref{sec:method}), set $P_{\mathrm{thr}} = 0.9999$ when evaluating $\mathcal{L}_{\mathrm{thr}}$, and keep a minimum of $N_{\mathrm{min}}=1,000$ points at each cycle to evaluate $\mathcal{L}_{N_{\mathrm{min}}}$ (Step 2 in Section~\ref{sec:method}).\footnote{As we do not accurately compute the likelihood for all the intrinsic samples~(see Eq.~\ref{eq:loglkl_optimistic}) we sample until the number of live points increase by $N_{\mathrm{min}}$ and count $N_{\mathrm{pts}}$ accordingly.} We terminate sampling once 8 cycles have been completed. Table~\ref{table:intrinsic_gw151226} lists the number of bins in the multi-dimensional histogram, the likelihood threshold and the number of effective samples over the cycles. Like previous examples, the likelihood threshold increases initially before reaching a final value. Figure~\ref{fig:bin_evoltion_intrinsic_gw151226} shows the evolution of the live volume in the $\mathcal{M}-q$ plane with the increase in cycle number.
\begin{table}[t!]
\begin{ruledtabular}
\begin{tabular}{cccccc}
   Cycle & $N_b$ & log Likelihood threshold & $n_{\mathrm{eff}}$\\
   \hline
1 & 8 & 65.6 & 133 \\
2 & 11 & 71.4 & 940 \\
4 & 13 & 71.4 & 3420 \\
8 & 15 & 71.4 & 9070 \\
\end{tabular}
\end{ruledtabular}
\caption{Output from VARAHA showing the evolution of the number of bins in each dimension $N_{\mathrm{bins}}$, the likelihood threshold in each cycle (either $\mathcal{L}_{N_{\mathrm{min}}}$ or $\mathcal{L}_{\mathrm{thr}}$ depending on the situation, see text for details), and the number of effective samples $n_{\mathrm{eff}}$. Here we estimate the intrinsic parameters for the observation of GW151226.}
\label{table:intrinsic_gw151226}
\end{table}

\subsection{Implementation}
\label{sec:gw_implementation}

To summarise, VARAHA analyses gravitational-wave signals as follows,

\begin{enumerate}

\item{\textbf{Obtain posterior on the extrinsic parameters}: Obtain the posterior distribution for the extrinsic parameters by following the steps detailed in Sec.~\ref{sec:implementation}. The full gravitational-wave likelihood is used but we fix the intrinsic parameters to the values reported by the \ac{GW} search pipelines.}

\item{\textbf{Construct a fiducial volume for the extrinsic parameters}: Retrieve luminosity distance, inclination angle, right ascension, and declination and augment with the remaining extrinsic parameters as defined in Sec.~\ref{subsec:intrinsic}.}

\item{\textbf{Obtain posterior on the intrinsic parameters}: Marginalise the full likelihood on the fiducial volume calculated in Step 2, and obtain the posterior distribution for the intrinsic parameters by following steps detailed in Sec.~\ref{sec:implementation}. We now use the marginalised likelihood given in Eq.~\ref{eq:reduced_likelihood_markov} and a smaller $P_{\mathrm{thr}}$ than in Step 1. The fiducial volume based on the extrinsic parameters remains fixed.}

\end{enumerate}

\subsection{Processing Time}

The faster processing times of this analysis are due to the following reasons
\begin{enumerate}
    \item A large dimensional problem has been broken into two small dimensional problems resulting in reduced computational requirement.
    \item The sampling method is entirely likelihood driven and swiftly converges to the relevant region of the parameter space.
    \item Taking inverse Fourier transform of Equation \ref{eq:mf} allows constraining bounds on $t_c$ and vectorised estimation of likelihood values at thousands of samples in the fiducial set of extrinsic parameters.
    \item The waveform morphology of the inferred templates is expected to be similar. This analysis does not match filter if the phase accumulated in the detector's sensitivity band~($\sim$ 20Hz--2000Hz) by the fiducial waveform and a template waveform differs by more than 30 radians~(approximately five cycles).  
    \item This analysis does not marginalise over extrinsic parameters if an approximate but optimistic estimate of marginalised likelihood given in Eq.~\ref{eq:loglkl_optimistic} is smaller than the likelihood threshold. 
    \item Analysis has been rigorously optimised and performs a judicious vectorised operation to save on computation times.
\end{enumerate}

\section{Results}
\label{sec:results}
\subsection{Example: GW151226}
\label{gw151226}

\begin{figure*}
    \includegraphics[width=.37\linewidth]{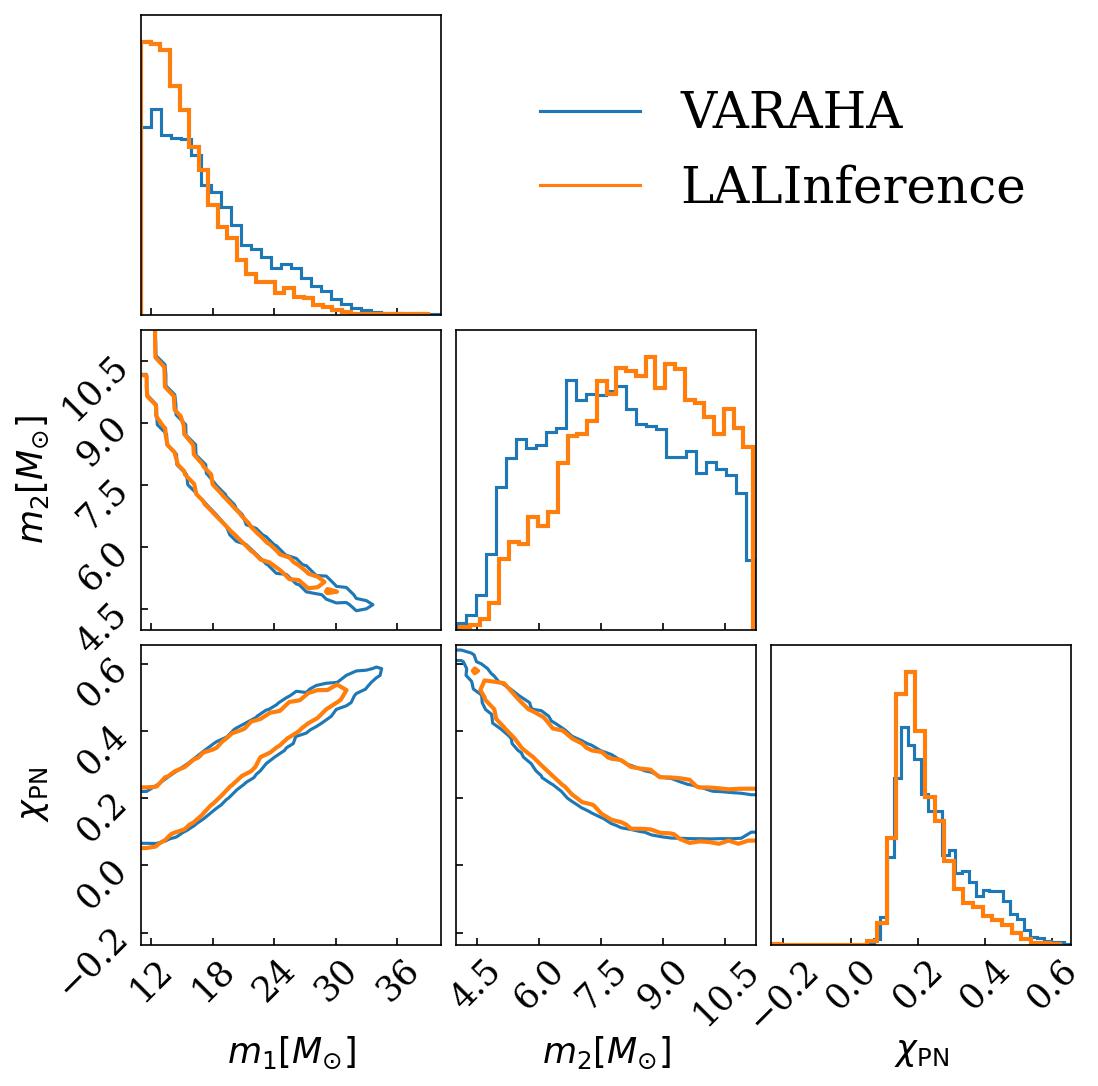}
    \includegraphics[width=.55\linewidth]{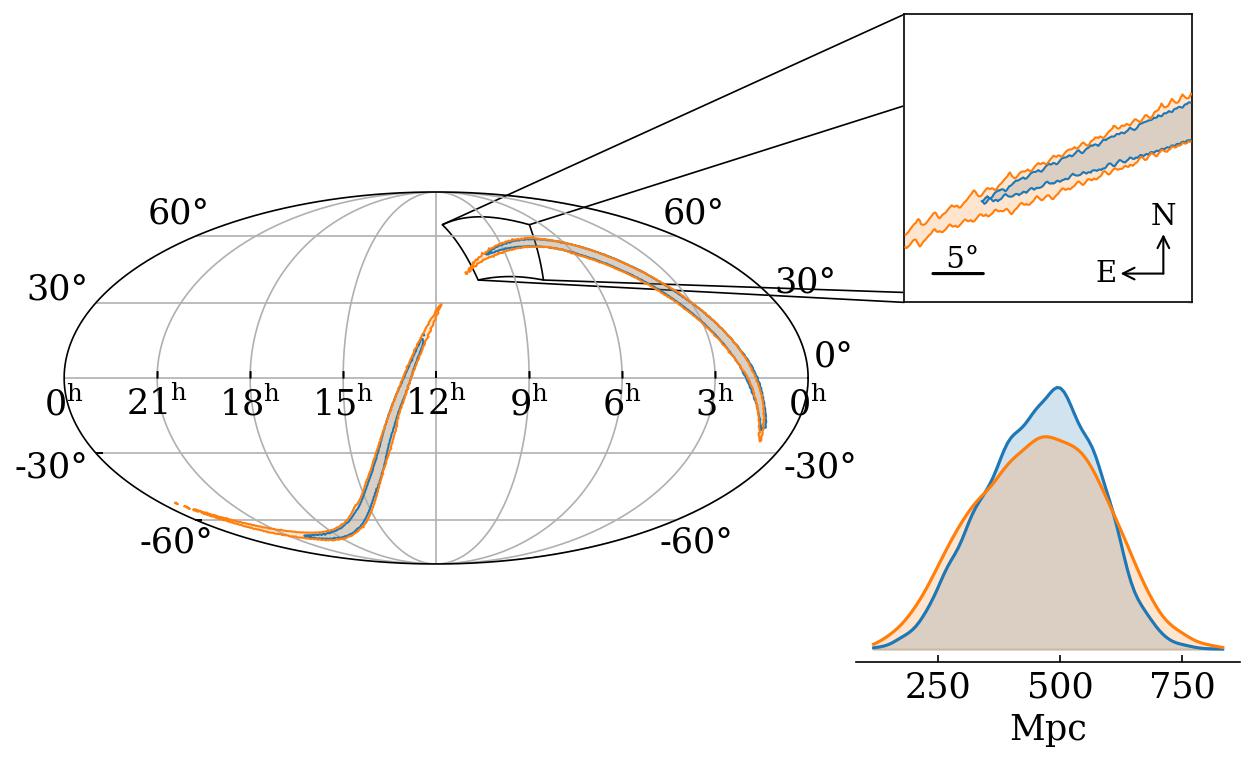}
    \caption{Plot comparing the posterior distributions obtained from VARAHA (blue) and LALInference (orange) when analysing GW151226. The left panel contains a corner plot for the primary and secondary masses as well as $\chi_{\mathrm{PN}}$(Equation 5 in \citep{2013PhRvD..87b4035B}). The right panel shows the most probable sky location of GW151226 as well as the inferred distance. In both panels, contours enclose 90\% probability mass.}
    \label{plt:lal_VARAHA_gw151226}
\end{figure*}

Figure~\ref{plt:lal_VARAHA_gw151226} compares the VARAHA's posterior with the posterior obtained using LALInference. Both analyses use IMRPhenomD \citep{2016PhRvD..93d4006H,phenomd2} for waveform generation and use almost equivalent priors on masses and spins. LALInference allows priors on the component masses, while VARAHA uses uniform priors on the chirp mass and mass ratio. We used a wide prior for the component masses in LALInference and then applied a chirp mass constraint to produce almost equivalent priors between the two algorithms. There is a good agreement in the two results; there are small differences in the marginalised one-dimensional posteriors, but they are consistent at the 90\% confidence level. This difference is likely a result of the slightly different priors assumed between the two codes. We also see good agreement in the recovered sky map and distance estimate, with any deviations likely a consequence of LALInference marginalising over the calibration uncertainty while VARAHA does not yet have the functionality to do so.
VARAHA obtained the posterior in less than one CPU hour. Based on the experience gained while running the two codes on different signals, we expect more than two orders of magnitude shorter computation times for VARAHA.

\subsection{Example: GW170817}
\label{gw170817}

VARAHA can rapidly estimate the origin of the observed gravitational wave. This can have significant implications on electromagnetic follow-up programs for \ac{BNS} observations. To demonstrate this, we analyse GW170817 using data from all three detectors and compare the estimated skymap with the location of the known host galaxy: NGC 4993~\cite{LIGOScientific:2017vwq}.

\begin{figure}[t!]
    \centering
    \includegraphics[width=.46\textwidth]{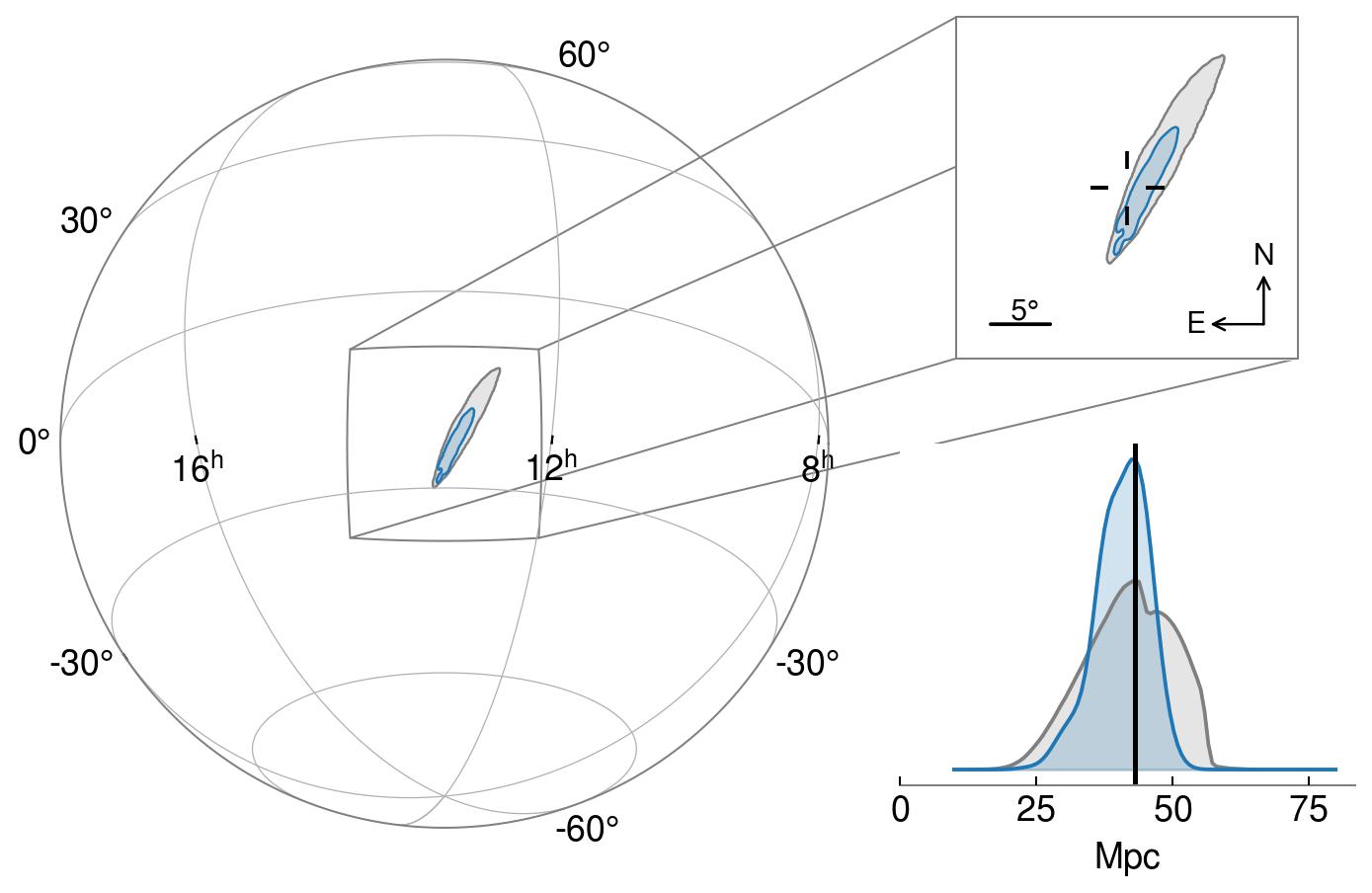}
    \caption{The most probable sky location of GW170817 when VARAHA samples only the extrinsic parameters (grey) and intrinsic plus extrinsic parameters (blue). The reticle marks in the top-right inset show the position of NGC 4993. The bottom-right panel shows the posterior distribution for the luminosity distance and the black vertical line shows the distance to NGC 4993. The contours show the 90\% confidence interval. The extrinsic-only analysis completed 8 cycles in less than 2 minutes using one CPU thread. The intrinsic plus extrinsic analysis was completed in 2 hours using 8 CPU threads.}
    \label{fig:gw170817_skymap}
\end{figure}

Fig.~\ref{fig:gw170817_skymap} shows two skymaps: the skymap produced when VARAHA samples only the extrinsic parameters, and the skymap produced when VARAHA samples both the intrinsic and extrinsic parameters. We see that within 2 CPU minutes, VARAHA is able to localise GW170817 to within 49 square degrees (at 90\% confidence) when sampling over only the extrinsic parameters. For this analysis, VARAHA used $N_{\mathrm{pts}} = 1,000,000$, $N_{\mathrm{min}} = 8,000$, $P_{\mathrm{thr}} = 0.999$ and we stopped sampling once 8 cycles were completed. The localisation area was reduced to 17 square degrees (at 90\% confidence) after 16 CPU hours when VARAHA samples over the intrinsic and extrinsic parameters. For this more detailed analysis, VARAHA used $N_{\mathrm{min}} = 1,000$, $P_{\mathrm{thr}} = 0.995$ and we terminated sampling once 8 cycles were completed. For comparison, BAYESTAR localises GW170817 to within 31 square degrees (at 90\% confidence) when analysing only the extrinsic parameters, and LALInference localises GW170817 to within 23 square degrees (at 90\% confidence)~\cite{LIGOScientific:2017vwq} when analysing both intrinsic and extrinsic parameters. Although the exact runtimes of BAYESTAR and LALInference are unknown for this case, we expect that BAYESTAR completed in $\sim 2$ CPU minutes (based on Fig.12 in Ref.~\cite{2016PhRvD..93b4013S}) and LALInference completed in $\sim 500$ CPU hours (based on Ref.~\cite{2015ApJ...804..114B})\footnote{The estimated runtimes of BAYESTAR and LALInference are based on results generated with, potentially, older CPUs than those used by VARAHA. Running on identical CPUs may decrease the expected runtime of BAYESTAR and LALInference. For the latest BAYESTAR runtimes, see \href{https://lscsoft.docs.ligo.org/ligo.skymap/performance.html}{https://lscsoft.docs.ligo.org/ligo.skymap/performance.html}}. Consequently, VARAHA matches the performance of BAYESTAR when sampling over only the extrinsic parameters, but, importantly, significantly improves upon LALInference when sampling over the intrinsic and extrinsic parameters. We note, however, that there has been recent work to reduce the runtime of LALInference by utilising reduced order quadrature models~\cite{Morisaki:2020oqk}, as well as using meshfree approximations~\cite{Pathak:2022ktt}. Since the inclusion of intrinsic parameters is preferred, as it reduces the 90\% contour for most cases, VARAHA may be pivotal for the rapid follow-up of binary neutron star observations.

\subsection{Population Level Test}
\label{sec:poptests}

\begin{figure}[t!]
    \centering
    \includegraphics[width=0.45\textwidth]{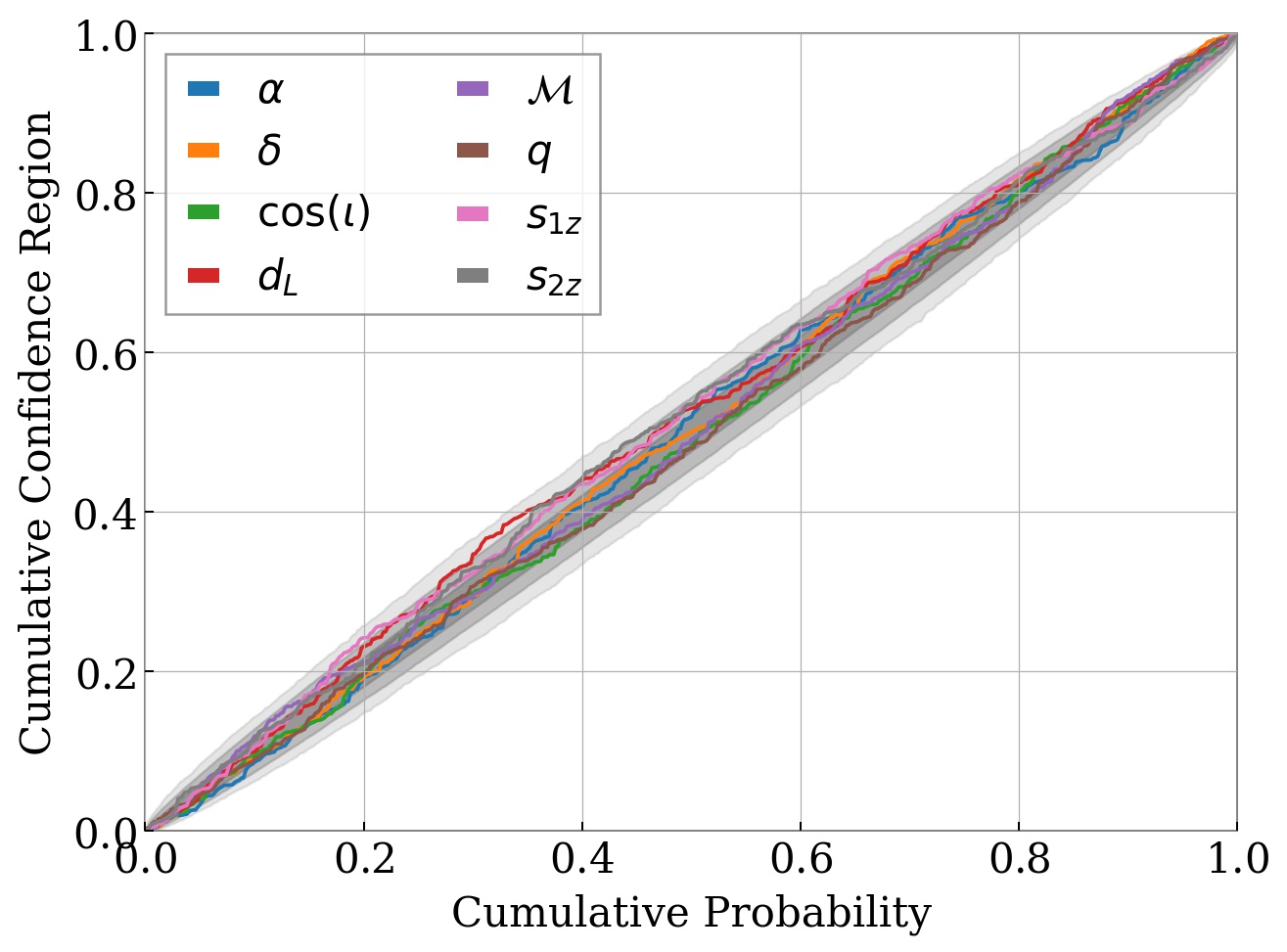}
    \caption{Percentile-Percentile (P-P) plot for 500 simulated injections. The 1-, 2- and 3-$\sigma$ confidence intervals are indicated by the grey bands. For results to be unbiased the trails are required to be enclosed by the bands~\cite{2014PhRvD..89h4060S}.}
    \label{fig:pp}
\end{figure}

We evaluate the population level accuracy of VARAHA by performing a Percentile-Percentile (P-P) test \cite{2014PhRvD..89h4060S}; this test involves performing hundreds of parameter estimation runs on synthetic signals embedded in simulated detector noise.  The P-P test investigates if the measured interval of parameters at a credibility $f\%$ also encloses $f\%$ of true values among all the measurements. We perform parameter estimation on 500 simulated signals and show the P-P plot in Figure~\ref{fig:pp}. The parameters of the synthetic signals are drawn from VARAHA's prior, and we only analyse signals that cross a chosen \ac{SNR} threshold. As described Section~\ref{sec:gw_implementation}, we first estimate the extrinsic parameters. We do this by fixing the intrinsic parameters to the true values used when generating the synthetic signals. This is a reasonable choice as we do not expect the detector noise to bias the measurement, and the inferred population should average out to the true population. We then construct the fiducial volume for the extrinsic parameters and use it to estimate the marginalised likelihood for sampling the intrinsic parameters, as well as, obtaining the 11-dimensional posterior on the full parameter space, as described in Sec.~\ref{subsec:intrinsic}.

The distribution of injection parameters is listed in Table~\ref{extr_ranges} and Table~\ref{intr_ranges}. The luminosity distance is uniform in volume and chirp mass is uniformly distributed between 10$M_\odot$ and 20$M_\odot$. Any injection that crosses a matched filter network \ac{SNR} of 10 is selected for estimating the parameters. In this analysis, the network is comprised of advanced LIGO Livingston/Hanford and the advanced Virgo detector \cite{2015CQGra..32g4001L, 2015CQGra..32b4001A}.

\begin{figure}[t!]
    \begin{center}
    \includegraphics[width=0.49\textwidth]{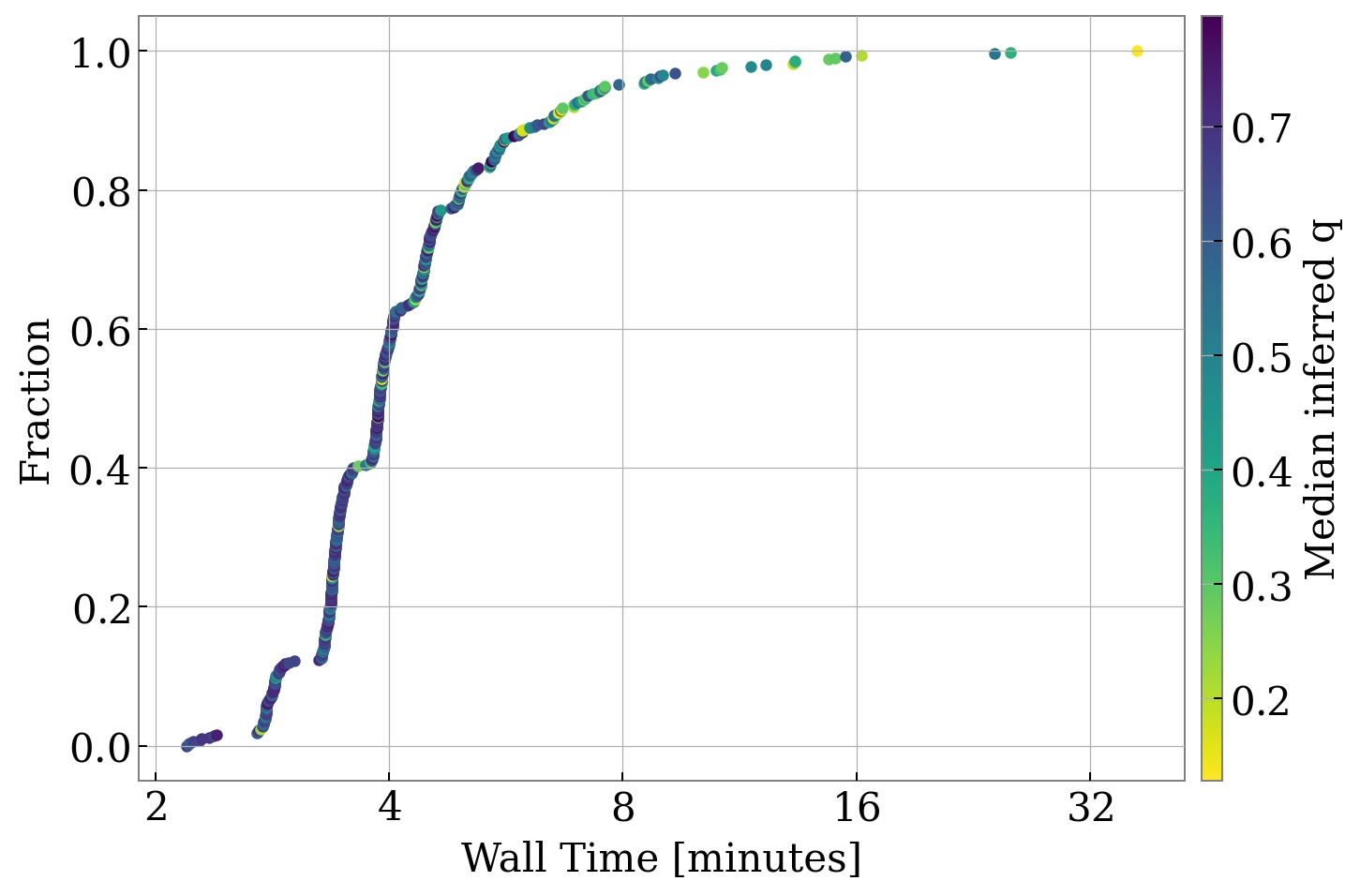}
    \caption{Plot showing the wall time taken to perform \ac{PE} on 500 synthetic signals embedded in simulated detector noise (see Fig.~\ref{fig:pp}). Each of the 500 individual \ac{PE} runs used 10 CPU threads and each point is colour coded by the median of the inferred mass ratio distribution. The median wall time is 4 minutes.}
    \label{fig:computation_time}
    \end{center}
\end{figure}

Most of the injections required 8 seconds of simulated noise to accommodate the duration of simulated signals last in the detector's sensitivity band ($\sim$20-2000Hz). Figure~\ref{fig:computation_time} shows the time taken by the analysis in performing \ac{PE}. The median time needed was less than 4 minutes using 10 threads. Almost all the \ac{PE} runs were completed in less than sixteen minutes. We see that, in general, VARAHA takes longer to analyse binaries with more assymetric masses, with the longest run time of 40 minutes arising from a binary with mass ratio $q=0.1$. Waveform generation and matched filtering consumed around 60\% of the time and calculation of the reduced likelihood consumed around 30\% of the time. 

\begin{figure}[t!]
    \begin{center}
    \includegraphics[width=0.49\textwidth]{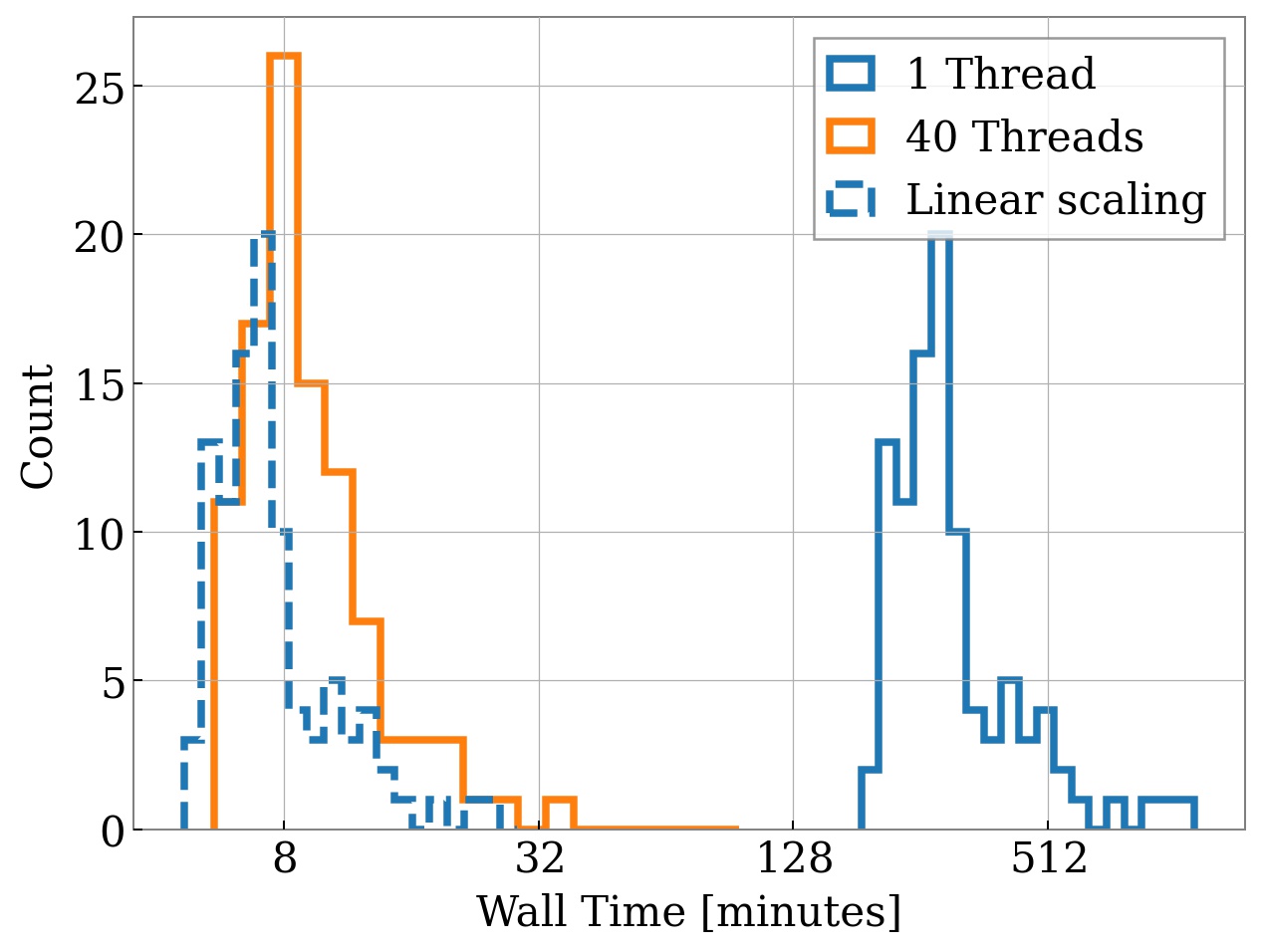}
    \caption{Plot showing the wall time taken to perform \ac{PE} on 100 synthetic signals embedded in simulated detector noise when the likelihood calculation is made expensive by including a delay in the waveform generation. The wall time taken when using a single CPU thread is shown in blue and 40 CPU threads in orange. Both analyses used identical settings. Assuming a linear CPU scaling from the 1 CPU thread analysis, the expected time taken to analyse 100 \ac{PE} runs on 40 CPU threads is shown by the blue dashed histogram. In reality, increasing the number of CPU threads to 40 reduces the computation time by a factor of 33.}
    \label{fig:computation_time_expensive}
    \end{center}
\end{figure}

In Figure~\ref{fig:computation_time_expensive}, we show the scalability of the analysis with the number of CPU threads. We perform two additional sets of runs each using an expensive likelihood calculation and performed using 1 and 40 CPU threads respectively. We make the likelihood calculation expensive, by including a one-tenth of a second delay in waveform generation, to reflect what we expect the scalability to be when VARAHA is extended to include additional physics (for example precession,  higher order multipoles and eccentricity) since these waveform models are more expensive to generate than the simple aligned-spin case. The median time when using 40 CPU threads was 508 seconds while the median time when using 1 CPU thread was 16910 seconds. Increasing the number of threads by a factor of 40 reduced the analysis time by a factor of 33. We expect this scaling to improve if the likelihood calculation is made more expensive.

\section{Discussion}

In this article, we introduced a new sampling method that estimates Bayesian posteriors by identifying the volume that encloses the posterior mass and calculating the likelihood values inside the identified volume. This approach significantly increases our ability to parallelise the analysis over multiple CPU threads and increases the efficacy to use vectorised likelihood calculations. In addition, we introduced the use of a likelihood threshold to judiciously populate the parts of the parameter space based on our prior understanding of the distribution at hand. Compared to Nested sampling, which focuses directly on the estimation of probability mass, this approach is less robust in estimating higher-dimensional multi-modal distributions. 

Our sampling method is ideally suited for estimating parameters of the compact binaries from their \ac{GW} signals. These parameters are inherently assumed to follow a uni-modal likelihood distribution. We show that a large one-dimensional sampling problem can be broken into two small-dimensional sampling problems. Using vectorised likelihood calculations, we first sample the extrinsic parameters and subsequently obtain posteriors on the intrinsic parameters. We employ several approximations of the likelihood functions to draw boundaries in the parameter space and calculate likelihood values in the parts that meaningfully contribute to the posterior distribution. Tests indicate that our analysis can estimate parameters for most of the \ac{BBH} signals in a few minutes.

There is significant scope to further decrease the computation time. The most expensive component of the analysis is waveform generation and matched filtering. Both of these can be significantly reduced by using existing proposals~\cite{2010arXiv1007.4820C, 2018arXiv180608792Z, 2023arXiv230205333W}. We reconstruct the volume that contains the probability mass by using a structured multi-dimensional grid. It results in the reconstructed volume being much larger than what is estimated from the \ac{MC} integral. A more efficient reconstruction can employ the use of unstructured grids reducing the number of cycles needed to obtain an effective sample size. Many of the calculation is done on the fly (time delay between detectors and antenna patterns) and can be pre-calculated to save computation time. 

\section{Conclusion}

The presented analysis offers significant improvement in processing time for estimating the parameters of a CBC while producing results comparable with the contemporary samplers. Although VARAHA is currently restricted to using only aligned-spin waveform models, it is trivial to extend this algorithm to include additional physics, such as precession, eccentricity or tidal deformability, without hindering performance. However, for most cases, we expect that \ac{GW} signals can be accurately modelled as produced from aligned spin binaries, since the degree of orbital precession is often difficult to measure~\cite[see e.g.][]{Green:2020ptm}. This means that often, the posterior distribution simply recovers the prior~\citep{2012PhRvD..86j4063S, 2019arXiv190805707F,Hoy:2021dqg}. Of course, for binaries that precess~\cite{2022Natur.610..652H}, this means that some information is lost.

We use a uniform prior on all the parameters but, as we calculate the likelihood for each of our samples, it is straightforward to re-calculate these likelihood values with a new prior by simply dividing the probability density of the new prior with the old one and multiplying it to the likelihood value \citep{2018ApJ...868..140T}. As we can also calculate the marginal likelihood the samples can naturally be used for model selection. Furthermore, owing to the reduced computational time compared to other samplers, VARAHA is a natural choice for data diagnostics in understanding the systematics or non-Gaussianity in data associated with a signal, as well as performing parameter estimation on gravitational-wave data collected by the Laser Interferometer Space Antenna (LISA)~\cite{bender1998lisa} and third generation detectors~\cite{Punturo:2010zz, reitze2019cosmic} where \ac{PE} is likely to be slow due to the large observation times.

This code is highly parallalisable as individual threads do not communicate. It also does not have to address any problem that a usual \ac{MCMC} encounters. This may include proper mixing of chains, tuning of the code and potential correlations due to low proposal acceptance rate.

\section{Acknowledgements}

We thank Christopher Berry, Mark Hannam and Vivien Raymond for constructive comments on this work. VT and SF thank STFC for support through the grants ST/V005618/1 and ST/N005430/1. 
CH thanks the UKRI Future Leaders Fellowship for support through the grant MR/T01881X/1. The authors are also grateful for computational resources provided by Cardiff University and LIGO Laboratory and supported by STFC grant ST/N000064/1 and National Science Foundation Grants PHY-0757058 and PHY-0823459.

This research made use of data, software and/or web tools obtained from the Gravitational Wave Open Science Center (\href{https://www.gw-openscience.org}{https://www.gw-openscience.org}), a service of LIGO Laboratory, the LIGO Scientific Collaboration and the Virgo Collaboration. LIGO is funded by the U.S. National Science Foundation. Virgo is funded by the French Centre National de Recherche Scientifique (CNRS), the Italian Istituto Nazionale della Fisica Nucleare (INFN) and the Dutch Nikhef, with contributions by Polish and Hungarian institutes. This material is based upon work supported by NSF's LIGO Laboratory which is a major facility fully funded by the National Science Foundation.

\emph{Software:} VARAHA is programmed in Python and implements modules from NumPy~\cite{harris2020array}, SciPy~\cite{2020SciPy-NMeth} and PyCBC~\cite{PyCBCSoftware}. It uses matplotlib~\cite{Hunter:2007}, seaborn~\cite{Waskom2021}, corner~\cite{corner} and ligo.skymap (\href{https://lscsoft.docs.ligo.org/ligo.skymap}{https://lscsoft.docs.ligo.org/ligo.skymap}) for producing result plots. 

\appendix
\section{Importance Sampling} \label{sec:importance_sampling}

Assuming the parameters of the statistical model, $\boldsymbol{\theta}$, are defined using the probability density $p(\boldsymbol{\theta})$, the mean and standard deviation of parameter $\theta \in \boldsymbol{\theta}$, when marginalising over the other parameters, is simply,

\begin{eqnarray}
    \langle\theta\rangle &=& \int \theta\,p(\boldsymbol{\theta})\,\mathrm{d}\boldsymbol{\theta} \nonumber \\
    \sigma_{\langle\theta\rangle} &=& \sqrt{\int \left(\theta - \langle\theta\rangle\right)^2\, p(\boldsymbol{\theta})\,\mathrm{d}\boldsymbol{\theta}}.
\end{eqnarray}
Often these integrals are intractable and a practical way to estimate these quantities is by drawing random samples $\boldsymbol{\theta}_i$ from the true distribution $p(\boldsymbol{\theta})$.
The mean and variance of $\theta$ can then be estimated from the values $\theta_i \in \boldsymbol{\theta}_i$ as
\begin{eqnarray}
\bar{\theta} &=& \sum_i \theta_i / N\,, \nonumber \\
\sigma_{\bar{\theta}} &=& \sqrt{\left( \theta_i - \bar{\theta}\right)^2 / N}\,,
\label{eq:musigma}
\end{eqnarray}
where $i$ indexes the $N$ samples drawn.

Alternatively, one can use importance sampling and estimate the integrals in Eq.~\ref{eq:musigma} by calculating the weighted mean and weighted standard deviation~\citep{held2020likelihood},
\begin{eqnarray}
\bar{\theta} &=& \frac{\sum_i w_i \theta_i}{\sum_i w_i}, \nonumber \\
\sigma_{\bar{\theta}} &=& \sqrt{\frac{w_i\left( \theta_i - \bar{\theta}\right)^2}{\sum_i w_i}},
\end{eqnarray}
where $i$ indexes the $N$ samples drawn from a proposal distribution $\pi(\theta)$ and $w_i$ is the sample weight,
\begin{equation}
    w_i = \frac{p(\boldsymbol{\theta_i})}{\pi(\boldsymbol{\theta_i})}.
\end{equation}

Calculating parameter expectation values and uncertainties using a limited number of samples inevitably introduces sampling errors. When performing importance sampling, these errors also depend on the choice of proposal distribution. The measurement of $\bar{\theta}$ depends on the values of the weights, and the standard error relative to the true mean behaves as, 
\begin{equation}
    \sigma_{\bar{\theta}} = \sigma_{\theta} / \sqrt{\frac{\sum_i \left( w_i ^ 2\right)}{\left(\sum_i w_i \right)^ 2}},
\end{equation}
where $\sigma_{\theta}$ is the standard deviation of the parameter $\theta$ in the distribution being sampled.
When samples are drawn from the true distribution, $p(\boldsymbol{\theta})$, then $w_{i} \equiv 1$ and the error reduces to the standard proportionality of $1/\sqrt{N}$~\cite{2018CQGra..35n5009T}. 

Consequently, an effective sample size for a given proposal distribution can be defined as~\citep{ESS}, 
\begin{equation}
    n_{\text{eff}} = \frac{\left(\sum_i w_i \right)^ 2}{\sum_i \left( w_i ^ 2\right)}.
\end{equation}
The effective number of samples, $n_{\text{eff}}$,  approximately represents the number of samples one would need to measure $\theta$ as accurately using samples from the true distribution.  Since $n_{\text{eff}} \le N$, with equality if only if samples are drawn from the distribution $p(\boldsymbol{\theta})$, it follows that we require a greater number of samples when drawing from an alternate distribution. A detailed discussion on effective sample size is provided in \citep{ESS} and the included references. 

Estimation of parameters using Bayes equation,
\begin{equation}
    p(\boldsymbol{\theta}|d) \propto p(d|\boldsymbol{\theta})\pi(\boldsymbol{\theta})
\end{equation}
is in essence importance sampling. The proposal distribution acts as the prior distribution and the weights are replaced by a likelihood function $\mathcal{L}\equiv p(d|\boldsymbol{\theta})$ conditioned on the observed data $d$. Thus, we use these terms interchangeably. The posterior distribution is just the weighted prior distribution,
\begin{equation}
    p(\boldsymbol{\theta}|d) \propto w\,\pi(\boldsymbol{\theta}),
\end{equation}
with the weights given by
\begin{equation}
    w = \exp(\ell),\;\ell = \log{(\mathcal{L})} - \log(\mathcal{L}_{\text{max}}),
    \label{eq:weightedbayes}
\end{equation}
where we have scaled $\mathcal{L}$, such that the maximum value of $\ell$ is zero. Such a scaling does not impact any discussion earlier as it gets absorbed when normalising Eq.~\ref{eq:weightedbayes}. It helps obtain equal-weight samples after performing rejection sampling on the weights.

A point to consider is that if rejection sampling is performed on the value of weights, it will result in a sample size of close to $\sum_i w_i$ with all the samples having equal weights of 1~\citep{10.2307/4356322}. However, such a procedure results in loss of information as $n_{\text{eff}}$ is always larger than $\sum_i w_{i}$. Often \ac{MCMC} methods are employed to sample from the posterior probability distribution. All the algorithms performing \ac{PE} using \ac{MCMC} methods implement some kind of rejection sampling. Although they produce equally weighted samples, they discard a good fraction of the likelihood information~\cite{vanRavenzwaaij2018}.

Using a proposal distribution that is significantly different from the true distribution when estimating parameters using importance sampling, leads to most of the samples having very small weights, resulting in a severely reduced $n_{\text{eff}}$ and a grossly inefficient analysis. If uniform priors are used, a naive estimation of likelihood inside an arbitrary big box leads to most samples having very small weights. Unlike \ac{MCMC} methods, importance sampling is severely impacted by the curse of dimensionality.

\bibliography{apsbib}

\begin{thebibliography}{127}%
\makeatletter
\providecommand \@ifxundefined [1]{%
 \@ifx{#1\undefined}
}%
\providecommand \@ifnum [1]{%
 \ifnum #1\expandafter \@firstoftwo
 \else \expandafter \@secondoftwo
 \fi
}%
\providecommand \@ifx [1]{%
 \ifx #1\expandafter \@firstoftwo
 \else \expandafter \@secondoftwo
 \fi
}%
\providecommand \natexlab [1]{#1}%
\providecommand \enquote  [1]{``#1''}%
\providecommand \bibnamefont  [1]{#1}%
\providecommand \bibfnamefont [1]{#1}%
\providecommand \citenamefont [1]{#1}%
\providecommand \href@noop [0]{\@secondoftwo}%
\providecommand \href [0]{\begingroup \@sanitize@url \@href}%
\providecommand \@href[1]{\@@startlink{#1}\@@href}%
\providecommand \@@href[1]{\endgroup#1\@@endlink}%
\providecommand \@sanitize@url [0]{\catcode `\\12\catcode `\$12\catcode
  `\&12\catcode `\#12\catcode `\^12\catcode `\_12\catcode `\%12\relax}%
\providecommand \@@startlink[1]{}%
\providecommand \@@endlink[0]{}%
\providecommand \url  [0]{\begingroup\@sanitize@url \@url }%
\providecommand \@url [1]{\endgroup\@href {#1}{\urlprefix }}%
\providecommand \urlprefix  [0]{URL }%
\providecommand \Eprint [0]{\href }%
\providecommand \doibase [0]{https://doi.org/}%
\providecommand \selectlanguage [0]{\@gobble}%
\providecommand \bibinfo  [0]{\@secondoftwo}%
\providecommand \bibfield  [0]{\@secondoftwo}%
\providecommand \translation [1]{[#1]}%
\providecommand \BibitemOpen [0]{}%
\providecommand \bibitemStop [0]{}%
\providecommand \bibitemNoStop [0]{.\EOS\space}%
\providecommand \EOS [0]{\spacefactor3000\relax}%
\providecommand \BibitemShut  [1]{\csname bibitem#1\endcsname}%
\let\auto@bib@innerbib\@empty
\bibitem [{\citenamefont {{Abbott}}\ \emph {et~al.}(2019)\citenamefont
  {{Abbott}}, \citenamefont {{Abbott}}, \citenamefont {{Abbott}} \emph
  {et~al.}}]{2019PhRvX...9c1040A}%
  \BibitemOpen
  \bibfield  {author} {\bibinfo {author} {\bibfnamefont {B.~P.}\ \bibnamefont
  {{Abbott}}}, \bibinfo {author} {\bibfnamefont {R.}~\bibnamefont {{Abbott}}},
  \bibinfo {author} {\bibfnamefont {T.~D.}\ \bibnamefont {{Abbott}}}, \emph
  {et~al.},\ }\bibfield  {title} {\bibinfo {title} {{GWTC-1: A
  Gravitational-Wave Transient Catalog of Compact Binary Mergers Observed by
  LIGO and Virgo during the First and Second Observing Runs}},\ }\href
  {https://doi.org/10.1103/PhysRevX.9.031040} {\bibfield  {journal} {\bibinfo
  {journal} {Physical Review X}\ }\textbf {\bibinfo {volume} {9}},\ \bibinfo
  {eid} {031040} (\bibinfo {year} {2019})},\ \Eprint
  {https://arxiv.org/abs/1811.12907} {arXiv:1811.12907 [astro-ph.HE]}
  \BibitemShut {NoStop}%
\bibitem [{\citenamefont {{Abbott}}\ \emph
  {et~al.}(2021{\natexlab{a}})\citenamefont {{Abbott}}, \citenamefont
  {{Abbott}}, \citenamefont {{Abraham}} \emph {et~al.}}]{2021PhRvX..11b1053A}%
  \BibitemOpen
  \bibfield  {author} {\bibinfo {author} {\bibfnamefont {R.}~\bibnamefont
  {{Abbott}}}, \bibinfo {author} {\bibfnamefont {T.~D.}\ \bibnamefont
  {{Abbott}}}, \bibinfo {author} {\bibfnamefont {S.}~\bibnamefont {{Abraham}}},
  \emph {et~al.},\ }\bibfield  {title} {\bibinfo {title} {{GWTC-2: Compact
  Binary Coalescences Observed by LIGO and Virgo during the First Half of the
  Third Observing Run}},\ }\href {https://doi.org/10.1103/PhysRevX.11.021053}
  {\bibfield  {journal} {\bibinfo  {journal} {Physical Review X}\ }\textbf
  {\bibinfo {volume} {11}},\ \bibinfo {eid} {021053} (\bibinfo {year}
  {2021}{\natexlab{a}})},\ \Eprint {https://arxiv.org/abs/2010.14527}
  {arXiv:2010.14527 [gr-qc]} \BibitemShut {NoStop}%
\bibitem [{\citenamefont {{Abbott}}\ \emph
  {et~al.}(2021{\natexlab{b}})\citenamefont {{Abbott}}, \citenamefont
  {{Abbott}}, \citenamefont {{Acernese}} \emph {et~al.}}]{2021arXiv210801045T}%
  \BibitemOpen
  \bibfield  {author} {\bibinfo {author} {\bibfnamefont {R.}~\bibnamefont
  {{Abbott}}}, \bibinfo {author} {\bibfnamefont {T.~D.}\ \bibnamefont
  {{Abbott}}}, \bibinfo {author} {\bibfnamefont {F.}~\bibnamefont
  {{Acernese}}}, \emph {et~al.},\ }\bibfield  {title} {\bibinfo {title}
  {{GWTC-2.1: Deep Extended Catalog of Compact Binary Coalescences Observed by
  LIGO and Virgo During the First Half of the Third Observing Run}},\
  }\href@noop {} {\bibfield  {journal} {\bibinfo  {journal} {arXiv e-prints}\
  ,\ \bibinfo {eid} {arXiv:2108.01045}} (\bibinfo {year}
  {2021}{\natexlab{b}})},\ \Eprint {https://arxiv.org/abs/2108.01045}
  {arXiv:2108.01045 [gr-qc]} \BibitemShut {NoStop}%
\bibitem [{\citenamefont {{Abbott}}\ \emph
  {et~al.}(2021{\natexlab{c}})\citenamefont {{Abbott}}, \citenamefont
  {{Abbott}}, \citenamefont {{Acernese}} \emph {et~al.}}]{o3b_cat}%
  \BibitemOpen
  \bibfield  {author} {\bibinfo {author} {\bibfnamefont {R.}~\bibnamefont
  {{Abbott}}}, \bibinfo {author} {\bibfnamefont {T.~D.}\ \bibnamefont
  {{Abbott}}}, \bibinfo {author} {\bibfnamefont {F.}~\bibnamefont
  {{Acernese}}}, \emph {et~al.},\ }\bibfield  {title} {\bibinfo {title}
  {{GWTC-3: Compact Binary Coalescences Observed by LIGO and Virgo During the
  Second Part of the Third Observing Run}},\ }\href@noop {} {\bibfield
  {journal} {\bibinfo  {journal} {arXiv e-prints}\ ,\ \bibinfo {eid}
  {arXiv:2111.03606}} (\bibinfo {year} {2021}{\natexlab{c}})},\ \Eprint
  {https://arxiv.org/abs/2111.03606} {arXiv:2111.03606 [gr-qc]} \BibitemShut
  {NoStop}%
\bibitem [{\citenamefont {Abbott}\ \emph {et~al.}(2021)\citenamefont {Abbott}
  \emph {et~al.}}]{LIGOScientific:2021nrg}%
  \BibitemOpen
  \bibfield  {author} {\bibinfo {author} {\bibfnamefont {R.}~\bibnamefont
  {Abbott}} \emph {et~al.} (\bibinfo {collaboration} {LIGO Scientific, Virgo,
  KAGRA}),\ }\bibfield  {title} {\bibinfo {title} {{Constraints on Cosmic
  Strings Using Data from the Third Advanced LIGO\textendash{}Virgo Observing
  Run}},\ }\href {https://doi.org/10.1103/PhysRevLett.126.241102} {\bibfield
  {journal} {\bibinfo  {journal} {Phys. Rev. Lett.}\ }\textbf {\bibinfo
  {volume} {126}},\ \bibinfo {pages} {241102} (\bibinfo {year} {2021})},\
  \Eprint {https://arxiv.org/abs/2101.12248} {arXiv:2101.12248 [gr-qc]}
  \BibitemShut {NoStop}%
\bibitem [{\citenamefont {Calder\'on~Bustillo}\ \emph
  {et~al.}(2021)\citenamefont {Calder\'on~Bustillo}, \citenamefont
  {Sanchis-Gual}, \citenamefont {Torres-Forn\'e}, \citenamefont {Font},
  \citenamefont {Vajpeyi}, \citenamefont {Smith}, \citenamefont {Herdeiro},
  \citenamefont {Radu},\ and\ \citenamefont
  {Leong}}]{CalderonBustillo:2020fyi}%
  \BibitemOpen
  \bibfield  {author} {\bibinfo {author} {\bibfnamefont {J.}~\bibnamefont
  {Calder\'on~Bustillo}}, \bibinfo {author} {\bibfnamefont {N.}~\bibnamefont
  {Sanchis-Gual}}, \bibinfo {author} {\bibfnamefont {A.}~\bibnamefont
  {Torres-Forn\'e}}, \bibinfo {author} {\bibfnamefont {J.~A.}\ \bibnamefont
  {Font}}, \bibinfo {author} {\bibfnamefont {A.}~\bibnamefont {Vajpeyi}},
  \bibinfo {author} {\bibfnamefont {R.}~\bibnamefont {Smith}}, \bibinfo
  {author} {\bibfnamefont {C.}~\bibnamefont {Herdeiro}}, \bibinfo {author}
  {\bibfnamefont {E.}~\bibnamefont {Radu}},\ and\ \bibinfo {author}
  {\bibfnamefont {S.~H.~W.}\ \bibnamefont {Leong}},\ }\bibfield  {title}
  {\bibinfo {title} {{GW190521 as a Merger of Proca Stars: A Potential New
  Vector Boson of $8.7\times 10^{-13}$ eV}},\ }\href
  {https://doi.org/10.1103/PhysRevLett.126.081101} {\bibfield  {journal}
  {\bibinfo  {journal} {Phys. Rev. Lett.}\ }\textbf {\bibinfo {volume} {126}},\
  \bibinfo {pages} {081101} (\bibinfo {year} {2021})},\ \Eprint
  {https://arxiv.org/abs/2009.05376} {arXiv:2009.05376 [gr-qc]} \BibitemShut
  {NoStop}%
\bibitem [{\citenamefont {{Calderon Bustillo}}\ \emph
  {et~al.}(2022)\citenamefont {{Calderon Bustillo}}, \citenamefont
  {{Sanchis-Gual}}, \citenamefont {{Leong}}, \citenamefont {{Chandra}},
  \citenamefont {{Torres-Forne}}, \citenamefont {{Font}}, \citenamefont
  {{Herdeiro}}, \citenamefont {{Radu}}, \citenamefont {{Wong}},\ and\
  \citenamefont {{Li}}}]{2022arXiv220602551C}%
  \BibitemOpen
  \bibfield  {author} {\bibinfo {author} {\bibfnamefont {J.}~\bibnamefont
  {{Calderon Bustillo}}}, \bibinfo {author} {\bibfnamefont {N.}~\bibnamefont
  {{Sanchis-Gual}}}, \bibinfo {author} {\bibfnamefont {S.~H.~W.}\ \bibnamefont
  {{Leong}}}, \bibinfo {author} {\bibfnamefont {K.}~\bibnamefont {{Chandra}}},
  \bibinfo {author} {\bibfnamefont {A.}~\bibnamefont {{Torres-Forne}}},
  \bibinfo {author} {\bibfnamefont {J.~A.}\ \bibnamefont {{Font}}}, \bibinfo
  {author} {\bibfnamefont {C.}~\bibnamefont {{Herdeiro}}}, \bibinfo {author}
  {\bibfnamefont {E.}~\bibnamefont {{Radu}}}, \bibinfo {author} {\bibfnamefont
  {I.~C.~F.}\ \bibnamefont {{Wong}}},\ and\ \bibinfo {author} {\bibfnamefont
  {T.~G.~F.}\ \bibnamefont {{Li}}},\ }\bibfield  {title} {\bibinfo {title}
  {{Searching for vector boson-star mergers within LIGO-Virgo intermediate-mass
  black-hole merger candidates}},\ }\href
  {https://doi.org/10.48550/arXiv.2206.02551} {\bibfield  {journal} {\bibinfo
  {journal} {arXiv e-prints}\ ,\ \bibinfo {eid} {arXiv:2206.02551}} (\bibinfo
  {year} {2022})},\ \Eprint {https://arxiv.org/abs/2206.02551}
  {arXiv:2206.02551 [gr-qc]} \BibitemShut {NoStop}%
\bibitem [{\citenamefont {Pratten}\ \emph {et~al.}(2021)\citenamefont {Pratten}
  \emph {et~al.}}]{Pratten:2020ceb}%
  \BibitemOpen
  \bibfield  {author} {\bibinfo {author} {\bibfnamefont {G.}~\bibnamefont
  {Pratten}} \emph {et~al.},\ }\bibfield  {title} {\bibinfo {title}
  {{Computationally efficient models for the dominant and subdominant harmonic
  modes of precessing binary black holes}},\ }\href
  {https://doi.org/10.1103/PhysRevD.103.104056} {\bibfield  {journal} {\bibinfo
   {journal} {Phys. Rev. D}\ }\textbf {\bibinfo {volume} {103}},\ \bibinfo
  {pages} {104056} (\bibinfo {year} {2021})},\ \Eprint
  {https://arxiv.org/abs/2004.06503} {arXiv:2004.06503 [gr-qc]} \BibitemShut
  {NoStop}%
\bibitem [{\citenamefont {Estell\'es}\ \emph {et~al.}(2022)\citenamefont
  {Estell\'es}, \citenamefont {Colleoni}, \citenamefont {Garc\'\i{}a-Quir\'os},
  \citenamefont {Husa}, \citenamefont {Keitel}, \citenamefont {Mateu-Lucena},
  \citenamefont {Planas},\ and\ \citenamefont
  {Ramos-Buades}}]{Estelles:2021gvs}%
  \BibitemOpen
  \bibfield  {author} {\bibinfo {author} {\bibfnamefont {H.}~\bibnamefont
  {Estell\'es}}, \bibinfo {author} {\bibfnamefont {M.}~\bibnamefont
  {Colleoni}}, \bibinfo {author} {\bibfnamefont {C.}~\bibnamefont
  {Garc\'\i{}a-Quir\'os}}, \bibinfo {author} {\bibfnamefont {S.}~\bibnamefont
  {Husa}}, \bibinfo {author} {\bibfnamefont {D.}~\bibnamefont {Keitel}},
  \bibinfo {author} {\bibfnamefont {M.}~\bibnamefont {Mateu-Lucena}}, \bibinfo
  {author} {\bibfnamefont {M.~d.~L.}\ \bibnamefont {Planas}},\ and\ \bibinfo
  {author} {\bibfnamefont {A.}~\bibnamefont {Ramos-Buades}},\ }\bibfield
  {title} {\bibinfo {title} {{New twists in compact binary waveform modeling: A
  fast time-domain model for precession}},\ }\href
  {https://doi.org/10.1103/PhysRevD.105.084040} {\bibfield  {journal} {\bibinfo
   {journal} {Phys. Rev. D}\ }\textbf {\bibinfo {volume} {105}},\ \bibinfo
  {pages} {084040} (\bibinfo {year} {2022})},\ \Eprint
  {https://arxiv.org/abs/2105.05872} {arXiv:2105.05872 [gr-qc]} \BibitemShut
  {NoStop}%
\bibitem [{\citenamefont {Hamilton}\ \emph {et~al.}(2021)\citenamefont
  {Hamilton}, \citenamefont {London}, \citenamefont {Thompson}, \citenamefont
  {Fauchon-Jones}, \citenamefont {Hannam}, \citenamefont {Kalaghatgi},
  \citenamefont {Khan}, \citenamefont {Pannarale},\ and\ \citenamefont
  {Vano-Vinuales}}]{Hamilton:2021pkf}%
  \BibitemOpen
  \bibfield  {author} {\bibinfo {author} {\bibfnamefont {E.}~\bibnamefont
  {Hamilton}}, \bibinfo {author} {\bibfnamefont {L.}~\bibnamefont {London}},
  \bibinfo {author} {\bibfnamefont {J.~E.}\ \bibnamefont {Thompson}}, \bibinfo
  {author} {\bibfnamefont {E.}~\bibnamefont {Fauchon-Jones}}, \bibinfo {author}
  {\bibfnamefont {M.}~\bibnamefont {Hannam}}, \bibinfo {author} {\bibfnamefont
  {C.}~\bibnamefont {Kalaghatgi}}, \bibinfo {author} {\bibfnamefont
  {S.}~\bibnamefont {Khan}}, \bibinfo {author} {\bibfnamefont {F.}~\bibnamefont
  {Pannarale}},\ and\ \bibinfo {author} {\bibfnamefont {A.}~\bibnamefont
  {Vano-Vinuales}},\ }\bibfield  {title} {\bibinfo {title} {{Model of
  gravitational waves from precessing black-hole binaries through merger and
  ringdown}},\ }\href {https://doi.org/10.1103/PhysRevD.104.124027} {\bibfield
  {journal} {\bibinfo  {journal} {Phys. Rev. D}\ }\textbf {\bibinfo {volume}
  {104}},\ \bibinfo {pages} {124027} (\bibinfo {year} {2021})},\ \Eprint
  {https://arxiv.org/abs/2107.08876} {arXiv:2107.08876 [gr-qc]} \BibitemShut
  {NoStop}%
\bibitem [{\citenamefont {Thompson}\ \emph {et~al.}(2020)\citenamefont
  {Thompson}, \citenamefont {Fauchon-Jones}, \citenamefont {Khan},
  \citenamefont {Nitoglia}, \citenamefont {Pannarale}, \citenamefont
  {Dietrich},\ and\ \citenamefont {Hannam}}]{Thompson:2020nei}%
  \BibitemOpen
  \bibfield  {author} {\bibinfo {author} {\bibfnamefont {J.~E.}\ \bibnamefont
  {Thompson}}, \bibinfo {author} {\bibfnamefont {E.}~\bibnamefont
  {Fauchon-Jones}}, \bibinfo {author} {\bibfnamefont {S.}~\bibnamefont {Khan}},
  \bibinfo {author} {\bibfnamefont {E.}~\bibnamefont {Nitoglia}}, \bibinfo
  {author} {\bibfnamefont {F.}~\bibnamefont {Pannarale}}, \bibinfo {author}
  {\bibfnamefont {T.}~\bibnamefont {Dietrich}},\ and\ \bibinfo {author}
  {\bibfnamefont {M.}~\bibnamefont {Hannam}},\ }\bibfield  {title} {\bibinfo
  {title} {{Modeling the gravitational wave signature of neutron star black
  hole coalescences}},\ }\href {https://doi.org/10.1103/PhysRevD.101.124059}
  {\bibfield  {journal} {\bibinfo  {journal} {Phys. Rev. D}\ }\textbf {\bibinfo
  {volume} {101}},\ \bibinfo {pages} {124059} (\bibinfo {year} {2020})},\
  \Eprint {https://arxiv.org/abs/2002.08383} {arXiv:2002.08383 [gr-qc]}
  \BibitemShut {NoStop}%
\bibitem [{\citenamefont {Ossokine}\ \emph {et~al.}(2020)\citenamefont
  {Ossokine} \emph {et~al.}}]{Ossokine:2020kjp}%
  \BibitemOpen
  \bibfield  {author} {\bibinfo {author} {\bibfnamefont {S.}~\bibnamefont
  {Ossokine}} \emph {et~al.},\ }\bibfield  {title} {\bibinfo {title}
  {{Multipolar Effective-One-Body Waveforms for Precessing Binary Black Holes:
  Construction and Validation}},\ }\href
  {https://doi.org/10.1103/PhysRevD.102.044055} {\bibfield  {journal} {\bibinfo
   {journal} {Phys. Rev. D}\ }\textbf {\bibinfo {volume} {102}},\ \bibinfo
  {pages} {044055} (\bibinfo {year} {2020})},\ \Eprint
  {https://arxiv.org/abs/2004.09442} {arXiv:2004.09442 [gr-qc]} \BibitemShut
  {NoStop}%
\bibitem [{\citenamefont {Matas}\ \emph {et~al.}(2020)\citenamefont {Matas}
  \emph {et~al.}}]{Matas:2020wab}%
  \BibitemOpen
  \bibfield  {author} {\bibinfo {author} {\bibfnamefont {A.}~\bibnamefont
  {Matas}} \emph {et~al.},\ }\bibfield  {title} {\bibinfo {title}
  {{Aligned-spin neutron-star\textendash{}black-hole waveform model based on
  the effective-one-body approach and numerical-relativity simulations}},\
  }\href {https://doi.org/10.1103/PhysRevD.102.043023} {\bibfield  {journal}
  {\bibinfo  {journal} {Phys. Rev. D}\ }\textbf {\bibinfo {volume} {102}},\
  \bibinfo {pages} {043023} (\bibinfo {year} {2020})},\ \Eprint
  {https://arxiv.org/abs/2004.10001} {arXiv:2004.10001 [gr-qc]} \BibitemShut
  {NoStop}%
\bibitem [{\citenamefont {Varma}\ \emph {et~al.}(2019)\citenamefont {Varma},
  \citenamefont {Field}, \citenamefont {Scheel}, \citenamefont {Blackman},
  \citenamefont {Gerosa}, \citenamefont {Stein}, \citenamefont {Kidder},\ and\
  \citenamefont {Pfeiffer}}]{Varma:2019csw}%
  \BibitemOpen
  \bibfield  {author} {\bibinfo {author} {\bibfnamefont {V.}~\bibnamefont
  {Varma}}, \bibinfo {author} {\bibfnamefont {S.~E.}\ \bibnamefont {Field}},
  \bibinfo {author} {\bibfnamefont {M.~A.}\ \bibnamefont {Scheel}}, \bibinfo
  {author} {\bibfnamefont {J.}~\bibnamefont {Blackman}}, \bibinfo {author}
  {\bibfnamefont {D.}~\bibnamefont {Gerosa}}, \bibinfo {author} {\bibfnamefont
  {L.~C.}\ \bibnamefont {Stein}}, \bibinfo {author} {\bibfnamefont {L.~E.}\
  \bibnamefont {Kidder}},\ and\ \bibinfo {author} {\bibfnamefont {H.~P.}\
  \bibnamefont {Pfeiffer}},\ }\bibfield  {title} {\bibinfo {title} {{Surrogate
  models for precessing binary black hole simulations with unequal masses}},\
  }\href {https://doi.org/10.1103/PhysRevResearch.1.033015} {\bibfield
  {journal} {\bibinfo  {journal} {Phys. Rev. Research.}\ }\textbf {\bibinfo
  {volume} {1}},\ \bibinfo {pages} {033015} (\bibinfo {year} {2019})},\ \Eprint
  {https://arxiv.org/abs/1905.09300} {arXiv:1905.09300 [gr-qc]} \BibitemShut
  {NoStop}%
\bibitem [{\citenamefont {Riemenschneider}\ \emph {et~al.}(2021)\citenamefont
  {Riemenschneider}, \citenamefont {Rettegno}, \citenamefont {Breschi},
  \citenamefont {Albertini}, \citenamefont {Gamba}, \citenamefont {Bernuzzi},\
  and\ \citenamefont {Nagar}}]{Riemenschneider:2021ppj}%
  \BibitemOpen
  \bibfield  {author} {\bibinfo {author} {\bibfnamefont {G.}~\bibnamefont
  {Riemenschneider}}, \bibinfo {author} {\bibfnamefont {P.}~\bibnamefont
  {Rettegno}}, \bibinfo {author} {\bibfnamefont {M.}~\bibnamefont {Breschi}},
  \bibinfo {author} {\bibfnamefont {A.}~\bibnamefont {Albertini}}, \bibinfo
  {author} {\bibfnamefont {R.}~\bibnamefont {Gamba}}, \bibinfo {author}
  {\bibfnamefont {S.}~\bibnamefont {Bernuzzi}},\ and\ \bibinfo {author}
  {\bibfnamefont {A.}~\bibnamefont {Nagar}},\ }\bibfield  {title} {\bibinfo
  {title} {{Assessment of consistent next-to-quasicircular corrections and
  postadiabatic approximation in effective-one-body multipolar waveforms for
  binary black hole coalescences}},\ }\href
  {https://doi.org/10.1103/PhysRevD.104.104045} {\bibfield  {journal} {\bibinfo
   {journal} {Phys. Rev. D}\ }\textbf {\bibinfo {volume} {104}},\ \bibinfo
  {pages} {104045} (\bibinfo {year} {2021})},\ \Eprint
  {https://arxiv.org/abs/2104.07533} {arXiv:2104.07533 [gr-qc]} \BibitemShut
  {NoStop}%
\bibitem [{\citenamefont {Dietrich}\ \emph {et~al.}(2019)\citenamefont
  {Dietrich}, \citenamefont {Samajdar}, \citenamefont {Khan}, \citenamefont
  {Johnson-McDaniel}, \citenamefont {Dudi},\ and\ \citenamefont
  {Tichy}}]{Dietrich:2019kaq}%
  \BibitemOpen
  \bibfield  {author} {\bibinfo {author} {\bibfnamefont {T.}~\bibnamefont
  {Dietrich}}, \bibinfo {author} {\bibfnamefont {A.}~\bibnamefont {Samajdar}},
  \bibinfo {author} {\bibfnamefont {S.}~\bibnamefont {Khan}}, \bibinfo {author}
  {\bibfnamefont {N.~K.}\ \bibnamefont {Johnson-McDaniel}}, \bibinfo {author}
  {\bibfnamefont {R.}~\bibnamefont {Dudi}},\ and\ \bibinfo {author}
  {\bibfnamefont {W.}~\bibnamefont {Tichy}},\ }\bibfield  {title} {\bibinfo
  {title} {{Improving the NRTidal model for binary neutron star systems}},\
  }\href {https://doi.org/10.1103/PhysRevD.100.044003} {\bibfield  {journal}
  {\bibinfo  {journal} {Phys. Rev. D}\ }\textbf {\bibinfo {volume} {100}},\
  \bibinfo {pages} {044003} (\bibinfo {year} {2019})},\ \Eprint
  {https://arxiv.org/abs/1905.06011} {arXiv:1905.06011 [gr-qc]} \BibitemShut
  {NoStop}%
\bibitem [{\citenamefont {Abbott}\ \emph {et~al.}(2016)\citenamefont {Abbott}
  \emph {et~al.}}]{first_monday_pe}%
  \BibitemOpen
  \bibfield  {author} {\bibinfo {author} {\bibfnamefont {B.~P.}\ \bibnamefont
  {Abbott}} \emph {et~al.} (\bibinfo {collaboration} {LIGO Scientific
  Collaboration, Virgo Collaboration}),\ }\bibfield  {title} {\bibinfo {title}
  {{Properties of the Binary Black Hole Merger GW150914}},\ }\href@noop {}
  {\bibfield  {journal} {\bibinfo  {journal} {Phys. Rev. Lett.}\ }\textbf
  {\bibinfo {volume} {116}},\ \bibinfo {pages} {241102} (\bibinfo {year}
  {2016})},\ \Eprint {https://arxiv.org/abs/1602.03840} {1602.03840}
  \BibitemShut {NoStop}%
\bibitem [{\citenamefont {Christensen}\ and\ \citenamefont
  {Meyer}(2022)}]{Christensen:2022bxb}%
  \BibitemOpen
  \bibfield  {author} {\bibinfo {author} {\bibfnamefont {N.}~\bibnamefont
  {Christensen}}\ and\ \bibinfo {author} {\bibfnamefont {R.}~\bibnamefont
  {Meyer}},\ }\bibfield  {title} {\bibinfo {title} {{Parameter estimation with
  gravitational waves}},\ }\href {https://doi.org/10.1103/RevModPhys.94.025001}
  {\bibfield  {journal} {\bibinfo  {journal} {Rev. Mod. Phys.}\ }\textbf
  {\bibinfo {volume} {94}},\ \bibinfo {pages} {025001} (\bibinfo {year}
  {2022})},\ \Eprint {https://arxiv.org/abs/2204.04449} {arXiv:2204.04449
  [gr-qc]} \BibitemShut {NoStop}%
\bibitem [{\citenamefont {{Abbott}}\ \emph
  {et~al.}(2021{\natexlab{a}})\citenamefont {{Abbott}}, \citenamefont
  {{Abbott}}, \citenamefont {{Abraham}} \emph {et~al.}}]{2021ApJ...913L...7A}%
  \BibitemOpen
  \bibfield  {author} {\bibinfo {author} {\bibfnamefont {R.}~\bibnamefont
  {{Abbott}}}, \bibinfo {author} {\bibfnamefont {T.~D.}\ \bibnamefont
  {{Abbott}}}, \bibinfo {author} {\bibfnamefont {S.}~\bibnamefont {{Abraham}}},
  \emph {et~al.},\ }\bibfield  {title} {\bibinfo {title} {{Population
  Properties of Compact Objects from the Second LIGO-Virgo Gravitational-Wave
  Transient Catalog}},\ }\href {https://doi.org/10.3847/2041-8213/abe949}
  {\bibfield  {journal} {\bibinfo  {journal} {Astrophys. J. Lett.}\ }\textbf
  {\bibinfo {volume} {913}},\ \bibinfo {eid} {L7} (\bibinfo {year}
  {2021}{\natexlab{a}})},\ \Eprint {https://arxiv.org/abs/2010.14533}
  {arXiv:2010.14533 [astro-ph.HE]} \BibitemShut {NoStop}%
\bibitem [{\citenamefont {{Tiwari}}\ and\ \citenamefont
  {{Fairhurst}}(2021)}]{2021ApJ...913L..19T}%
  \BibitemOpen
  \bibfield  {author} {\bibinfo {author} {\bibfnamefont {V.}~\bibnamefont
  {{Tiwari}}}\ and\ \bibinfo {author} {\bibfnamefont {S.}~\bibnamefont
  {{Fairhurst}}},\ }\bibfield  {title} {\bibinfo {title} {{The Emergence of
  Structure in the Binary Black Hole Mass Distribution}},\ }\href
  {https://doi.org/10.3847/2041-8213/abfbe7} {\bibfield  {journal} {\bibinfo
  {journal} {Astrophys. J. Lett.}\ }\textbf {\bibinfo {volume} {913}},\
  \bibinfo {eid} {L19} (\bibinfo {year} {2021})},\ \Eprint
  {https://arxiv.org/abs/2011.04502} {arXiv:2011.04502 [astro-ph.HE]}
  \BibitemShut {NoStop}%
\bibitem [{\citenamefont {{Abbott}}\ \emph
  {et~al.}(2021{\natexlab{b}})\citenamefont {{Abbott}} \emph
  {et~al.}}]{o3b_rnp}%
  \BibitemOpen
  \bibfield  {author} {\bibinfo {author} {\bibfnamefont {R.}~\bibnamefont
  {{Abbott}}} \emph {et~al.},\ }\bibfield  {title} {\bibinfo {title} {{The
  population of merging compact binaries inferred using gravitational waves
  through GWTC-3}},\ }\href@noop {} {\bibfield  {journal} {\bibinfo  {journal}
  {arXiv e-prints}\ ,\ \bibinfo {eid} {arXiv:2111.03634}} (\bibinfo {year}
  {2021}{\natexlab{b}})},\ \Eprint {https://arxiv.org/abs/2111.03634}
  {arXiv:2111.03634 [gr-qc]} \BibitemShut {NoStop}%
\bibitem [{\citenamefont {{Tiwari}}(2022)}]{2022ApJ...928..155T}%
  \BibitemOpen
  \bibfield  {author} {\bibinfo {author} {\bibfnamefont {V.}~\bibnamefont
  {{Tiwari}}},\ }\bibfield  {title} {\bibinfo {title} {{Exploring Features in
  the Binary Black Hole Population}},\ }\href
  {https://doi.org/10.3847/1538-4357/ac589a} {\bibfield  {journal} {\bibinfo
  {journal} {\apj}\ }\textbf {\bibinfo {volume} {928}},\ \bibinfo {eid} {155}
  (\bibinfo {year} {2022})},\ \Eprint {https://arxiv.org/abs/2111.13991}
  {arXiv:2111.13991 [astro-ph.HE]} \BibitemShut {NoStop}%
\bibitem [{\citenamefont {Callister}\ \emph {et~al.}(2021)\citenamefont
  {Callister}, \citenamefont {Haster}, \citenamefont {Ng}, \citenamefont
  {Vitale},\ and\ \citenamefont {Farr}}]{Callister:2021fpo}%
  \BibitemOpen
  \bibfield  {author} {\bibinfo {author} {\bibfnamefont {T.~A.}\ \bibnamefont
  {Callister}}, \bibinfo {author} {\bibfnamefont {C.-J.}\ \bibnamefont
  {Haster}}, \bibinfo {author} {\bibfnamefont {K.~K.~Y.}\ \bibnamefont {Ng}},
  \bibinfo {author} {\bibfnamefont {S.}~\bibnamefont {Vitale}},\ and\ \bibinfo
  {author} {\bibfnamefont {W.~M.}\ \bibnamefont {Farr}},\ }\bibfield  {title}
  {\bibinfo {title} {{Who Ordered That? Unequal-mass Binary Black Hole Mergers
  Have Larger Effective Spins}},\ }\href
  {https://doi.org/10.3847/2041-8213/ac2ccc} {\bibfield  {journal} {\bibinfo
  {journal} {Astrophys. J. Lett.}\ }\textbf {\bibinfo {volume} {922}},\
  \bibinfo {pages} {L5} (\bibinfo {year} {2021})},\ \Eprint
  {https://arxiv.org/abs/2106.00521} {arXiv:2106.00521 [astro-ph.HE]}
  \BibitemShut {NoStop}%
\bibitem [{\citenamefont {Roulet}\ and\ \citenamefont
  {Zaldarriaga}(2019)}]{Roulet:2018jbe}%
  \BibitemOpen
  \bibfield  {author} {\bibinfo {author} {\bibfnamefont {J.}~\bibnamefont
  {Roulet}}\ and\ \bibinfo {author} {\bibfnamefont {M.}~\bibnamefont
  {Zaldarriaga}},\ }\bibfield  {title} {\bibinfo {title} {{Constraints on
  binary black hole populations from LIGO\textendash{}Virgo detections}},\
  }\href {https://doi.org/10.1093/mnras/stz226} {\bibfield  {journal} {\bibinfo
   {journal} {Mon. Not. Roy. Astron. Soc.}\ }\textbf {\bibinfo {volume}
  {484}},\ \bibinfo {pages} {4216} (\bibinfo {year} {2019})},\ \Eprint
  {https://arxiv.org/abs/1806.10610} {arXiv:1806.10610 [astro-ph.HE]}
  \BibitemShut {NoStop}%
\bibitem [{\citenamefont {Vitale}\ \emph {et~al.}(2022)\citenamefont {Vitale},
  \citenamefont {Biscoveanu},\ and\ \citenamefont {Talbot}}]{Vitale:2022pmu}%
  \BibitemOpen
  \bibfield  {author} {\bibinfo {author} {\bibfnamefont {S.}~\bibnamefont
  {Vitale}}, \bibinfo {author} {\bibfnamefont {S.}~\bibnamefont {Biscoveanu}},\
  and\ \bibinfo {author} {\bibfnamefont {C.}~\bibnamefont {Talbot}},\
  }\bibfield  {title} {\bibinfo {title} {{The orientations of the binary black
  holes in GWTC-3}},\ }\href@noop {} {\bibfield  {journal} {\bibinfo  {journal}
  {arXiv}\ } (\bibinfo {year} {2022})},\ \Eprint
  {https://arxiv.org/abs/2204.00968} {arXiv:2204.00968 [gr-qc]} \BibitemShut
  {NoStop}%
\bibitem [{\citenamefont {Edelman}\ \emph {et~al.}(2022)\citenamefont
  {Edelman}, \citenamefont {Farr},\ and\ \citenamefont
  {Doctor}}]{Edelman:2022ydv}%
  \BibitemOpen
  \bibfield  {author} {\bibinfo {author} {\bibfnamefont {B.}~\bibnamefont
  {Edelman}}, \bibinfo {author} {\bibfnamefont {B.}~\bibnamefont {Farr}},\ and\
  \bibinfo {author} {\bibfnamefont {Z.}~\bibnamefont {Doctor}},\ }\bibfield
  {title} {\bibinfo {title} {{Cover Your Basis: Comprehensive Data-Driven
  Characterization of the Binary Black Hole Population}},\ }\href@noop {}
  {\bibfield  {journal} {\bibinfo  {journal} {arXiv}\ } (\bibinfo {year}
  {2022})},\ \Eprint {https://arxiv.org/abs/2210.12834} {arXiv:2210.12834
  [astro-ph.HE]} \BibitemShut {NoStop}%
\bibitem [{\citenamefont {{The LIGO Scientific Collaboration and the Virgo
  Collaboration}}(2019)}]{2019PhRvL.122f1104A}%
  \BibitemOpen
  \bibfield  {author} {\bibinfo {author} {\bibnamefont {{The LIGO Scientific
  Collaboration and the Virgo Collaboration}}},\ }\bibfield  {title} {\bibinfo
  {title} {{Constraining the p -Mode-g -Mode Tidal Instability with
  GW170817}},\ }\href {https://doi.org/10.1103/PhysRevLett.122.061104}
  {\bibfield  {journal} {\bibinfo  {journal} {\prl}\ }\textbf {\bibinfo
  {volume} {122}},\ \bibinfo {eid} {061104} (\bibinfo {year} {2019})},\ \Eprint
  {https://arxiv.org/abs/1808.08676} {arXiv:1808.08676 [astro-ph.HE]}
  \BibitemShut {NoStop}%
\bibitem [{\citenamefont {{Agathos}}\ \emph {et~al.}(2015)\citenamefont
  {{Agathos}}, \citenamefont {{Meidam}}, \citenamefont {{Del Pozzo}},
  \citenamefont {{Li}}, \citenamefont {{Tompitak}}, \citenamefont {{Veitch}},
  \citenamefont {{Vitale}},\ and\ \citenamefont {{Van Den
  Broeck}}}]{2015PhRvD..92b3012A}%
  \BibitemOpen
  \bibfield  {author} {\bibinfo {author} {\bibfnamefont {M.}~\bibnamefont
  {{Agathos}}}, \bibinfo {author} {\bibfnamefont {J.}~\bibnamefont {{Meidam}}},
  \bibinfo {author} {\bibfnamefont {W.}~\bibnamefont {{Del Pozzo}}}, \bibinfo
  {author} {\bibfnamefont {T.~G.~F.}\ \bibnamefont {{Li}}}, \bibinfo {author}
  {\bibfnamefont {M.}~\bibnamefont {{Tompitak}}}, \bibinfo {author}
  {\bibfnamefont {J.}~\bibnamefont {{Veitch}}}, \bibinfo {author}
  {\bibfnamefont {S.}~\bibnamefont {{Vitale}}},\ and\ \bibinfo {author}
  {\bibfnamefont {C.}~\bibnamefont {{Van Den Broeck}}},\ }\bibfield  {title}
  {\bibinfo {title} {{Constraining the neutron star equation of state with
  gravitational wave signals from coalescing binary neutron stars}},\ }\href
  {https://doi.org/10.1103/PhysRevD.92.023012} {\bibfield  {journal} {\bibinfo
  {journal} {\prd}\ }\textbf {\bibinfo {volume} {92}},\ \bibinfo {eid} {023012}
  (\bibinfo {year} {2015})},\ \Eprint {https://arxiv.org/abs/1503.05405}
  {arXiv:1503.05405 [gr-qc]} \BibitemShut {NoStop}%
\bibitem [{\citenamefont {Abbott}\ \emph {et~al.}(2021)\citenamefont {Abbott}
  \emph {et~al.}}]{LIGOScientific:2021aug}%
  \BibitemOpen
  \bibfield  {author} {\bibinfo {author} {\bibfnamefont {R.}~\bibnamefont
  {Abbott}} \emph {et~al.} (\bibinfo {collaboration} {LIGO Scientific, VIRGO,
  KAGRA}),\ }\bibfield  {title} {\bibinfo {title} {{Constraints on the cosmic
  expansion history from GWTC-3}},\ }\href@noop {} {\bibfield  {journal}
  {\bibinfo  {journal} {arXiv}\ } (\bibinfo {year} {2021})},\ \Eprint
  {https://arxiv.org/abs/2111.03604} {arXiv:2111.03604 [astro-ph.CO]}
  \BibitemShut {NoStop}%
\bibitem [{\citenamefont {{Apostolatos}}\ \emph {et~al.}(1994)\citenamefont
  {{Apostolatos}}, \citenamefont {{Cutler}}, \citenamefont {{Sussman}},\ and\
  \citenamefont {{Thorne}}}]{1994PhRvD..49.6274A}%
  \BibitemOpen
  \bibfield  {author} {\bibinfo {author} {\bibfnamefont {T.~A.}\ \bibnamefont
  {{Apostolatos}}}, \bibinfo {author} {\bibfnamefont {C.}~\bibnamefont
  {{Cutler}}}, \bibinfo {author} {\bibfnamefont {G.~J.}\ \bibnamefont
  {{Sussman}}},\ and\ \bibinfo {author} {\bibfnamefont {K.~S.}\ \bibnamefont
  {{Thorne}}},\ }\bibfield  {title} {\bibinfo {title} {{Spin-induced orbital
  precession and its modulation of the gravitational waveforms from merging
  binaries}},\ }\href {https://doi.org/10.1103/PhysRevD.49.6274} {\bibfield
  {journal} {\bibinfo  {journal} {\prd}\ }\textbf {\bibinfo {volume} {49}},\
  \bibinfo {pages} {6274} (\bibinfo {year} {1994})}\BibitemShut {NoStop}%
\bibitem [{\citenamefont {{Yunes}}\ \emph {et~al.}(2009)\citenamefont
  {{Yunes}}, \citenamefont {{Arun}}, \citenamefont {{Berti}},\ and\
  \citenamefont {{Will}}}]{2009PhRvD..80h4001Y}%
  \BibitemOpen
  \bibfield  {author} {\bibinfo {author} {\bibfnamefont {N.}~\bibnamefont
  {{Yunes}}}, \bibinfo {author} {\bibfnamefont {K.~G.}\ \bibnamefont {{Arun}}},
  \bibinfo {author} {\bibfnamefont {E.}~\bibnamefont {{Berti}}},\ and\ \bibinfo
  {author} {\bibfnamefont {C.~M.}\ \bibnamefont {{Will}}},\ }\bibfield  {title}
  {\bibinfo {title} {{Post-circular expansion of eccentric binary inspirals:
  Fourier-domain waveforms in the stationary phase approximation}},\ }\href
  {https://doi.org/10.1103/PhysRevD.80.08400110.48550/arXiv.0906.0313}
  {\bibfield  {journal} {\bibinfo  {journal} {\prd}\ }\textbf {\bibinfo
  {volume} {80}},\ \bibinfo {eid} {084001} (\bibinfo {year} {2009})},\ \Eprint
  {https://arxiv.org/abs/0906.0313} {arXiv:0906.0313 [gr-qc]} \BibitemShut
  {NoStop}%
\bibitem [{\citenamefont {{Owen}}(1996)}]{1996PhRvD..53.6749O}%
  \BibitemOpen
  \bibfield  {author} {\bibinfo {author} {\bibfnamefont {B.~J.}\ \bibnamefont
  {{Owen}}},\ }\bibfield  {title} {\bibinfo {title} {{Search templates for
  gravitational waves from inspiraling binaries: Choice of template spacing}},\
  }\href {https://doi.org/10.1103/PhysRevD.53.6749} {\bibfield  {journal}
  {\bibinfo  {journal} {\prd}\ }\textbf {\bibinfo {volume} {53}},\ \bibinfo
  {pages} {6749} (\bibinfo {year} {1996})},\ \Eprint
  {https://arxiv.org/abs/gr-qc/9511032} {arXiv:gr-qc/9511032 [gr-qc]}
  \BibitemShut {NoStop}%
\bibitem [{\citenamefont {{Cutler}}\ and\ \citenamefont
  {{Flanagan}}(1994)}]{1994PhRvD..49.2658C}%
  \BibitemOpen
  \bibfield  {author} {\bibinfo {author} {\bibfnamefont {C.}~\bibnamefont
  {{Cutler}}}\ and\ \bibinfo {author} {\bibfnamefont {{\'E}.~E.}\ \bibnamefont
  {{Flanagan}}},\ }\bibfield  {title} {\bibinfo {title} {{Gravitational waves
  from merging compact binaries: How accurately can one extract the binary's
  parameters from the inspiral waveform?}},\ }\href
  {https://doi.org/10.1103/PhysRevD.49.2658} {\bibfield  {journal} {\bibinfo
  {journal} {\prd}\ }\textbf {\bibinfo {volume} {49}},\ \bibinfo {pages} {2658}
  (\bibinfo {year} {1994})},\ \Eprint {https://arxiv.org/abs/gr-qc/9402014}
  {arXiv:gr-qc/9402014 [gr-qc]} \BibitemShut {NoStop}%
\bibitem [{\citenamefont {{Poisson}}\ and\ \citenamefont
  {{Will}}(1995)}]{1995PhRvD..52..848P}%
  \BibitemOpen
  \bibfield  {author} {\bibinfo {author} {\bibfnamefont {E.}~\bibnamefont
  {{Poisson}}}\ and\ \bibinfo {author} {\bibfnamefont {C.~M.}\ \bibnamefont
  {{Will}}},\ }\bibfield  {title} {\bibinfo {title} {{Gravitational waves from
  inspiraling compact binaries: Parameter estimation using
  second-post-Newtonian waveforms}},\ }\href
  {https://doi.org/10.1103/PhysRevD.52.848} {\bibfield  {journal} {\bibinfo
  {journal} {\prd}\ }\textbf {\bibinfo {volume} {52}},\ \bibinfo {pages} {848}
  (\bibinfo {year} {1995})},\ \Eprint {https://arxiv.org/abs/gr-qc/9502040}
  {arXiv:gr-qc/9502040 [gr-qc]} \BibitemShut {NoStop}%
\bibitem [{\citenamefont {{Singer}}\ and\ \citenamefont
  {{Price}}(2016)}]{2016PhRvD..93b4013S}%
  \BibitemOpen
  \bibfield  {author} {\bibinfo {author} {\bibfnamefont {L.~P.}\ \bibnamefont
  {{Singer}}}\ and\ \bibinfo {author} {\bibfnamefont {L.~R.}\ \bibnamefont
  {{Price}}},\ }\bibfield  {title} {\bibinfo {title} {{Rapid Bayesian position
  reconstruction for gravitational-wave transients}},\ }\href
  {https://doi.org/10.1103/PhysRevD.93.024013} {\bibfield  {journal} {\bibinfo
  {journal} {\prd}\ }\textbf {\bibinfo {volume} {93}},\ \bibinfo {eid} {024013}
  (\bibinfo {year} {2016})},\ \Eprint {https://arxiv.org/abs/1508.03634}
  {arXiv:1508.03634 [gr-qc]} \BibitemShut {NoStop}%
\bibitem [{\citenamefont {{Veitch}}\ \emph {et~al.}(2015)\citenamefont
  {{Veitch}}, \citenamefont {{Raymond}}, \citenamefont {{Farr}}, \citenamefont
  {{Farr}}, \citenamefont {{Graff}}, \citenamefont {{Vitale}}, \citenamefont
  {{Aylott}}, \citenamefont {{Blackburn}}, \citenamefont {{Christensen}},
  \citenamefont {{Coughlin}}, \citenamefont {{Del Pozzo}}, \citenamefont
  {{Feroz}}, \citenamefont {{Gair}}, \citenamefont {{Haster}}, \citenamefont
  {{Kalogera}}, \citenamefont {{Littenberg}}, \citenamefont {{Mandel}},
  \citenamefont {{O'Shaughnessy}}, \citenamefont {{Pitkin}}, \citenamefont
  {{Rodriguez}}, \citenamefont {{R{\"o}ver}}, \citenamefont {{Sidery}},
  \citenamefont {{Smith}}, \citenamefont {{Van Der Sluys}}, \citenamefont
  {{Vecchio}}, \citenamefont {{Vousden}},\ and\ \citenamefont
  {{Wade}}}]{2015PhRvD..91d2003V}%
  \BibitemOpen
  \bibfield  {author} {\bibinfo {author} {\bibfnamefont {J.}~\bibnamefont
  {{Veitch}}}, \bibinfo {author} {\bibfnamefont {V.}~\bibnamefont {{Raymond}}},
  \bibinfo {author} {\bibfnamefont {B.}~\bibnamefont {{Farr}}}, \bibinfo
  {author} {\bibfnamefont {W.}~\bibnamefont {{Farr}}}, \bibinfo {author}
  {\bibfnamefont {P.}~\bibnamefont {{Graff}}}, \bibinfo {author} {\bibfnamefont
  {S.}~\bibnamefont {{Vitale}}}, \bibinfo {author} {\bibfnamefont
  {B.}~\bibnamefont {{Aylott}}}, \bibinfo {author} {\bibfnamefont
  {K.}~\bibnamefont {{Blackburn}}}, \bibinfo {author} {\bibfnamefont
  {N.}~\bibnamefont {{Christensen}}}, \bibinfo {author} {\bibfnamefont
  {M.}~\bibnamefont {{Coughlin}}}, \bibinfo {author} {\bibfnamefont
  {W.}~\bibnamefont {{Del Pozzo}}}, \bibinfo {author} {\bibfnamefont
  {F.}~\bibnamefont {{Feroz}}}, \bibinfo {author} {\bibfnamefont
  {J.}~\bibnamefont {{Gair}}}, \bibinfo {author} {\bibfnamefont {C.~J.}\
  \bibnamefont {{Haster}}}, \bibinfo {author} {\bibfnamefont {V.}~\bibnamefont
  {{Kalogera}}}, \bibinfo {author} {\bibfnamefont {T.}~\bibnamefont
  {{Littenberg}}}, \bibinfo {author} {\bibfnamefont {I.}~\bibnamefont
  {{Mandel}}}, \bibinfo {author} {\bibfnamefont {R.}~\bibnamefont
  {{O'Shaughnessy}}}, \bibinfo {author} {\bibfnamefont {M.}~\bibnamefont
  {{Pitkin}}}, \bibinfo {author} {\bibfnamefont {C.}~\bibnamefont
  {{Rodriguez}}}, \bibinfo {author} {\bibfnamefont {C.}~\bibnamefont
  {{R{\"o}ver}}}, \bibinfo {author} {\bibfnamefont {T.}~\bibnamefont
  {{Sidery}}}, \bibinfo {author} {\bibfnamefont {R.}~\bibnamefont {{Smith}}},
  \bibinfo {author} {\bibfnamefont {M.}~\bibnamefont {{Van Der Sluys}}},
  \bibinfo {author} {\bibfnamefont {A.}~\bibnamefont {{Vecchio}}}, \bibinfo
  {author} {\bibfnamefont {W.}~\bibnamefont {{Vousden}}},\ and\ \bibinfo
  {author} {\bibfnamefont {L.}~\bibnamefont {{Wade}}},\ }\bibfield  {title}
  {\bibinfo {title} {{Parameter estimation for compact binaries with
  ground-based gravitational-wave observations using the LALInference software
  library}},\ }\href {https://doi.org/10.1103/PhysRevD.91.042003} {\bibfield
  {journal} {\bibinfo  {journal} {\prd}\ }\textbf {\bibinfo {volume} {91}},\
  \bibinfo {eid} {042003} (\bibinfo {year} {2015})},\ \Eprint
  {https://arxiv.org/abs/1409.7215} {arXiv:1409.7215 [gr-qc]} \BibitemShut
  {NoStop}%
\bibitem [{\citenamefont {{Biwer}}\ \emph {et~al.}(2019)\citenamefont
  {{Biwer}}, \citenamefont {{Capano}}, \citenamefont {{De}}, \citenamefont
  {{Cabero}}, \citenamefont {{Brown}}, \citenamefont {{Nitz}},\ and\
  \citenamefont {{Raymond}}}]{2019PASP..131b4503B}%
  \BibitemOpen
  \bibfield  {author} {\bibinfo {author} {\bibfnamefont {C.~M.}\ \bibnamefont
  {{Biwer}}}, \bibinfo {author} {\bibfnamefont {C.~D.}\ \bibnamefont
  {{Capano}}}, \bibinfo {author} {\bibfnamefont {S.}~\bibnamefont {{De}}},
  \bibinfo {author} {\bibfnamefont {M.}~\bibnamefont {{Cabero}}}, \bibinfo
  {author} {\bibfnamefont {D.~A.}\ \bibnamefont {{Brown}}}, \bibinfo {author}
  {\bibfnamefont {A.~H.}\ \bibnamefont {{Nitz}}},\ and\ \bibinfo {author}
  {\bibfnamefont {V.}~\bibnamefont {{Raymond}}},\ }\bibfield  {title} {\bibinfo
  {title} {{PyCBC Inference: A Python-based Parameter Estimation Toolkit for
  Compact Binary Coalescence Signal}},\ }\href
  {https://doi.org/10.1088/1538-3873/aaef0b} {\bibfield  {journal} {\bibinfo
  {journal} {Publ. Astron. Soc. Pac.}\ }\textbf {\bibinfo {volume} {131}},\
  \bibinfo {pages} {024503} (\bibinfo {year} {2019})},\ \Eprint
  {https://arxiv.org/abs/1807.10312} {arXiv:1807.10312 [astro-ph.IM]}
  \BibitemShut {NoStop}%
\bibitem [{\citenamefont {{Ashton}}\ \emph {et~al.}(2019)\citenamefont
  {{Ashton}}, \citenamefont {{H{\"u}bner}}, \citenamefont {{Lasky}},
  \citenamefont {{Talbot}}, \citenamefont {{Ackley}}, \citenamefont
  {{Biscoveanu}}, \citenamefont {{Chu}}, \citenamefont {{Divakarla}},
  \citenamefont {{Easter}}, \citenamefont {{Goncharov}}, \citenamefont
  {{Hernandez Vivanco}}, \citenamefont {{Harms}}, \citenamefont {{Lower}},
  \citenamefont {{Meadors}}, \citenamefont {{Melchor}}, \citenamefont
  {{Payne}}, \citenamefont {{Pitkin}}, \citenamefont {{Powell}}, \citenamefont
  {{Sarin}}, \citenamefont {{Smith}},\ and\ \citenamefont
  {{Thrane}}}]{2019ApJS..241...27A}%
  \BibitemOpen
  \bibfield  {author} {\bibinfo {author} {\bibfnamefont {G.}~\bibnamefont
  {{Ashton}}}, \bibinfo {author} {\bibfnamefont {M.}~\bibnamefont
  {{H{\"u}bner}}}, \bibinfo {author} {\bibfnamefont {P.~D.}\ \bibnamefont
  {{Lasky}}}, \bibinfo {author} {\bibfnamefont {C.}~\bibnamefont {{Talbot}}},
  \bibinfo {author} {\bibfnamefont {K.}~\bibnamefont {{Ackley}}}, \bibinfo
  {author} {\bibfnamefont {S.}~\bibnamefont {{Biscoveanu}}}, \bibinfo {author}
  {\bibfnamefont {Q.}~\bibnamefont {{Chu}}}, \bibinfo {author} {\bibfnamefont
  {A.}~\bibnamefont {{Divakarla}}}, \bibinfo {author} {\bibfnamefont {P.~J.}\
  \bibnamefont {{Easter}}}, \bibinfo {author} {\bibfnamefont {B.}~\bibnamefont
  {{Goncharov}}}, \bibinfo {author} {\bibfnamefont {F.}~\bibnamefont
  {{Hernandez Vivanco}}}, \bibinfo {author} {\bibfnamefont {J.}~\bibnamefont
  {{Harms}}}, \bibinfo {author} {\bibfnamefont {M.~E.}\ \bibnamefont
  {{Lower}}}, \bibinfo {author} {\bibfnamefont {G.~D.}\ \bibnamefont
  {{Meadors}}}, \bibinfo {author} {\bibfnamefont {D.}~\bibnamefont
  {{Melchor}}}, \bibinfo {author} {\bibfnamefont {E.}~\bibnamefont {{Payne}}},
  \bibinfo {author} {\bibfnamefont {M.~D.}\ \bibnamefont {{Pitkin}}}, \bibinfo
  {author} {\bibfnamefont {J.}~\bibnamefont {{Powell}}}, \bibinfo {author}
  {\bibfnamefont {N.}~\bibnamefont {{Sarin}}}, \bibinfo {author} {\bibfnamefont
  {R.~J.~E.}\ \bibnamefont {{Smith}}},\ and\ \bibinfo {author} {\bibfnamefont
  {E.}~\bibnamefont {{Thrane}}},\ }\bibfield  {title} {\bibinfo {title}
  {{BILBY: A User-friendly Bayesian Inference Library for Gravitational-wave
  Astronomy}},\ }\href {https://doi.org/10.3847/1538-4365/ab06fc} {\bibfield
  {journal} {\bibinfo  {journal} {The Astrophysical Journal Supplement Series}\
  }\textbf {\bibinfo {volume} {241}},\ \bibinfo {eid} {27} (\bibinfo {year}
  {2019})},\ \Eprint {https://arxiv.org/abs/1811.02042} {arXiv:1811.02042
  [astro-ph.IM]} \BibitemShut {NoStop}%
\bibitem [{\citenamefont {Romero-Shaw}\ \emph {et~al.}(2020)\citenamefont
  {Romero-Shaw} \emph {et~al.}}]{Romero-Shaw:2020owr}%
  \BibitemOpen
  \bibfield  {author} {\bibinfo {author} {\bibfnamefont {I.~M.}\ \bibnamefont
  {Romero-Shaw}} \emph {et~al.},\ }\bibfield  {title} {\bibinfo {title}
  {{Bayesian inference for compact binary coalescences with bilby: validation
  and application to the first LIGO\textendash{}Virgo gravitational-wave
  transient catalogue}},\ }\href {https://doi.org/10.1093/mnras/staa2850}
  {\bibfield  {journal} {\bibinfo  {journal} {Mon. Not. Roy. Astron. Soc.}\
  }\textbf {\bibinfo {volume} {499}},\ \bibinfo {pages} {3295} (\bibinfo {year}
  {2020})},\ \Eprint {https://arxiv.org/abs/2006.00714} {arXiv:2006.00714
  [astro-ph.IM]} \BibitemShut {NoStop}%
\bibitem [{\citenamefont {Smith}\ \emph {et~al.}(2020)\citenamefont {Smith},
  \citenamefont {Ashton}, \citenamefont {Vajpeyi},\ and\ \citenamefont
  {Talbot}}]{Smith:2019ucc}%
  \BibitemOpen
  \bibfield  {author} {\bibinfo {author} {\bibfnamefont {R.~J.~E.}\
  \bibnamefont {Smith}}, \bibinfo {author} {\bibfnamefont {G.}~\bibnamefont
  {Ashton}}, \bibinfo {author} {\bibfnamefont {A.}~\bibnamefont {Vajpeyi}},\
  and\ \bibinfo {author} {\bibfnamefont {C.}~\bibnamefont {Talbot}},\
  }\bibfield  {title} {\bibinfo {title} {{Massively parallel Bayesian inference
  for transient gravitational-wave astronomy}},\ }\href
  {https://doi.org/10.1093/mnras/staa2483} {\bibfield  {journal} {\bibinfo
  {journal} {Mon. Not. Roy. Astron. Soc.}\ }\textbf {\bibinfo {volume} {498}},\
  \bibinfo {pages} {4492} (\bibinfo {year} {2020})},\ \Eprint
  {https://arxiv.org/abs/1909.11873} {arXiv:1909.11873 [gr-qc]} \BibitemShut
  {NoStop}%
\bibitem [{\citenamefont {Ashton}\ and\ \citenamefont
  {Talbot}(2021)}]{Ashton:2021anp}%
  \BibitemOpen
  \bibfield  {author} {\bibinfo {author} {\bibfnamefont {G.}~\bibnamefont
  {Ashton}}\ and\ \bibinfo {author} {\bibfnamefont {C.}~\bibnamefont
  {Talbot}},\ }\bibfield  {title} {\bibinfo {title} {{B\,ilby-MCMC: an MCMC
  sampler for gravitational-wave inference}},\ }\href
  {https://doi.org/10.1093/mnras/stab2236} {\bibfield  {journal} {\bibinfo
  {journal} {Mon. Not. Roy. Astron. Soc.}\ }\textbf {\bibinfo {volume} {507}},\
  \bibinfo {pages} {2037} (\bibinfo {year} {2021})},\ \Eprint
  {https://arxiv.org/abs/2106.08730} {arXiv:2106.08730 [gr-qc]} \BibitemShut
  {NoStop}%
\bibitem [{\citenamefont {Skilling}(2006)}]{10.1214/06-BA127}%
  \BibitemOpen
  \bibfield  {author} {\bibinfo {author} {\bibfnamefont {J.}~\bibnamefont
  {Skilling}},\ }\bibfield  {title} {\bibinfo {title} {{Nested sampling for
  general Bayesian computation}},\ }\href {https://doi.org/10.1214/06-BA127}
  {\bibfield  {journal} {\bibinfo  {journal} {Bayesian Analysis}\ }\textbf
  {\bibinfo {volume} {1}},\ \bibinfo {pages} {833 } (\bibinfo {year}
  {2006})}\BibitemShut {NoStop}%
\bibitem [{\citenamefont {Metropolis}\ and\ \citenamefont
  {Ulam}(1949)}]{metropolis1949monte}%
  \BibitemOpen
  \bibfield  {author} {\bibinfo {author} {\bibfnamefont {N.}~\bibnamefont
  {Metropolis}}\ and\ \bibinfo {author} {\bibfnamefont {S.}~\bibnamefont
  {Ulam}},\ }\bibfield  {title} {\bibinfo {title} {The monte carlo method},\
  }\href@noop {} {\bibfield  {journal} {\bibinfo  {journal} {Journal of the
  American statistical association}\ }\textbf {\bibinfo {volume} {44}},\
  \bibinfo {pages} {335} (\bibinfo {year} {1949})}\BibitemShut {NoStop}%
\bibitem [{\citenamefont {{Pankow}}\ \emph {et~al.}(2015)\citenamefont
  {{Pankow}}, \citenamefont {{Brady}}, \citenamefont {{Ochsner}},\ and\
  \citenamefont {{O'Shaughnessy}}}]{2015PhRvD..92b3002P}%
  \BibitemOpen
  \bibfield  {author} {\bibinfo {author} {\bibfnamefont {C.}~\bibnamefont
  {{Pankow}}}, \bibinfo {author} {\bibfnamefont {P.}~\bibnamefont {{Brady}}},
  \bibinfo {author} {\bibfnamefont {E.}~\bibnamefont {{Ochsner}}},\ and\
  \bibinfo {author} {\bibfnamefont {R.}~\bibnamefont {{O'Shaughnessy}}},\
  }\bibfield  {title} {\bibinfo {title} {{Novel scheme for rapid parallel
  parameter estimation of gravitational waves from compact binary
  coalescences}},\ }\href {https://doi.org/10.1103/PhysRevD.92.023002}
  {\bibfield  {journal} {\bibinfo  {journal} {\prd}\ }\textbf {\bibinfo
  {volume} {92}},\ \bibinfo {eid} {023002} (\bibinfo {year} {2015})},\ \Eprint
  {https://arxiv.org/abs/1502.04370} {arXiv:1502.04370 [gr-qc]} \BibitemShut
  {NoStop}%
\bibitem [{\citenamefont {{Lange}}\ \emph {et~al.}(2018)\citenamefont
  {{Lange}}, \citenamefont {{O'Shaughnessy}},\ and\ \citenamefont
  {{Rizzo}}}]{2018arXiv180510457L}%
  \BibitemOpen
  \bibfield  {author} {\bibinfo {author} {\bibfnamefont {J.}~\bibnamefont
  {{Lange}}}, \bibinfo {author} {\bibfnamefont {R.}~\bibnamefont
  {{O'Shaughnessy}}},\ and\ \bibinfo {author} {\bibfnamefont {M.}~\bibnamefont
  {{Rizzo}}},\ }\bibfield  {title} {\bibinfo {title} {{Rapid and accurate
  parameter inference for coalescing, precessing compact binaries}},\
  }\href@noop {} {\bibfield  {journal} {\bibinfo  {journal} {arXiv e-prints}\
  ,\ \bibinfo {eid} {arXiv:1805.10457}} (\bibinfo {year} {2018})},\ \Eprint
  {https://arxiv.org/abs/1805.10457} {arXiv:1805.10457 [gr-qc]} \BibitemShut
  {NoStop}%
\bibitem [{\citenamefont {{Wofford}}\ \emph {et~al.}(2022)\citenamefont
  {{Wofford}}, \citenamefont {{Yelikar}}, \citenamefont {{Gallagher}},
  \citenamefont {{Champion}}, \citenamefont {{Wysocki}}, \citenamefont
  {{Delfavero}}, \citenamefont {{Lange}}, \citenamefont {{Rose}}, \citenamefont
  {{Valsan}}, \citenamefont {{Morisaki}}, \citenamefont {{Read}}, \citenamefont
  {{Henshaw}},\ and\ \citenamefont {{O'Shaughnessy}}}]{Wofford:2022ykb}%
  \BibitemOpen
  \bibfield  {author} {\bibinfo {author} {\bibfnamefont {J.}~\bibnamefont
  {{Wofford}}}, \bibinfo {author} {\bibfnamefont {A.}~\bibnamefont
  {{Yelikar}}}, \bibinfo {author} {\bibfnamefont {H.}~\bibnamefont
  {{Gallagher}}}, \bibinfo {author} {\bibfnamefont {E.}~\bibnamefont
  {{Champion}}}, \bibinfo {author} {\bibfnamefont {D.}~\bibnamefont
  {{Wysocki}}}, \bibinfo {author} {\bibfnamefont {V.}~\bibnamefont
  {{Delfavero}}}, \bibinfo {author} {\bibfnamefont {J.}~\bibnamefont
  {{Lange}}}, \bibinfo {author} {\bibfnamefont {C.}~\bibnamefont {{Rose}}},
  \bibinfo {author} {\bibfnamefont {V.}~\bibnamefont {{Valsan}}}, \bibinfo
  {author} {\bibfnamefont {S.}~\bibnamefont {{Morisaki}}}, \bibinfo {author}
  {\bibfnamefont {J.}~\bibnamefont {{Read}}}, \bibinfo {author} {\bibfnamefont
  {C.}~\bibnamefont {{Henshaw}}},\ and\ \bibinfo {author} {\bibfnamefont
  {R.}~\bibnamefont {{O'Shaughnessy}}},\ }\bibfield  {title} {\bibinfo {title}
  {{Expanding RIFT: Improving performance for GW parameter inference}},\
  }\href@noop {} {\bibfield  {journal} {\bibinfo  {journal} {arXiv e-prints}\
  ,\ \bibinfo {eid} {arXiv:2210.07912}} (\bibinfo {year} {2022})},\ \Eprint
  {https://arxiv.org/abs/2210.07912} {arXiv:2210.07912 [gr-qc]} \BibitemShut
  {NoStop}%
\bibitem [{\citenamefont {Ashton}\ and\ \citenamefont
  {Dietrich}(2022)}]{Ashton:2021cub}%
  \BibitemOpen
  \bibfield  {author} {\bibinfo {author} {\bibfnamefont {G.}~\bibnamefont
  {Ashton}}\ and\ \bibinfo {author} {\bibfnamefont {T.}~\bibnamefont
  {Dietrich}},\ }\bibfield  {title} {\bibinfo {title} {{The use of hypermodels
  to understand binary neutron star collisions}},\ }\href
  {https://doi.org/10.1038/s41550-022-01707-x} {\bibfield  {journal} {\bibinfo
  {journal} {Nature Astron.}\ }\textbf {\bibinfo {volume} {6}},\ \bibinfo
  {pages} {961} (\bibinfo {year} {2022})},\ \Eprint
  {https://arxiv.org/abs/2111.09214} {arXiv:2111.09214 [gr-qc]} \BibitemShut
  {NoStop}%
\bibitem [{\citenamefont {Hoy}(2022)}]{Hoy:2022tst}%
  \BibitemOpen
  \bibfield  {author} {\bibinfo {author} {\bibfnamefont {C.}~\bibnamefont
  {Hoy}},\ }\bibfield  {title} {\bibinfo {title} {{Accelerating multimodel
  Bayesian inference, model selection, and systematic studies for gravitational
  wave astronomy}},\ }\href {https://doi.org/10.1103/PhysRevD.106.083003}
  {\bibfield  {journal} {\bibinfo  {journal} {Phys. Rev. D}\ }\textbf {\bibinfo
  {volume} {106}},\ \bibinfo {pages} {083003} (\bibinfo {year} {2022})},\
  \Eprint {https://arxiv.org/abs/2208.00106} {arXiv:2208.00106 [gr-qc]}
  \BibitemShut {NoStop}%
\bibitem [{\citenamefont {{Tiwari}}\ \emph {et~al.}(2016)\citenamefont
  {{Tiwari}}, \citenamefont {{Klimenko}}, \citenamefont {{Necula}},\ and\
  \citenamefont {{Mitselmakher}}}]{2016CQGra..33aLT01T}%
  \BibitemOpen
  \bibfield  {author} {\bibinfo {author} {\bibfnamefont {V.}~\bibnamefont
  {{Tiwari}}}, \bibinfo {author} {\bibfnamefont {S.}~\bibnamefont
  {{Klimenko}}}, \bibinfo {author} {\bibfnamefont {V.}~\bibnamefont
  {{Necula}}},\ and\ \bibinfo {author} {\bibfnamefont {G.}~\bibnamefont
  {{Mitselmakher}}},\ }\bibfield  {title} {\bibinfo {title} {{Reconstruction of
  chirp mass in searches for gravitational wave transients}},\ }\href
  {https://doi.org/10.1088/0264-9381/33/1/01LT01} {\bibfield  {journal}
  {\bibinfo  {journal} {Classical and Quantum Gravity}\ }\textbf {\bibinfo
  {volume} {33}},\ \bibinfo {eid} {01LT01} (\bibinfo {year} {2016})},\ \Eprint
  {https://arxiv.org/abs/1510.02426} {arXiv:1510.02426 [astro-ph.IM]}
  \BibitemShut {NoStop}%
\bibitem [{\citenamefont {Gabbard}\ \emph {et~al.}(2022)\citenamefont
  {Gabbard}, \citenamefont {Messenger}, \citenamefont {Heng}, \citenamefont
  {Tonolini},\ and\ \citenamefont {Murray-Smith}}]{Gabbard:2019rde}%
  \BibitemOpen
  \bibfield  {author} {\bibinfo {author} {\bibfnamefont {H.}~\bibnamefont
  {Gabbard}}, \bibinfo {author} {\bibfnamefont {C.}~\bibnamefont {Messenger}},
  \bibinfo {author} {\bibfnamefont {I.~S.}\ \bibnamefont {Heng}}, \bibinfo
  {author} {\bibfnamefont {F.}~\bibnamefont {Tonolini}},\ and\ \bibinfo
  {author} {\bibfnamefont {R.}~\bibnamefont {Murray-Smith}},\ }\bibfield
  {title} {\bibinfo {title} {{Bayesian parameter estimation using conditional
  variational autoencoders for gravitational-wave astronomy}},\ }\href
  {https://doi.org/10.1038/s41567-021-01425-7} {\bibfield  {journal} {\bibinfo
  {journal} {Nature Phys.}\ }\textbf {\bibinfo {volume} {18}},\ \bibinfo
  {pages} {112} (\bibinfo {year} {2022})},\ \Eprint
  {https://arxiv.org/abs/1909.06296} {arXiv:1909.06296 [astro-ph.IM]}
  \BibitemShut {NoStop}%
\bibitem [{\citenamefont {Green}\ and\ \citenamefont
  {Gair}(2021)}]{Green:2020dnx}%
  \BibitemOpen
  \bibfield  {author} {\bibinfo {author} {\bibfnamefont {S.~R.}\ \bibnamefont
  {Green}}\ and\ \bibinfo {author} {\bibfnamefont {J.}~\bibnamefont {Gair}},\
  }\bibfield  {title} {\bibinfo {title} {{Complete parameter inference for
  GW150914 using deep learning}},\ }\href
  {https://doi.org/10.1088/2632-2153/abfaed} {\bibfield  {journal} {\bibinfo
  {journal} {Mach. Learn. Sci. Tech.}\ }\textbf {\bibinfo {volume} {2}},\
  \bibinfo {pages} {03LT01} (\bibinfo {year} {2021})},\ \Eprint
  {https://arxiv.org/abs/2008.03312} {arXiv:2008.03312 [astro-ph.IM]}
  \BibitemShut {NoStop}%
\bibitem [{\citenamefont {Green}\ \emph {et~al.}(2020)\citenamefont {Green},
  \citenamefont {Simpson},\ and\ \citenamefont {Gair}}]{Green:2020hst}%
  \BibitemOpen
  \bibfield  {author} {\bibinfo {author} {\bibfnamefont {S.~R.}\ \bibnamefont
  {Green}}, \bibinfo {author} {\bibfnamefont {C.}~\bibnamefont {Simpson}},\
  and\ \bibinfo {author} {\bibfnamefont {J.}~\bibnamefont {Gair}},\ }\bibfield
  {title} {\bibinfo {title} {{Gravitational-wave parameter estimation with
  autoregressive neural network flows}},\ }\href
  {https://doi.org/10.1103/PhysRevD.102.104057} {\bibfield  {journal} {\bibinfo
   {journal} {Phys. Rev. D}\ }\textbf {\bibinfo {volume} {102}},\ \bibinfo
  {pages} {104057} (\bibinfo {year} {2020})},\ \Eprint
  {https://arxiv.org/abs/2002.07656} {arXiv:2002.07656 [astro-ph.IM]}
  \BibitemShut {NoStop}%
\bibitem [{\citenamefont {Dax}\ \emph {et~al.}(2021)\citenamefont {Dax},
  \citenamefont {Green}, \citenamefont {Gair}, \citenamefont {Macke},
  \citenamefont {Buonanno},\ and\ \citenamefont {Sch\"olkopf}}]{Dax:2021tsq}%
  \BibitemOpen
  \bibfield  {author} {\bibinfo {author} {\bibfnamefont {M.}~\bibnamefont
  {Dax}}, \bibinfo {author} {\bibfnamefont {S.~R.}\ \bibnamefont {Green}},
  \bibinfo {author} {\bibfnamefont {J.}~\bibnamefont {Gair}}, \bibinfo {author}
  {\bibfnamefont {J.~H.}\ \bibnamefont {Macke}}, \bibinfo {author}
  {\bibfnamefont {A.}~\bibnamefont {Buonanno}},\ and\ \bibinfo {author}
  {\bibfnamefont {B.}~\bibnamefont {Sch\"olkopf}},\ }\bibfield  {title}
  {\bibinfo {title} {{Real-Time Gravitational Wave Science with Neural
  Posterior Estimation}},\ }\href
  {https://doi.org/10.1103/PhysRevLett.127.241103} {\bibfield  {journal}
  {\bibinfo  {journal} {Phys. Rev. Lett.}\ }\textbf {\bibinfo {volume} {127}},\
  \bibinfo {pages} {241103} (\bibinfo {year} {2021})},\ \Eprint
  {https://arxiv.org/abs/2106.12594} {arXiv:2106.12594 [gr-qc]} \BibitemShut
  {NoStop}%
\bibitem [{\citenamefont {{Dax}}\ \emph {et~al.}(2022)\citenamefont {{Dax}},
  \citenamefont {{Green}}, \citenamefont {{Gair}}, \citenamefont
  {{P{\"u}rrer}}, \citenamefont {{Wildberger}}, \citenamefont {{Macke}},
  \citenamefont {{Buonanno}},\ and\ \citenamefont
  {{Sch{\"o}lkopf}}}]{Dax:2022pxd}%
  \BibitemOpen
  \bibfield  {author} {\bibinfo {author} {\bibfnamefont {M.}~\bibnamefont
  {{Dax}}}, \bibinfo {author} {\bibfnamefont {S.~R.}\ \bibnamefont {{Green}}},
  \bibinfo {author} {\bibfnamefont {J.}~\bibnamefont {{Gair}}}, \bibinfo
  {author} {\bibfnamefont {M.}~\bibnamefont {{P{\"u}rrer}}}, \bibinfo {author}
  {\bibfnamefont {J.}~\bibnamefont {{Wildberger}}}, \bibinfo {author}
  {\bibfnamefont {J.~H.}\ \bibnamefont {{Macke}}}, \bibinfo {author}
  {\bibfnamefont {A.}~\bibnamefont {{Buonanno}}},\ and\ \bibinfo {author}
  {\bibfnamefont {B.}~\bibnamefont {{Sch{\"o}lkopf}}},\ }\bibfield  {title}
  {\bibinfo {title} {{Neural Importance Sampling for Rapid and Reliable
  Gravitational-Wave Inference}},\ }\href@noop {} {\bibfield  {journal}
  {\bibinfo  {journal} {arXiv e-prints}\ ,\ \bibinfo {eid} {arXiv:2210.05686}}
  (\bibinfo {year} {2022})},\ \Eprint {https://arxiv.org/abs/2210.05686}
  {arXiv:2210.05686 [gr-qc]} \BibitemShut {NoStop}%
\bibitem [{\citenamefont {Shen}\ \emph {et~al.}(2022)\citenamefont {Shen},
  \citenamefont {Huerta}, \citenamefont {O'Shea}, \citenamefont {Kumar},\ and\
  \citenamefont {Zhao}}]{Shen:2019vep}%
  \BibitemOpen
  \bibfield  {author} {\bibinfo {author} {\bibfnamefont {H.}~\bibnamefont
  {Shen}}, \bibinfo {author} {\bibfnamefont {E.~A.}\ \bibnamefont {Huerta}},
  \bibinfo {author} {\bibfnamefont {E.}~\bibnamefont {O'Shea}}, \bibinfo
  {author} {\bibfnamefont {P.}~\bibnamefont {Kumar}},\ and\ \bibinfo {author}
  {\bibfnamefont {Z.}~\bibnamefont {Zhao}},\ }\bibfield  {title} {\bibinfo
  {title} {{Statistically-informed deep learning for gravitational wave
  parameter estimation}},\ }\href {https://doi.org/10.1088/2632-2153/ac3843}
  {\bibfield  {journal} {\bibinfo  {journal} {Mach. Learn. Sci. Tech.}\
  }\textbf {\bibinfo {volume} {3}},\ \bibinfo {pages} {015007} (\bibinfo {year}
  {2022})},\ \Eprint {https://arxiv.org/abs/1903.01998} {arXiv:1903.01998
  [gr-qc]} \BibitemShut {NoStop}%
\bibitem [{\citenamefont {Williams}\ \emph {et~al.}(2021)\citenamefont
  {Williams}, \citenamefont {Veitch},\ and\ \citenamefont
  {Messenger}}]{Williams:2021qyt}%
  \BibitemOpen
  \bibfield  {author} {\bibinfo {author} {\bibfnamefont {M.~J.}\ \bibnamefont
  {Williams}}, \bibinfo {author} {\bibfnamefont {J.}~\bibnamefont {Veitch}},\
  and\ \bibinfo {author} {\bibfnamefont {C.}~\bibnamefont {Messenger}},\
  }\bibfield  {title} {\bibinfo {title} {{Nested sampling with normalizing
  flows for gravitational-wave inference}},\ }\href
  {https://doi.org/10.1103/PhysRevD.103.103006} {\bibfield  {journal} {\bibinfo
   {journal} {Phys. Rev. D}\ }\textbf {\bibinfo {volume} {103}},\ \bibinfo
  {pages} {103006} (\bibinfo {year} {2021})},\ \Eprint
  {https://arxiv.org/abs/2102.11056} {arXiv:2102.11056 [gr-qc]} \BibitemShut
  {NoStop}%
\bibitem [{\citenamefont {{Farr}}\ \emph {et~al.}(2016)\citenamefont {{Farr}},
  \citenamefont {{Berry}}, \citenamefont {{Farr}}, \citenamefont {{Haster}},
  \citenamefont {{Middleton}}, \citenamefont {{Cannon}}, \citenamefont
  {{Graff}}, \citenamefont {{Hanna}}, \citenamefont {{Mandel}}, \citenamefont
  {{Pankow}}, \citenamefont {{Price}}, \citenamefont {{Sidery}}, \citenamefont
  {{Singer}}, \citenamefont {{Urban}}, \citenamefont {{Vecchio}}, \citenamefont
  {{Veitch}},\ and\ \citenamefont {{Vitale}}}]{2016ApJ...825..116F}%
  \BibitemOpen
  \bibfield  {author} {\bibinfo {author} {\bibfnamefont {B.}~\bibnamefont
  {{Farr}}}, \bibinfo {author} {\bibfnamefont {C.~P.~L.}\ \bibnamefont
  {{Berry}}}, \bibinfo {author} {\bibfnamefont {W.~M.}\ \bibnamefont {{Farr}}},
  \bibinfo {author} {\bibfnamefont {C.-J.}\ \bibnamefont {{Haster}}}, \bibinfo
  {author} {\bibfnamefont {H.}~\bibnamefont {{Middleton}}}, \bibinfo {author}
  {\bibfnamefont {K.}~\bibnamefont {{Cannon}}}, \bibinfo {author}
  {\bibfnamefont {P.~B.}\ \bibnamefont {{Graff}}}, \bibinfo {author}
  {\bibfnamefont {C.}~\bibnamefont {{Hanna}}}, \bibinfo {author} {\bibfnamefont
  {I.}~\bibnamefont {{Mandel}}}, \bibinfo {author} {\bibfnamefont
  {C.}~\bibnamefont {{Pankow}}}, \bibinfo {author} {\bibfnamefont {L.~R.}\
  \bibnamefont {{Price}}}, \bibinfo {author} {\bibfnamefont {T.}~\bibnamefont
  {{Sidery}}}, \bibinfo {author} {\bibfnamefont {L.~P.}\ \bibnamefont
  {{Singer}}}, \bibinfo {author} {\bibfnamefont {A.~L.}\ \bibnamefont
  {{Urban}}}, \bibinfo {author} {\bibfnamefont {A.}~\bibnamefont {{Vecchio}}},
  \bibinfo {author} {\bibfnamefont {J.}~\bibnamefont {{Veitch}}},\ and\
  \bibinfo {author} {\bibfnamefont {S.}~\bibnamefont {{Vitale}}},\ }\bibfield
  {title} {\bibinfo {title} {{Parameter Estimation on Gravitational Waves from
  Neutron-star Binaries with Spinning Components}},\ }\href
  {https://doi.org/10.3847/0004-637X/825/2/116} {\bibfield  {journal} {\bibinfo
   {journal} {\apj}\ }\textbf {\bibinfo {volume} {825}},\ \bibinfo {eid} {116}
  (\bibinfo {year} {2016})},\ \Eprint {https://arxiv.org/abs/1508.05336}
  {arXiv:1508.05336 [astro-ph.HE]} \BibitemShut {NoStop}%
\bibitem [{\citenamefont {{Berry}}\ \emph {et~al.}(2015)\citenamefont
  {{Berry}}, \citenamefont {{Mandel}}, \citenamefont {{Middleton}},
  \citenamefont {{Singer}}, \citenamefont {{Urban}}, \citenamefont {{Vecchio}},
  \citenamefont {{Vitale}}, \citenamefont {{Cannon}}, \citenamefont {{Farr}},
  \citenamefont {{Farr}}, \citenamefont {{Graff}}, \citenamefont {{Hanna}},
  \citenamefont {{Haster}}, \citenamefont {{Mohapatra}}, \citenamefont
  {{Pankow}}, \citenamefont {{Price}}, \citenamefont {{Sidery}},\ and\
  \citenamefont {{Veitch}}}]{2015ApJ...804..114B}%
  \BibitemOpen
  \bibfield  {author} {\bibinfo {author} {\bibfnamefont {C.~P.~L.}\
  \bibnamefont {{Berry}}}, \bibinfo {author} {\bibfnamefont {I.}~\bibnamefont
  {{Mandel}}}, \bibinfo {author} {\bibfnamefont {H.}~\bibnamefont
  {{Middleton}}}, \bibinfo {author} {\bibfnamefont {L.~P.}\ \bibnamefont
  {{Singer}}}, \bibinfo {author} {\bibfnamefont {A.~L.}\ \bibnamefont
  {{Urban}}}, \bibinfo {author} {\bibfnamefont {A.}~\bibnamefont {{Vecchio}}},
  \bibinfo {author} {\bibfnamefont {S.}~\bibnamefont {{Vitale}}}, \bibinfo
  {author} {\bibfnamefont {K.}~\bibnamefont {{Cannon}}}, \bibinfo {author}
  {\bibfnamefont {B.}~\bibnamefont {{Farr}}}, \bibinfo {author} {\bibfnamefont
  {W.~M.}\ \bibnamefont {{Farr}}}, \bibinfo {author} {\bibfnamefont {P.~B.}\
  \bibnamefont {{Graff}}}, \bibinfo {author} {\bibfnamefont {C.}~\bibnamefont
  {{Hanna}}}, \bibinfo {author} {\bibfnamefont {C.-J.}\ \bibnamefont
  {{Haster}}}, \bibinfo {author} {\bibfnamefont {S.}~\bibnamefont
  {{Mohapatra}}}, \bibinfo {author} {\bibfnamefont {C.}~\bibnamefont
  {{Pankow}}}, \bibinfo {author} {\bibfnamefont {L.~R.}\ \bibnamefont
  {{Price}}}, \bibinfo {author} {\bibfnamefont {T.}~\bibnamefont {{Sidery}}},\
  and\ \bibinfo {author} {\bibfnamefont {J.}~\bibnamefont {{Veitch}}},\
  }\bibfield  {title} {\bibinfo {title} {{Parameter Estimation for Binary
  Neutron-star Coalescences with Realistic Noise during the Advanced LIGO
  Era}},\ }\href {https://doi.org/10.1088/0004-637X/804/2/114} {\bibfield
  {journal} {\bibinfo  {journal} {\apj}\ }\textbf {\bibinfo {volume} {804}},\
  \bibinfo {eid} {114} (\bibinfo {year} {2015})},\ \Eprint
  {https://arxiv.org/abs/1411.6934} {arXiv:1411.6934 [astro-ph.HE]}
  \BibitemShut {NoStop}%
\bibitem [{\citenamefont {{Abbott}}\ \emph {et~al.}(2020)\citenamefont
  {{Abbott}} \emph {et~al.}}]{2020LRR....23....3A}%
  \BibitemOpen
  \bibfield  {author} {\bibinfo {author} {\bibfnamefont {R.}~\bibnamefont
  {{Abbott}}} \emph {et~al.},\ }\bibfield  {title} {\bibinfo {title}
  {{Prospects for observing and localizing gravitational-wave transients with
  Advanced LIGO, Advanced Virgo and KAGRA}},\ }\href
  {https://doi.org/10.1007/s41114-020-00026-9} {\bibfield  {journal} {\bibinfo
  {journal} {Living Reviews in Relativity}\ }\textbf {\bibinfo {volume} {23}},\
  \bibinfo {eid} {3} (\bibinfo {year} {2020})}\BibitemShut {NoStop}%
\bibitem [{\citenamefont {{P{\"u}rrer}}(2014)}]{2014CQGra..31s5010P}%
  \BibitemOpen
  \bibfield  {author} {\bibinfo {author} {\bibfnamefont {M.}~\bibnamefont
  {{P{\"u}rrer}}},\ }\bibfield  {title} {\bibinfo {title} {{Frequency-domain
  reduced order models for gravitational waves from aligned-spin compact
  binaries}},\ }\href {https://doi.org/10.1088/0264-9381/31/19/195010}
  {\bibfield  {journal} {\bibinfo  {journal} {Classical and Quantum Gravity}\
  }\textbf {\bibinfo {volume} {31}},\ \bibinfo {eid} {195010} (\bibinfo {year}
  {2014})},\ \Eprint {https://arxiv.org/abs/1402.4146} {arXiv:1402.4146
  [gr-qc]} \BibitemShut {NoStop}%
\bibitem [{\citenamefont {{Canizares}}\ \emph {et~al.}(2015)\citenamefont
  {{Canizares}}, \citenamefont {{Field}}, \citenamefont {{Gair}}, \citenamefont
  {{Raymond}}, \citenamefont {{Smith}},\ and\ \citenamefont
  {{Tiglio}}}]{2015PhRvL.114g1104C}%
  \BibitemOpen
  \bibfield  {author} {\bibinfo {author} {\bibfnamefont {P.}~\bibnamefont
  {{Canizares}}}, \bibinfo {author} {\bibfnamefont {S.~E.}\ \bibnamefont
  {{Field}}}, \bibinfo {author} {\bibfnamefont {J.}~\bibnamefont {{Gair}}},
  \bibinfo {author} {\bibfnamefont {V.}~\bibnamefont {{Raymond}}}, \bibinfo
  {author} {\bibfnamefont {R.}~\bibnamefont {{Smith}}},\ and\ \bibinfo {author}
  {\bibfnamefont {M.}~\bibnamefont {{Tiglio}}},\ }\bibfield  {title} {\bibinfo
  {title} {{Accelerated Gravitational Wave Parameter Estimation with Reduced
  Order Modeling}},\ }\href {https://doi.org/10.1103/PhysRevLett.114.071104}
  {\bibfield  {journal} {\bibinfo  {journal} {\prl}\ }\textbf {\bibinfo
  {volume} {114}},\ \bibinfo {eid} {071104} (\bibinfo {year} {2015})},\ \Eprint
  {https://arxiv.org/abs/1404.6284} {arXiv:1404.6284 [gr-qc]} \BibitemShut
  {NoStop}%
\bibitem [{\citenamefont {{Vinciguerra}}\ \emph {et~al.}(2017)\citenamefont
  {{Vinciguerra}}, \citenamefont {{Veitch}},\ and\ \citenamefont
  {{Mandel}}}]{2017CQGra..34k5006V}%
  \BibitemOpen
  \bibfield  {author} {\bibinfo {author} {\bibfnamefont {S.}~\bibnamefont
  {{Vinciguerra}}}, \bibinfo {author} {\bibfnamefont {J.}~\bibnamefont
  {{Veitch}}},\ and\ \bibinfo {author} {\bibfnamefont {I.}~\bibnamefont
  {{Mandel}}},\ }\bibfield  {title} {\bibinfo {title} {{Accelerating
  gravitational wave parameter estimation with multi-band template
  interpolation}},\ }\href {https://doi.org/10.1088/1361-6382/aa6d44}
  {\bibfield  {journal} {\bibinfo  {journal} {Classical and Quantum Gravity}\
  }\textbf {\bibinfo {volume} {34}},\ \bibinfo {eid} {115006} (\bibinfo {year}
  {2017})},\ \Eprint {https://arxiv.org/abs/1703.02062} {arXiv:1703.02062
  [gr-qc]} \BibitemShut {NoStop}%
\bibitem [{\citenamefont {{Setyawati}}\ \emph {et~al.}(2019)\citenamefont
  {{Setyawati}}, \citenamefont {{P{\"u}rrer}},\ and\ \citenamefont
  {{Ohme}}}]{2019arXiv190910986S}%
  \BibitemOpen
  \bibfield  {author} {\bibinfo {author} {\bibfnamefont {Y.}~\bibnamefont
  {{Setyawati}}}, \bibinfo {author} {\bibfnamefont {M.}~\bibnamefont
  {{P{\"u}rrer}}},\ and\ \bibinfo {author} {\bibfnamefont {F.}~\bibnamefont
  {{Ohme}}},\ }\bibfield  {title} {\bibinfo {title} {{Regression methods in
  waveform modeling: a comparative study}},\ }\href@noop {} {\bibfield
  {journal} {\bibinfo  {journal} {arXiv e-prints}\ ,\ \bibinfo {eid}
  {arXiv:1909.10986}} (\bibinfo {year} {2019})},\ \Eprint
  {https://arxiv.org/abs/1909.10986} {arXiv:1909.10986 [astro-ph.IM]}
  \BibitemShut {NoStop}%
\bibitem [{\citenamefont {Cornish}(2021)}]{Cornish:2021lje}%
  \BibitemOpen
  \bibfield  {author} {\bibinfo {author} {\bibfnamefont {N.~J.}\ \bibnamefont
  {Cornish}},\ }\bibfield  {title} {\bibinfo {title} {{Heterodyned likelihood
  for rapid gravitational wave parameter inference}},\ }\href
  {https://doi.org/10.1103/PhysRevD.104.104054} {\bibfield  {journal} {\bibinfo
   {journal} {Phys. Rev. D}\ }\textbf {\bibinfo {volume} {104}},\ \bibinfo
  {pages} {104054} (\bibinfo {year} {2021})},\ \Eprint
  {https://arxiv.org/abs/2109.02728} {arXiv:2109.02728 [gr-qc]} \BibitemShut
  {NoStop}%
\bibitem [{\citenamefont {{Abbott}}\ \emph {et~al.}(2016)\citenamefont
  {{Abbott}}, \citenamefont {{Abbott}}, \citenamefont {{Abbott}} \emph
  {et~al.}}]{2016PhRvL.116x1103A}%
  \BibitemOpen
  \bibfield  {author} {\bibinfo {author} {\bibfnamefont {B.~P.}\ \bibnamefont
  {{Abbott}}}, \bibinfo {author} {\bibfnamefont {R.}~\bibnamefont {{Abbott}}},
  \bibinfo {author} {\bibfnamefont {T.~D.}\ \bibnamefont {{Abbott}}}, \emph
  {et~al.},\ }\bibfield  {title} {\bibinfo {title} {{GW151226: Observation of
  Gravitational Waves from a 22-Solar-Mass Binary Black Hole Coalescence}},\
  }\href {https://doi.org/10.1103/PhysRevLett.116.241103} {\bibfield  {journal}
  {\bibinfo  {journal} {\prl}\ }\textbf {\bibinfo {volume} {116}},\ \bibinfo
  {eid} {241103} (\bibinfo {year} {2016})},\ \Eprint
  {https://arxiv.org/abs/1606.04855} {arXiv:1606.04855 [gr-qc]} \BibitemShut
  {NoStop}%
\bibitem [{\citenamefont {Abbott}\ \emph {et~al.}(2017)\citenamefont {Abbott}
  \emph {et~al.}}]{LIGOScientific:2017vwq}%
  \BibitemOpen
  \bibfield  {author} {\bibinfo {author} {\bibfnamefont {B.~P.}\ \bibnamefont
  {Abbott}} \emph {et~al.} (\bibinfo {collaboration} {LIGO Scientific,
  Virgo}),\ }\bibfield  {title} {\bibinfo {title} {{GW170817: Observation of
  Gravitational Waves from a Binary Neutron Star Inspiral}},\ }\href
  {https://doi.org/10.1103/PhysRevLett.119.161101} {\bibfield  {journal}
  {\bibinfo  {journal} {Phys. Rev. Lett.}\ }\textbf {\bibinfo {volume} {119}},\
  \bibinfo {pages} {161101} (\bibinfo {year} {2017})},\ \Eprint
  {https://arxiv.org/abs/1710.05832} {arXiv:1710.05832 [gr-qc]} \BibitemShut
  {NoStop}%
\bibitem [{\citenamefont {Hastings}(1970)}]{10.1093/biomet/57.1.97}%
  \BibitemOpen
  \bibfield  {author} {\bibinfo {author} {\bibfnamefont {W.~K.}\ \bibnamefont
  {Hastings}},\ }\bibfield  {title} {\bibinfo {title} {{Monte Carlo sampling
  methods using Markov chains and their applications}},\ }\href
  {https://doi.org/10.1093/biomet/57.1.97} {\bibfield  {journal} {\bibinfo
  {journal} {Biometrika}\ }\textbf {\bibinfo {volume} {57}},\ \bibinfo {pages}
  {97} (\bibinfo {year} {1970})},\ \Eprint
  {https://arxiv.org/abs/https://academic.oup.com/biomet/article-pdf/57/1/97/23940249/57-1-97.pdf}
  {https://academic.oup.com/biomet/article-pdf/57/1/97/23940249/57-1-97.pdf}
  \BibitemShut {NoStop}%
\bibitem [{\citenamefont {{Skilling}}(2004)}]{2004AIPC..735..395S}%
  \BibitemOpen
  \bibfield  {author} {\bibinfo {author} {\bibfnamefont {J.}~\bibnamefont
  {{Skilling}}},\ }\bibfield  {title} {\bibinfo {title} {{Nested Sampling}},\
  }in\ \href {https://doi.org/10.1063/1.1835238} {\emph {\bibinfo {booktitle}
  {American Institute of Physics Conference Series}}},\ Vol.\ \bibinfo {volume}
  {735},\ \bibinfo {editor} {edited by\ \bibinfo {editor} {\bibfnamefont
  {R.}~\bibnamefont {{Fischer}}}, \bibinfo {editor} {\bibfnamefont
  {R.}~\bibnamefont {{Preuss}}},\ and\ \bibinfo {editor} {\bibfnamefont
  {U.~V.}\ \bibnamefont {{Toussaint}}}}\ (\bibinfo {year} {2004})\ pp.\
  \bibinfo {pages} {395--405}\BibitemShut {NoStop}%
\bibitem [{\citenamefont {Higson}\ \emph {et~al.}(2019)\citenamefont {Higson},
  \citenamefont {Handley}, \citenamefont {Hobson},\ and\ \citenamefont
  {Lasenby}}]{higson2019dynamic}%
  \BibitemOpen
  \bibfield  {author} {\bibinfo {author} {\bibfnamefont {E.}~\bibnamefont
  {Higson}}, \bibinfo {author} {\bibfnamefont {W.}~\bibnamefont {Handley}},
  \bibinfo {author} {\bibfnamefont {M.}~\bibnamefont {Hobson}},\ and\ \bibinfo
  {author} {\bibfnamefont {A.}~\bibnamefont {Lasenby}},\ }\bibfield  {title}
  {\bibinfo {title} {Dynamic nested sampling: an improved algorithm for
  parameter estimation and evidence calculation},\ }\href@noop {} {\bibfield
  {journal} {\bibinfo  {journal} {Statistics and Computing}\ }\textbf {\bibinfo
  {volume} {29}},\ \bibinfo {pages} {891} (\bibinfo {year} {2019})}\BibitemShut
  {NoStop}%
\bibitem [{\citenamefont {Speagle}(2020)}]{Speagle:2019ivv}%
  \BibitemOpen
  \bibfield  {author} {\bibinfo {author} {\bibfnamefont {J.~S.}\ \bibnamefont
  {Speagle}},\ }\bibfield  {title} {\bibinfo {title} {{dynesty: a dynamic
  nested sampling package for estimating Bayesian posteriors and evidences}},\
  }\href {https://doi.org/10.1093/mnras/staa278} {\bibfield  {journal}
  {\bibinfo  {journal} {Mon. Not. Roy. Astron. Soc.}\ }\textbf {\bibinfo
  {volume} {493}},\ \bibinfo {pages} {3132} (\bibinfo {year} {2020})},\ \Eprint
  {https://arxiv.org/abs/1904.02180} {arXiv:1904.02180 [astro-ph.IM]}
  \BibitemShut {NoStop}%
\bibitem [{\citenamefont {Dvoretzky}\ \emph {et~al.}(1956)\citenamefont
  {Dvoretzky}, \citenamefont {Kiefer},\ and\ \citenamefont
  {Wolfowitz}}]{10.1214/aoms/1177728174}%
  \BibitemOpen
  \bibfield  {author} {\bibinfo {author} {\bibfnamefont {A.}~\bibnamefont
  {Dvoretzky}}, \bibinfo {author} {\bibfnamefont {J.}~\bibnamefont {Kiefer}},\
  and\ \bibinfo {author} {\bibfnamefont {J.}~\bibnamefont {Wolfowitz}},\
  }\bibfield  {title} {\bibinfo {title} {{Asymptotic Minimax Character of the
  Sample Distribution Function and of the Classical Multinomial Estimator}},\
  }\href {https://doi.org/10.1214/aoms/1177728174} {\bibfield  {journal}
  {\bibinfo  {journal} {The Annals of Mathematical Statistics}\ }\textbf
  {\bibinfo {volume} {27}},\ \bibinfo {pages} {642 } (\bibinfo {year}
  {1956})}\BibitemShut {NoStop}%
\bibitem [{\citenamefont {Casella}\ \emph {et~al.}(2004)\citenamefont
  {Casella}, \citenamefont {Robert},\ and\ \citenamefont
  {Wells}}]{10.2307/4356322}%
  \BibitemOpen
  \bibfield  {author} {\bibinfo {author} {\bibfnamefont {G.}~\bibnamefont
  {Casella}}, \bibinfo {author} {\bibfnamefont {C.~P.}\ \bibnamefont
  {Robert}},\ and\ \bibinfo {author} {\bibfnamefont {M.~T.}\ \bibnamefont
  {Wells}},\ }\bibfield  {title} {\bibinfo {title} {Generalized accept-reject
  sampling schemes},\ }\href {http://www.jstor.org/stable/4356322} {\bibfield
  {journal} {\bibinfo  {journal} {Lecture Notes-Monograph Series}\ }\textbf
  {\bibinfo {volume} {45}},\ \bibinfo {pages} {342} (\bibinfo {year}
  {2004})}\BibitemShut {NoStop}%
\bibitem [{\citenamefont {van Ravenzwaaij}\ \emph {et~al.}(2018)\citenamefont
  {van Ravenzwaaij}, \citenamefont {Cassey},\ and\ \citenamefont
  {Brown}}]{vanRavenzwaaij2018}%
  \BibitemOpen
  \bibfield  {author} {\bibinfo {author} {\bibfnamefont {D.}~\bibnamefont {van
  Ravenzwaaij}}, \bibinfo {author} {\bibfnamefont {P.}~\bibnamefont {Cassey}},\
  and\ \bibinfo {author} {\bibfnamefont {S.~D.}\ \bibnamefont {Brown}},\
  }\bibfield  {title} {\bibinfo {title} {A simple introduction to markov chain
  monte--carlo sampling},\ }\href {https://doi.org/10.3758/s13423-016-1015-8}
  {\bibfield  {journal} {\bibinfo  {journal} {Psychonomic Bulletin {\&}
  Review}\ }\textbf {\bibinfo {volume} {25}},\ \bibinfo {pages} {143} (\bibinfo
  {year} {2018})}\BibitemShut {NoStop}%
\bibitem [{\citenamefont {Martino}\ \emph {et~al.}(2017)\citenamefont
  {Martino}, \citenamefont {Victor},\ and\ \citenamefont {Carlos}}]{ESS}%
  \BibitemOpen
  \bibfield  {author} {\bibinfo {author} {\bibfnamefont {L.}~\bibnamefont
  {Martino}}, \bibinfo {author} {\bibfnamefont {E.}~\bibnamefont {Victor}},\
  and\ \bibinfo {author} {\bibfnamefont {S.}~\bibnamefont {Carlos}},\
  }\bibfield  {title} {\bibinfo {title} {{Effective sample size for importance
  sampling based on discrepancy measures}},\ }\href
  {https://doi.org/https://doi.org/10.1016/j.sigpro.2016.08.025} {\bibfield
  {journal} {\bibinfo  {journal} {Signal Processing-Elsevier}\ ,\ \bibinfo
  {pages} {386}} (\bibinfo {year} {2017})},\ \Eprint
  {https://arxiv.org/abs/1602.03572v5} {arXiv:1602.03572v5 [Computation
  (stat.CO)]} \BibitemShut {NoStop}%
\bibitem [{\citenamefont {Nydick}(2012)}]{nydick2012wishart}%
  \BibitemOpen
  \bibfield  {author} {\bibinfo {author} {\bibfnamefont {S.~W.}\ \bibnamefont
  {Nydick}},\ }\bibfield  {title} {\bibinfo {title} {The wishart and inverse
  wishart distributions},\ }\href@noop {} {\bibfield  {journal} {\bibinfo
  {journal} {Electronic Journal of Statistics}\ }\textbf {\bibinfo {volume}
  {6}} (\bibinfo {year} {2012})}\BibitemShut {NoStop}%
\bibitem [{\citenamefont {Foreman-Mackey}(2016)}]{corner}%
  \BibitemOpen
  \bibfield  {author} {\bibinfo {author} {\bibfnamefont {D.}~\bibnamefont
  {Foreman-Mackey}},\ }\bibfield  {title} {\bibinfo {title} {corner.py:
  Scatterplot matrices in python},\ }\href
  {https://doi.org/10.21105/joss.00024} {\bibfield  {journal} {\bibinfo
  {journal} {The Journal of Open Source Software}\ }\textbf {\bibinfo {volume}
  {1}},\ \bibinfo {pages} {24} (\bibinfo {year} {2016})}\BibitemShut {NoStop}%
\bibitem [{\citenamefont {{Zackay}}\ \emph {et~al.}(2018)\citenamefont
  {{Zackay}}, \citenamefont {{Dai}},\ and\ \citenamefont
  {{Venumadhav}}}]{2018arXiv180608792Z}%
  \BibitemOpen
  \bibfield  {author} {\bibinfo {author} {\bibfnamefont {B.}~\bibnamefont
  {{Zackay}}}, \bibinfo {author} {\bibfnamefont {L.}~\bibnamefont {{Dai}}},\
  and\ \bibinfo {author} {\bibfnamefont {T.}~\bibnamefont {{Venumadhav}}},\
  }\bibfield  {title} {\bibinfo {title} {{Relative Binning and Fast Likelihood
  Evaluation for Gravitational Wave Parameter Estimation}},\ }\href@noop {}
  {\bibfield  {journal} {\bibinfo  {journal} {arXiv e-prints}\ ,\ \bibinfo
  {eid} {arXiv:1806.08792}} (\bibinfo {year} {2018})},\ \Eprint
  {https://arxiv.org/abs/1806.08792} {arXiv:1806.08792 [astro-ph.IM]}
  \BibitemShut {NoStop}%
\bibitem [{\citenamefont {Leslie}\ \emph {et~al.}(2021)\citenamefont {Leslie},
  \citenamefont {Dai},\ and\ \citenamefont {Pratten}}]{Leslie:2021ssu}%
  \BibitemOpen
  \bibfield  {author} {\bibinfo {author} {\bibfnamefont {N.}~\bibnamefont
  {Leslie}}, \bibinfo {author} {\bibfnamefont {L.}~\bibnamefont {Dai}},\ and\
  \bibinfo {author} {\bibfnamefont {G.}~\bibnamefont {Pratten}},\ }\bibfield
  {title} {\bibinfo {title} {{Mode-by-mode relative binning: Fast likelihood
  estimation for gravitational waveforms with spin-orbit precession and
  multiple harmonics}},\ }\href {https://doi.org/10.1103/PhysRevD.104.123030}
  {\bibfield  {journal} {\bibinfo  {journal} {Phys. Rev. D}\ }\textbf {\bibinfo
  {volume} {104}},\ \bibinfo {pages} {123030} (\bibinfo {year} {2021})},\
  \Eprint {https://arxiv.org/abs/2109.09872} {arXiv:2109.09872 [astro-ph.IM]}
  \BibitemShut {NoStop}%
\bibitem [{\citenamefont {{Mills}}\ and\ \citenamefont
  {{Fairhurst}}(2021)}]{2021PhRvD.103b4042M}%
  \BibitemOpen
  \bibfield  {author} {\bibinfo {author} {\bibfnamefont {C.}~\bibnamefont
  {{Mills}}}\ and\ \bibinfo {author} {\bibfnamefont {S.}~\bibnamefont
  {{Fairhurst}}},\ }\bibfield  {title} {\bibinfo {title} {{Measuring
  gravitational-wave higher-order multipoles}},\ }\href
  {https://doi.org/10.1103/PhysRevD.103.024042} {\bibfield  {journal} {\bibinfo
   {journal} {\prd}\ }\textbf {\bibinfo {volume} {103}},\ \bibinfo {eid}
  {024042} (\bibinfo {year} {2021})},\ \Eprint
  {https://arxiv.org/abs/2007.04313} {arXiv:2007.04313 [gr-qc]} \BibitemShut
  {NoStop}%
\bibitem [{\citenamefont {{Husa}}\ \emph {et~al.}(2016)\citenamefont {{Husa}},
  \citenamefont {{Khan}}, \citenamefont {{Hannam}}, \citenamefont
  {{P{\"u}rrer}}, \citenamefont {{Ohme}}, \citenamefont {{Forteza}},\ and\
  \citenamefont {{Boh{\'e}}}}]{2016PhRvD..93d4006H}%
  \BibitemOpen
  \bibfield  {author} {\bibinfo {author} {\bibfnamefont {S.}~\bibnamefont
  {{Husa}}}, \bibinfo {author} {\bibfnamefont {S.}~\bibnamefont {{Khan}}},
  \bibinfo {author} {\bibfnamefont {M.}~\bibnamefont {{Hannam}}}, \bibinfo
  {author} {\bibfnamefont {M.}~\bibnamefont {{P{\"u}rrer}}}, \bibinfo {author}
  {\bibfnamefont {F.}~\bibnamefont {{Ohme}}}, \bibinfo {author} {\bibfnamefont
  {X.~J.}\ \bibnamefont {{Forteza}}},\ and\ \bibinfo {author} {\bibfnamefont
  {A.}~\bibnamefont {{Boh{\'e}}}},\ }\bibfield  {title} {\bibinfo {title}
  {{Frequency-domain gravitational waves from nonprecessing black-hole
  binaries. I. New numerical waveforms and anatomy of the signal}},\ }\href
  {https://doi.org/10.1103/PhysRevD.93.044006} {\bibfield  {journal} {\bibinfo
  {journal} {\prd}\ }\textbf {\bibinfo {volume} {93}},\ \bibinfo {eid} {044006}
  (\bibinfo {year} {2016})},\ \Eprint {https://arxiv.org/abs/1508.07250}
  {arXiv:1508.07250 [gr-qc]} \BibitemShut {NoStop}%
\bibitem [{\citenamefont {Khan}\ \emph {et~al.}(2016)\citenamefont {Khan} \emph
  {et~al.}}]{phenomd2}%
  \BibitemOpen
  \bibfield  {author} {\bibinfo {author} {\bibfnamefont {S.}~\bibnamefont
  {Khan}} \emph {et~al.},\ }\bibfield  {title} {\bibinfo {title}
  {{Frequency-domain gravitational waves from nonprecessing black-hole
  binaries. II. A phenomenological model for the advanced detector era}},\
  }\href {https://journals.aps.org/prd/abstract/10.1103/PhysRevD.93.044007}
  {\bibfield  {journal} {\bibinfo  {journal} {Phys. Rev. D}\ }\textbf {\bibinfo
  {volume} {93}} (\bibinfo {year} {2016})}\BibitemShut {NoStop}%
\bibitem [{\citenamefont {Taracchini}\ \emph {et~al.}(2012)\citenamefont
  {Taracchini}, \citenamefont {Pan}, \citenamefont {Buonanno}, \citenamefont
  {Barausse}, \citenamefont {Boyle}, \citenamefont {Chu}, \citenamefont
  {Lovelace}, \citenamefont {Pfeiffer},\ and\ \citenamefont
  {Scheel}}]{Taracchini:2012ig}%
  \BibitemOpen
  \bibfield  {author} {\bibinfo {author} {\bibfnamefont {A.}~\bibnamefont
  {Taracchini}}, \bibinfo {author} {\bibfnamefont {Y.}~\bibnamefont {Pan}},
  \bibinfo {author} {\bibfnamefont {A.}~\bibnamefont {Buonanno}}, \bibinfo
  {author} {\bibfnamefont {E.}~\bibnamefont {Barausse}}, \bibinfo {author}
  {\bibfnamefont {M.}~\bibnamefont {Boyle}}, \bibinfo {author} {\bibfnamefont
  {T.}~\bibnamefont {Chu}}, \bibinfo {author} {\bibfnamefont {G.}~\bibnamefont
  {Lovelace}}, \bibinfo {author} {\bibfnamefont {H.~P.}\ \bibnamefont
  {Pfeiffer}},\ and\ \bibinfo {author} {\bibfnamefont {M.~A.}\ \bibnamefont
  {Scheel}},\ }\bibfield  {title} {\bibinfo {title} {{Prototype
  effective-one-body model for nonprecessing spinning inspiral-merger-ringdown
  waveforms}},\ }\href {https://doi.org/10.1103/PhysRevD.86.024011} {\bibfield
  {journal} {\bibinfo  {journal} {Phys. Rev. D}\ }\textbf {\bibinfo {volume}
  {86}},\ \bibinfo {pages} {024011} (\bibinfo {year} {2012})},\ \Eprint
  {https://arxiv.org/abs/1202.0790} {arXiv:1202.0790 [gr-qc]} \BibitemShut
  {NoStop}%
\bibitem [{\citenamefont {Taracchini}\ \emph {et~al.}(2014)\citenamefont
  {Taracchini} \emph {et~al.}}]{Taracchini:2013rva}%
  \BibitemOpen
  \bibfield  {author} {\bibinfo {author} {\bibfnamefont {A.}~\bibnamefont
  {Taracchini}} \emph {et~al.},\ }\bibfield  {title} {\bibinfo {title}
  {{Effective-one-body model for black-hole binaries with generic mass ratios
  and spins}},\ }\href {https://doi.org/10.1103/PhysRevD.89.061502} {\bibfield
  {journal} {\bibinfo  {journal} {Phys. Rev. D}\ }\textbf {\bibinfo {volume}
  {89}},\ \bibinfo {pages} {061502} (\bibinfo {year} {2014})},\ \Eprint
  {https://arxiv.org/abs/1311.2544} {arXiv:1311.2544 [gr-qc]} \BibitemShut
  {NoStop}%
\bibitem [{\citenamefont {Boh\'e}\ \emph {et~al.}(2017)\citenamefont {Boh\'e}
  \emph {et~al.}}]{Bohe:2016gbl}%
  \BibitemOpen
  \bibfield  {author} {\bibinfo {author} {\bibfnamefont {A.}~\bibnamefont
  {Boh\'e}} \emph {et~al.},\ }\bibfield  {title} {\bibinfo {title} {{Improved
  effective-one-body model of spinning, nonprecessing binary black holes for
  the era of gravitational-wave astrophysics with advanced detectors}},\ }\href
  {https://doi.org/10.1103/PhysRevD.95.044028} {\bibfield  {journal} {\bibinfo
  {journal} {Phys. Rev. D}\ }\textbf {\bibinfo {volume} {95}},\ \bibinfo
  {pages} {044028} (\bibinfo {year} {2017})},\ \Eprint
  {https://arxiv.org/abs/1611.03703} {arXiv:1611.03703 [gr-qc]} \BibitemShut
  {NoStop}%
\bibitem [{\citenamefont {Pratten}\ \emph {et~al.}(2020)\citenamefont
  {Pratten}, \citenamefont {Husa}, \citenamefont {Garcia-Quiros}, \citenamefont
  {Colleoni}, \citenamefont {Ramos-Buades}, \citenamefont {Estelles},\ and\
  \citenamefont {Jaume}}]{Pratten:2020fqn}%
  \BibitemOpen
  \bibfield  {author} {\bibinfo {author} {\bibfnamefont {G.}~\bibnamefont
  {Pratten}}, \bibinfo {author} {\bibfnamefont {S.}~\bibnamefont {Husa}},
  \bibinfo {author} {\bibfnamefont {C.}~\bibnamefont {Garcia-Quiros}}, \bibinfo
  {author} {\bibfnamefont {M.}~\bibnamefont {Colleoni}}, \bibinfo {author}
  {\bibfnamefont {A.}~\bibnamefont {Ramos-Buades}}, \bibinfo {author}
  {\bibfnamefont {H.}~\bibnamefont {Estelles}},\ and\ \bibinfo {author}
  {\bibfnamefont {R.}~\bibnamefont {Jaume}},\ }\bibfield  {title} {\bibinfo
  {title} {{Setting the cornerstone for a family of models for gravitational
  waves from compact binaries: The dominant harmonic for nonprecessing
  quasicircular black holes}},\ }\href
  {https://doi.org/10.1103/PhysRevD.102.064001} {\bibfield  {journal} {\bibinfo
   {journal} {Phys. Rev. D}\ }\textbf {\bibinfo {volume} {102}},\ \bibinfo
  {pages} {064001} (\bibinfo {year} {2020})},\ \Eprint
  {https://arxiv.org/abs/2001.11412} {arXiv:2001.11412 [gr-qc]} \BibitemShut
  {NoStop}%
\bibitem [{\citenamefont {Kalaghatgi}\ \emph {et~al.}(2020)\citenamefont
  {Kalaghatgi}, \citenamefont {Hannam},\ and\ \citenamefont
  {Raymond}}]{Kalaghatgi:2019log}%
  \BibitemOpen
  \bibfield  {author} {\bibinfo {author} {\bibfnamefont {C.}~\bibnamefont
  {Kalaghatgi}}, \bibinfo {author} {\bibfnamefont {M.}~\bibnamefont {Hannam}},\
  and\ \bibinfo {author} {\bibfnamefont {V.}~\bibnamefont {Raymond}},\
  }\bibfield  {title} {\bibinfo {title} {{Parameter estimation with a spinning
  multimode waveform model}},\ }\href
  {https://doi.org/10.1103/PhysRevD.101.103004} {\bibfield  {journal} {\bibinfo
   {journal} {Phys. Rev. D}\ }\textbf {\bibinfo {volume} {101}},\ \bibinfo
  {pages} {103004} (\bibinfo {year} {2020})},\ \Eprint
  {https://arxiv.org/abs/1909.10010} {arXiv:1909.10010 [gr-qc]} \BibitemShut
  {NoStop}%
\bibitem [{\citenamefont {Shaik}\ \emph {et~al.}(2020)\citenamefont {Shaik},
  \citenamefont {Lange}, \citenamefont {Field}, \citenamefont {O'Shaughnessy},
  \citenamefont {Varma}, \citenamefont {Kidder}, \citenamefont {Pfeiffer},\
  and\ \citenamefont {Wysocki}}]{Shaik:2019dym}%
  \BibitemOpen
  \bibfield  {author} {\bibinfo {author} {\bibfnamefont {F.~H.}\ \bibnamefont
  {Shaik}}, \bibinfo {author} {\bibfnamefont {J.}~\bibnamefont {Lange}},
  \bibinfo {author} {\bibfnamefont {S.~E.}\ \bibnamefont {Field}}, \bibinfo
  {author} {\bibfnamefont {R.}~\bibnamefont {O'Shaughnessy}}, \bibinfo {author}
  {\bibfnamefont {V.}~\bibnamefont {Varma}}, \bibinfo {author} {\bibfnamefont
  {L.~E.}\ \bibnamefont {Kidder}}, \bibinfo {author} {\bibfnamefont {H.~P.}\
  \bibnamefont {Pfeiffer}},\ and\ \bibinfo {author} {\bibfnamefont
  {D.}~\bibnamefont {Wysocki}},\ }\bibfield  {title} {\bibinfo {title} {{Impact
  of subdominant modes on the interpretation of gravitational-wave signals from
  heavy binary black hole systems}},\ }\href
  {https://doi.org/10.1103/PhysRevD.101.124054} {\bibfield  {journal} {\bibinfo
   {journal} {Phys. Rev. D}\ }\textbf {\bibinfo {volume} {101}},\ \bibinfo
  {pages} {124054} (\bibinfo {year} {2020})},\ \Eprint
  {https://arxiv.org/abs/1911.02693} {arXiv:1911.02693 [gr-qc]} \BibitemShut
  {NoStop}%
\bibitem [{\citenamefont {Abbott}\ \emph
  {et~al.}(2020{\natexlab{a}})\citenamefont {Abbott} \emph
  {et~al.}}]{LIGOScientific:2020ufj}%
  \BibitemOpen
  \bibfield  {author} {\bibinfo {author} {\bibfnamefont {R.}~\bibnamefont
  {Abbott}} \emph {et~al.} (\bibinfo {collaboration} {LIGO Scientific,
  Virgo}),\ }\bibfield  {title} {\bibinfo {title} {{Properties and
  Astrophysical Implications of the 150 M$_\odot$ Binary Black Hole Merger
  GW190521}},\ }\href {https://doi.org/10.3847/2041-8213/aba493} {\bibfield
  {journal} {\bibinfo  {journal} {Astrophys. J. Lett.}\ }\textbf {\bibinfo
  {volume} {900}},\ \bibinfo {pages} {L13} (\bibinfo {year}
  {2020}{\natexlab{a}})},\ \Eprint {https://arxiv.org/abs/2009.01190}
  {arXiv:2009.01190 [astro-ph.HE]} \BibitemShut {NoStop}%
\bibitem [{\citenamefont {Abbott}\ \emph
  {et~al.}(2020{\natexlab{b}})\citenamefont {Abbott} \emph
  {et~al.}}]{LIGOScientific:2020stg}%
  \BibitemOpen
  \bibfield  {author} {\bibinfo {author} {\bibfnamefont {R.}~\bibnamefont
  {Abbott}} \emph {et~al.} (\bibinfo {collaboration} {LIGO Scientific,
  Virgo}),\ }\bibfield  {title} {\bibinfo {title} {{GW190412: Observation of a
  Binary-Black-Hole Coalescence with Asymmetric Masses}},\ }\href
  {https://doi.org/10.1103/PhysRevD.102.043015} {\bibfield  {journal} {\bibinfo
   {journal} {Phys. Rev. D}\ }\textbf {\bibinfo {volume} {102}},\ \bibinfo
  {pages} {043015} (\bibinfo {year} {2020}{\natexlab{b}})},\ \Eprint
  {https://arxiv.org/abs/2004.08342} {arXiv:2004.08342 [astro-ph.HE]}
  \BibitemShut {NoStop}%
\bibitem [{\citenamefont {Abbott}\ \emph
  {et~al.}(2020{\natexlab{c}})\citenamefont {Abbott} \emph
  {et~al.}}]{LIGOScientific:2020zkf}%
  \BibitemOpen
  \bibfield  {author} {\bibinfo {author} {\bibfnamefont {R.}~\bibnamefont
  {Abbott}} \emph {et~al.} (\bibinfo {collaboration} {LIGO Scientific,
  Virgo}),\ }\bibfield  {title} {\bibinfo {title} {{GW190814: Gravitational
  Waves from the Coalescence of a 23 Solar Mass Black Hole with a 2.6 Solar
  Mass Compact Object}},\ }\href {https://doi.org/10.3847/2041-8213/ab960f}
  {\bibfield  {journal} {\bibinfo  {journal} {Astrophys. J. Lett.}\ }\textbf
  {\bibinfo {volume} {896}},\ \bibinfo {pages} {L44} (\bibinfo {year}
  {2020}{\natexlab{c}})},\ \Eprint {https://arxiv.org/abs/2006.12611}
  {arXiv:2006.12611 [astro-ph.HE]} \BibitemShut {NoStop}%
\bibitem [{\citenamefont {{Hannam}}\ \emph {et~al.}(2022)\citenamefont
  {{Hannam}}, \citenamefont {{Hoy}}, \citenamefont {{Thompson}}, \citenamefont
  {{Fairhurst}}, \citenamefont {{Raymond}}, \citenamefont {{Colleoni}},
  \citenamefont {{Davis}}, \citenamefont {{Estell{\'e}s}}, \citenamefont
  {{Haster}}, \citenamefont {{Helmling-Cornell}}, \citenamefont {{Husa}},
  \citenamefont {{Keitel}}, \citenamefont {{Massinger}}, \citenamefont
  {{Men{\'e}ndez-V{\'a}zquez}}, \citenamefont {{Mogushi}}, \citenamefont
  {{Ossokine}}, \citenamefont {{Payne}}, \citenamefont {{Pratten}},
  \citenamefont {{Romero-Shaw}}, \citenamefont {{Sadiq}}, \citenamefont
  {{Schmidt}}, \citenamefont {{Tenorio}}, \citenamefont {{Udall}},
  \citenamefont {{Veitch}}, \citenamefont {{Williams}}, \citenamefont
  {{Yelikar}},\ and\ \citenamefont {{Zimmerman}}}]{2022Natur.610..652H}%
  \BibitemOpen
  \bibfield  {author} {\bibinfo {author} {\bibfnamefont {M.}~\bibnamefont
  {{Hannam}}}, \bibinfo {author} {\bibfnamefont {C.}~\bibnamefont {{Hoy}}},
  \bibinfo {author} {\bibfnamefont {J.~E.}\ \bibnamefont {{Thompson}}},
  \bibinfo {author} {\bibfnamefont {S.}~\bibnamefont {{Fairhurst}}}, \bibinfo
  {author} {\bibfnamefont {V.}~\bibnamefont {{Raymond}}}, \bibinfo {author}
  {\bibfnamefont {M.}~\bibnamefont {{Colleoni}}}, \bibinfo {author}
  {\bibfnamefont {D.}~\bibnamefont {{Davis}}}, \bibinfo {author} {\bibfnamefont
  {H.}~\bibnamefont {{Estell{\'e}s}}}, \bibinfo {author} {\bibfnamefont
  {C.-J.}\ \bibnamefont {{Haster}}}, \bibinfo {author} {\bibfnamefont
  {A.}~\bibnamefont {{Helmling-Cornell}}}, \bibinfo {author} {\bibfnamefont
  {S.}~\bibnamefont {{Husa}}}, \bibinfo {author} {\bibfnamefont
  {D.}~\bibnamefont {{Keitel}}}, \bibinfo {author} {\bibfnamefont {T.~J.}\
  \bibnamefont {{Massinger}}}, \bibinfo {author} {\bibfnamefont
  {A.}~\bibnamefont {{Men{\'e}ndez-V{\'a}zquez}}}, \bibinfo {author}
  {\bibfnamefont {K.}~\bibnamefont {{Mogushi}}}, \bibinfo {author}
  {\bibfnamefont {S.}~\bibnamefont {{Ossokine}}}, \bibinfo {author}
  {\bibfnamefont {E.}~\bibnamefont {{Payne}}}, \bibinfo {author} {\bibfnamefont
  {G.}~\bibnamefont {{Pratten}}}, \bibinfo {author} {\bibfnamefont
  {I.}~\bibnamefont {{Romero-Shaw}}}, \bibinfo {author} {\bibfnamefont
  {J.}~\bibnamefont {{Sadiq}}}, \bibinfo {author} {\bibfnamefont
  {P.}~\bibnamefont {{Schmidt}}}, \bibinfo {author} {\bibfnamefont
  {R.}~\bibnamefont {{Tenorio}}}, \bibinfo {author} {\bibfnamefont
  {R.}~\bibnamefont {{Udall}}}, \bibinfo {author} {\bibfnamefont
  {J.}~\bibnamefont {{Veitch}}}, \bibinfo {author} {\bibfnamefont
  {D.}~\bibnamefont {{Williams}}}, \bibinfo {author} {\bibfnamefont {A.~B.}\
  \bibnamefont {{Yelikar}}},\ and\ \bibinfo {author} {\bibfnamefont
  {A.}~\bibnamefont {{Zimmerman}}},\ }\bibfield  {title} {\bibinfo {title}
  {{General-relativistic precession in a black-hole binary}},\ }\href
  {https://doi.org/10.1038/s41586-022-05212-z} {\bibfield  {journal} {\bibinfo
  {journal} {\nat}\ }\textbf {\bibinfo {volume} {610}},\ \bibinfo {pages} {652}
  (\bibinfo {year} {2022})},\ \Eprint {https://arxiv.org/abs/2112.11300}
  {arXiv:2112.11300 [gr-qc]} \BibitemShut {NoStop}%
\bibitem [{\citenamefont {Krishnendu}\ and\ \citenamefont
  {Ohme}(2022)}]{Krishnendu:2021cyi}%
  \BibitemOpen
  \bibfield  {author} {\bibinfo {author} {\bibfnamefont {N.~V.}\ \bibnamefont
  {Krishnendu}}\ and\ \bibinfo {author} {\bibfnamefont {F.}~\bibnamefont
  {Ohme}},\ }\bibfield  {title} {\bibinfo {title} {{Interplay of
  spin-precession and higher harmonics in the parameter estimation of binary
  black holes}},\ }\href {https://doi.org/10.1103/PhysRevD.105.064012}
  {\bibfield  {journal} {\bibinfo  {journal} {Phys. Rev. D}\ }\textbf {\bibinfo
  {volume} {105}},\ \bibinfo {pages} {064012} (\bibinfo {year} {2022})},\
  \Eprint {https://arxiv.org/abs/2110.00766} {arXiv:2110.00766 [gr-qc]}
  \BibitemShut {NoStop}%
\bibitem [{\citenamefont {Aasi}\ \emph {et~al.}(2015)\citenamefont {Aasi} \emph
  {et~al.}}]{LIGOScientific:2014pky}%
  \BibitemOpen
  \bibfield  {author} {\bibinfo {author} {\bibfnamefont {J.}~\bibnamefont
  {Aasi}} \emph {et~al.} (\bibinfo {collaboration} {LIGO Scientific}),\
  }\bibfield  {title} {\bibinfo {title} {{Advanced LIGO}},\ }\href
  {https://doi.org/10.1088/0264-9381/32/7/074001} {\bibfield  {journal}
  {\bibinfo  {journal} {Class. Quant. Grav.}\ }\textbf {\bibinfo {volume}
  {32}},\ \bibinfo {pages} {074001} (\bibinfo {year} {2015})},\ \Eprint
  {https://arxiv.org/abs/1411.4547} {arXiv:1411.4547 [gr-qc]} \BibitemShut
  {NoStop}%
\bibitem [{\citenamefont {Acernese}\ \emph {et~al.}(2015)\citenamefont
  {Acernese} \emph {et~al.}}]{VIRGO:2014yos}%
  \BibitemOpen
  \bibfield  {author} {\bibinfo {author} {\bibfnamefont {F.}~\bibnamefont
  {Acernese}} \emph {et~al.} (\bibinfo {collaboration} {VIRGO}),\ }\bibfield
  {title} {\bibinfo {title} {{Advanced Virgo: a second-generation
  interferometric gravitational wave detector}},\ }\href
  {https://doi.org/10.1088/0264-9381/32/2/024001} {\bibfield  {journal}
  {\bibinfo  {journal} {Class. Quant. Grav.}\ }\textbf {\bibinfo {volume}
  {32}},\ \bibinfo {pages} {024001} (\bibinfo {year} {2015})},\ \Eprint
  {https://arxiv.org/abs/1408.3978} {arXiv:1408.3978 [gr-qc]} \BibitemShut
  {NoStop}%
\bibitem [{\citenamefont {Akutsu}\ \emph {et~al.}(2021)\citenamefont {Akutsu}
  \emph {et~al.}}]{KAGRA:2020agh}%
  \BibitemOpen
  \bibfield  {author} {\bibinfo {author} {\bibfnamefont {T.}~\bibnamefont
  {Akutsu}} \emph {et~al.} (\bibinfo {collaboration} {KAGRA}),\ }\bibfield
  {title} {\bibinfo {title} {{Overview of KAGRA: Calibration, detector
  characterization, physical environmental monitors, and the geophysics
  interferometer}},\ }\href {https://doi.org/10.1093/ptep/ptab018} {\bibfield
  {journal} {\bibinfo  {journal} {PTEP}\ }\textbf {\bibinfo {volume} {2021}},\
  \bibinfo {pages} {05A102} (\bibinfo {year} {2021})},\ \Eprint
  {https://arxiv.org/abs/2009.09305} {arXiv:2009.09305 [gr-qc]} \BibitemShut
  {NoStop}%
\bibitem [{\citenamefont {{Babak}}\ \emph {et~al.}(2013)\citenamefont
  {{Babak}}, \citenamefont {{Biswas}}, \citenamefont {{Brady}}, \citenamefont
  {{Brown}}, \citenamefont {{Cannon}}, \citenamefont {{Capano}}, \citenamefont
  {{Clayton}}, \citenamefont {{Cokelaer}}, \citenamefont {{Creighton}},
  \citenamefont {{Dent}}, \citenamefont {{Dietz}}, \citenamefont {{Fairhurst}},
  \citenamefont {{Fotopoulos}}, \citenamefont {{Gonz{\'a}lez}}, \citenamefont
  {{Hanna}}, \citenamefont {{Harry}}, \citenamefont {{Jones}}, \citenamefont
  {{Keppel}}, \citenamefont {{McKechan}}, \citenamefont {{Pekowsky}},
  \citenamefont {{Privitera}}, \citenamefont {{Robinson}}, \citenamefont
  {{Rodriguez}}, \citenamefont {{Sathyaprakash}}, \citenamefont {{Sengupta}},
  \citenamefont {{Vallisneri}}, \citenamefont {{Vaulin}},\ and\ \citenamefont
  {{Weinstein}}}]{2013PhRvD..87b4033B}%
  \BibitemOpen
  \bibfield  {author} {\bibinfo {author} {\bibfnamefont {S.}~\bibnamefont
  {{Babak}}}, \bibinfo {author} {\bibfnamefont {R.}~\bibnamefont {{Biswas}}},
  \bibinfo {author} {\bibfnamefont {P.~R.}\ \bibnamefont {{Brady}}}, \bibinfo
  {author} {\bibfnamefont {D.~A.}\ \bibnamefont {{Brown}}}, \bibinfo {author}
  {\bibfnamefont {K.}~\bibnamefont {{Cannon}}}, \bibinfo {author}
  {\bibfnamefont {C.~D.}\ \bibnamefont {{Capano}}}, \bibinfo {author}
  {\bibfnamefont {J.~H.}\ \bibnamefont {{Clayton}}}, \bibinfo {author}
  {\bibfnamefont {T.}~\bibnamefont {{Cokelaer}}}, \bibinfo {author}
  {\bibfnamefont {J.~D.~E.}\ \bibnamefont {{Creighton}}}, \bibinfo {author}
  {\bibfnamefont {T.}~\bibnamefont {{Dent}}}, \bibinfo {author} {\bibfnamefont
  {A.}~\bibnamefont {{Dietz}}}, \bibinfo {author} {\bibfnamefont
  {S.}~\bibnamefont {{Fairhurst}}}, \bibinfo {author} {\bibfnamefont
  {N.}~\bibnamefont {{Fotopoulos}}}, \bibinfo {author} {\bibfnamefont
  {G.}~\bibnamefont {{Gonz{\'a}lez}}}, \bibinfo {author} {\bibfnamefont
  {C.}~\bibnamefont {{Hanna}}}, \bibinfo {author} {\bibfnamefont {I.~W.}\
  \bibnamefont {{Harry}}}, \bibinfo {author} {\bibfnamefont {G.}~\bibnamefont
  {{Jones}}}, \bibinfo {author} {\bibfnamefont {D.}~\bibnamefont {{Keppel}}},
  \bibinfo {author} {\bibfnamefont {D.~J.~A.}\ \bibnamefont {{McKechan}}},
  \bibinfo {author} {\bibfnamefont {L.}~\bibnamefont {{Pekowsky}}}, \bibinfo
  {author} {\bibfnamefont {S.}~\bibnamefont {{Privitera}}}, \bibinfo {author}
  {\bibfnamefont {C.}~\bibnamefont {{Robinson}}}, \bibinfo {author}
  {\bibfnamefont {A.~C.}\ \bibnamefont {{Rodriguez}}}, \bibinfo {author}
  {\bibfnamefont {B.~S.}\ \bibnamefont {{Sathyaprakash}}}, \bibinfo {author}
  {\bibfnamefont {A.~S.}\ \bibnamefont {{Sengupta}}}, \bibinfo {author}
  {\bibfnamefont {M.}~\bibnamefont {{Vallisneri}}}, \bibinfo {author}
  {\bibfnamefont {R.}~\bibnamefont {{Vaulin}}},\ and\ \bibinfo {author}
  {\bibfnamefont {A.~J.}\ \bibnamefont {{Weinstein}}},\ }\bibfield  {title}
  {\bibinfo {title} {{Searching for gravitational waves from binary
  coalescence}},\ }\href {https://doi.org/10.1103/PhysRevD.87.024033}
  {\bibfield  {journal} {\bibinfo  {journal} {\prd}\ }\textbf {\bibinfo
  {volume} {87}},\ \bibinfo {eid} {024033} (\bibinfo {year} {2013})},\ \Eprint
  {https://arxiv.org/abs/1208.3491} {arXiv:1208.3491 [gr-qc]} \BibitemShut
  {NoStop}%
\bibitem [{\citenamefont {{Schutz}}(2011)}]{2011CQGra..28l5023S}%
  \BibitemOpen
  \bibfield  {author} {\bibinfo {author} {\bibfnamefont {B.~F.}\ \bibnamefont
  {{Schutz}}},\ }\bibfield  {title} {\bibinfo {title} {{Networks of
  gravitational wave detectors and three figures of merit}},\ }\href
  {https://doi.org/10.1088/0264-9381/28/12/125023} {\bibfield  {journal}
  {\bibinfo  {journal} {Classical and Quantum Gravity}\ }\textbf {\bibinfo
  {volume} {28}},\ \bibinfo {eid} {125023} (\bibinfo {year} {2011})},\ \Eprint
  {https://arxiv.org/abs/1102.5421} {arXiv:1102.5421 [astro-ph.IM]}
  \BibitemShut {NoStop}%
\bibitem [{\citenamefont {{Allen}}\ \emph {et~al.}(2012)\citenamefont
  {{Allen}}, \citenamefont {{Anderson}}, \citenamefont {{Brady}}, \citenamefont
  {{Brown}},\ and\ \citenamefont {{Creighton}}}]{2012PhRvD..85l2006A}%
  \BibitemOpen
  \bibfield  {author} {\bibinfo {author} {\bibfnamefont {B.}~\bibnamefont
  {{Allen}}}, \bibinfo {author} {\bibfnamefont {W.~G.}\ \bibnamefont
  {{Anderson}}}, \bibinfo {author} {\bibfnamefont {P.~R.}\ \bibnamefont
  {{Brady}}}, \bibinfo {author} {\bibfnamefont {D.~A.}\ \bibnamefont
  {{Brown}}},\ and\ \bibinfo {author} {\bibfnamefont {J.~D.~E.}\ \bibnamefont
  {{Creighton}}},\ }\bibfield  {title} {\bibinfo {title} {{FINDCHIRP: An
  algorithm for detection of gravitational waves from inspiraling compact
  binaries}},\ }\href {https://doi.org/10.1103/PhysRevD.85.122006} {\bibfield
  {journal} {\bibinfo  {journal} {\prd}\ }\textbf {\bibinfo {volume} {85}},\
  \bibinfo {eid} {122006} (\bibinfo {year} {2012})},\ \Eprint
  {https://arxiv.org/abs/gr-qc/0509116} {arXiv:gr-qc/0509116 [gr-qc]}
  \BibitemShut {NoStop}%
\bibitem [{\citenamefont {{Fairhurst}}(2018)}]{2018CQGra..35j5002F}%
  \BibitemOpen
  \bibfield  {author} {\bibinfo {author} {\bibfnamefont {S.}~\bibnamefont
  {{Fairhurst}}},\ }\bibfield  {title} {\bibinfo {title} {{Localization of
  transient gravitational wave sources: beyond triangulation}},\ }\href
  {https://doi.org/10.1088/1361-6382/aab675} {\bibfield  {journal} {\bibinfo
  {journal} {Classical and Quantum Gravity}\ }\textbf {\bibinfo {volume}
  {35}},\ \bibinfo {eid} {105002} (\bibinfo {year} {2018})},\ \Eprint
  {https://arxiv.org/abs/1712.04724} {arXiv:1712.04724 [gr-qc]} \BibitemShut
  {NoStop}%
\bibitem [{\citenamefont {Usman}\ \emph {et~al.}(2016)\citenamefont {Usman}
  \emph {et~al.}}]{pycbc}%
  \BibitemOpen
  \bibfield  {author} {\bibinfo {author} {\bibfnamefont {S.~A.}\ \bibnamefont
  {Usman}} \emph {et~al.},\ }\bibfield  {title} {\bibinfo {title} {{The PyCBC
  search for gravitational waves from compact binary coalescence}},\ }\href
  {https://doi.org/10.1088/0264-9381/33/21/215004} {\bibfield  {journal}
  {\bibinfo  {journal} {Class. Quantum Grav.}\ }\textbf {\bibinfo {volume}
  {33}} (\bibinfo {year} {2016})}\BibitemShut {NoStop}%
\bibitem [{\citenamefont {Messick}\ \emph {et~al.}(2017)\citenamefont {Messick}
  \emph {et~al.}}]{gstlal}%
  \BibitemOpen
  \bibfield  {author} {\bibinfo {author} {\bibfnamefont {C.}~\bibnamefont
  {Messick}} \emph {et~al.},\ }\bibfield  {title} {\bibinfo {title} {{Analysis
  framework for the prompt discovery of compact binary mergers in
  gravitational-wave data}},\ }\href
  {https://doi.org/10.1103/PhysRevD.95.042001} {\bibfield  {journal} {\bibinfo
  {journal} {Phys. Rev. D}\ }\textbf {\bibinfo {volume} {95}} (\bibinfo {year}
  {2017})}\BibitemShut {NoStop}%
\bibitem [{\citenamefont {{Adams}}\ \emph {et~al.}(2016)\citenamefont
  {{Adams}}, \citenamefont {{Buskulic}}, \citenamefont {{Germain}},
  \citenamefont {{Guidi}}, \citenamefont {{Marion}}, \citenamefont {{Montani}},
  \citenamefont {{Mours}}, \citenamefont {{Piergiovanni}},\ and\ \citenamefont
  {{Wang}}}]{2016CQGra..33q5012A}%
  \BibitemOpen
  \bibfield  {author} {\bibinfo {author} {\bibfnamefont {T.}~\bibnamefont
  {{Adams}}}, \bibinfo {author} {\bibfnamefont {D.}~\bibnamefont {{Buskulic}}},
  \bibinfo {author} {\bibfnamefont {V.}~\bibnamefont {{Germain}}}, \bibinfo
  {author} {\bibfnamefont {G.~M.}\ \bibnamefont {{Guidi}}}, \bibinfo {author}
  {\bibfnamefont {F.}~\bibnamefont {{Marion}}}, \bibinfo {author}
  {\bibfnamefont {M.}~\bibnamefont {{Montani}}}, \bibinfo {author}
  {\bibfnamefont {B.}~\bibnamefont {{Mours}}}, \bibinfo {author} {\bibfnamefont
  {F.}~\bibnamefont {{Piergiovanni}}},\ and\ \bibinfo {author} {\bibfnamefont
  {G.}~\bibnamefont {{Wang}}},\ }\bibfield  {title} {\bibinfo {title}
  {{Low-latency analysis pipeline for compact binary coalescences in the
  advanced gravitational wave detector era}},\ }\href
  {https://doi.org/10.1088/0264-9381/33/17/175012} {\bibfield  {journal}
  {\bibinfo  {journal} {Classical and Quantum Gravity}\ }\textbf {\bibinfo
  {volume} {33}},\ \bibinfo {eid} {175012} (\bibinfo {year} {2016})},\ \Eprint
  {https://arxiv.org/abs/1512.02864} {arXiv:1512.02864 [gr-qc]} \BibitemShut
  {NoStop}%
\bibitem [{\citenamefont {{Chu}}\ \emph {et~al.}(2022)\citenamefont {{Chu}},
  \citenamefont {{Kovalam}}, \citenamefont {{Wen}}, \citenamefont
  {{Slaven-Blair}}, \citenamefont {{Bosveld}}, \citenamefont {{Chen}},
  \citenamefont {{Clearwater}}, \citenamefont {{Codoreanu}}, \citenamefont
  {{Du}}, \citenamefont {{Guo}}, \citenamefont {{Guo}}, \citenamefont {{Kim}},
  \citenamefont {{Li}}, \citenamefont {{Oloworaran}}, \citenamefont
  {{Panther}}, \citenamefont {{Powell}}, \citenamefont {{Sengupta}},
  \citenamefont {{Wette}},\ and\ \citenamefont {{Zhu}}}]{2022PhRvD.105b4023C}%
  \BibitemOpen
  \bibfield  {author} {\bibinfo {author} {\bibfnamefont {Q.}~\bibnamefont
  {{Chu}}}, \bibinfo {author} {\bibfnamefont {M.}~\bibnamefont {{Kovalam}}},
  \bibinfo {author} {\bibfnamefont {L.}~\bibnamefont {{Wen}}}, \bibinfo
  {author} {\bibfnamefont {T.}~\bibnamefont {{Slaven-Blair}}}, \bibinfo
  {author} {\bibfnamefont {J.}~\bibnamefont {{Bosveld}}}, \bibinfo {author}
  {\bibfnamefont {Y.}~\bibnamefont {{Chen}}}, \bibinfo {author} {\bibfnamefont
  {P.}~\bibnamefont {{Clearwater}}}, \bibinfo {author} {\bibfnamefont
  {A.}~\bibnamefont {{Codoreanu}}}, \bibinfo {author} {\bibfnamefont
  {Z.}~\bibnamefont {{Du}}}, \bibinfo {author} {\bibfnamefont {X.}~\bibnamefont
  {{Guo}}}, \bibinfo {author} {\bibfnamefont {X.}~\bibnamefont {{Guo}}},
  \bibinfo {author} {\bibfnamefont {K.}~\bibnamefont {{Kim}}}, \bibinfo
  {author} {\bibfnamefont {T.~G.~F.}\ \bibnamefont {{Li}}}, \bibinfo {author}
  {\bibfnamefont {V.}~\bibnamefont {{Oloworaran}}}, \bibinfo {author}
  {\bibfnamefont {F.}~\bibnamefont {{Panther}}}, \bibinfo {author}
  {\bibfnamefont {J.}~\bibnamefont {{Powell}}}, \bibinfo {author}
  {\bibfnamefont {A.~S.}\ \bibnamefont {{Sengupta}}}, \bibinfo {author}
  {\bibfnamefont {K.}~\bibnamefont {{Wette}}},\ and\ \bibinfo {author}
  {\bibfnamefont {X.}~\bibnamefont {{Zhu}}},\ }\bibfield  {title} {\bibinfo
  {title} {{SPIIR online coherent pipeline to search for gravitational waves
  from compact binary coalescences}},\ }\href
  {https://doi.org/10.1103/PhysRevD.105.024023} {\bibfield  {journal} {\bibinfo
   {journal} {\prd}\ }\textbf {\bibinfo {volume} {105}},\ \bibinfo {eid}
  {024023} (\bibinfo {year} {2022})},\ \Eprint
  {https://arxiv.org/abs/2109.14183} {arXiv:2109.14183 [gr-qc]} \BibitemShut
  {NoStop}%
\bibitem [{\citenamefont {Fairhurst}(2009)}]{Fairhurst:2009tc}%
  \BibitemOpen
  \bibfield  {author} {\bibinfo {author} {\bibfnamefont {S.}~\bibnamefont
  {Fairhurst}},\ }\bibfield  {title} {\bibinfo {title} {{Triangulation of
  gravitational wave sources with a network of detectors}},\ }\href
  {https://doi.org/10.1088/1367-2630/11/12/123006} {\bibfield  {journal}
  {\bibinfo  {journal} {New J. Phys.}\ }\textbf {\bibinfo {volume} {11}},\
  \bibinfo {pages} {123006} (\bibinfo {year} {2009})},\ \bibinfo {note}
  {[Erratum: New J.Phys. 13, 069602 (2011)]},\ \Eprint
  {https://arxiv.org/abs/0908.2356} {arXiv:0908.2356 [gr-qc]} \BibitemShut
  {NoStop}%
\bibitem [{\citenamefont {{Baird}}\ \emph {et~al.}(2013)\citenamefont
  {{Baird}}, \citenamefont {{Fairhurst}}, \citenamefont {{Hannam}},\ and\
  \citenamefont {{Murphy}}}]{2013PhRvD..87b4035B}%
  \BibitemOpen
  \bibfield  {author} {\bibinfo {author} {\bibfnamefont {E.}~\bibnamefont
  {{Baird}}}, \bibinfo {author} {\bibfnamefont {S.}~\bibnamefont
  {{Fairhurst}}}, \bibinfo {author} {\bibfnamefont {M.}~\bibnamefont
  {{Hannam}}},\ and\ \bibinfo {author} {\bibfnamefont {P.}~\bibnamefont
  {{Murphy}}},\ }\bibfield  {title} {\bibinfo {title} {{Degeneracy between mass
  and spin in black-hole-binary waveforms}},\ }\href
  {https://doi.org/10.1103/PhysRevD.87.024035} {\bibfield  {journal} {\bibinfo
  {journal} {\prd}\ }\textbf {\bibinfo {volume} {87}},\ \bibinfo {eid} {024035}
  (\bibinfo {year} {2013})},\ \Eprint {https://arxiv.org/abs/1211.0546}
  {arXiv:1211.0546 [gr-qc]} \BibitemShut {NoStop}%
\bibitem [{\citenamefont {Morisaki}\ and\ \citenamefont
  {Raymond}(2020)}]{Morisaki:2020oqk}%
  \BibitemOpen
  \bibfield  {author} {\bibinfo {author} {\bibfnamefont {S.}~\bibnamefont
  {Morisaki}}\ and\ \bibinfo {author} {\bibfnamefont {V.}~\bibnamefont
  {Raymond}},\ }\bibfield  {title} {\bibinfo {title} {{Rapid Parameter
  Estimation of Gravitational Waves from Binary Neutron Star Coalescence using
  Focused Reduced Order Quadrature}},\ }\href
  {https://doi.org/10.1103/PhysRevD.102.104020} {\bibfield  {journal} {\bibinfo
   {journal} {Phys. Rev. D}\ }\textbf {\bibinfo {volume} {102}},\ \bibinfo
  {pages} {104020} (\bibinfo {year} {2020})},\ \Eprint
  {https://arxiv.org/abs/2007.09108} {arXiv:2007.09108 [gr-qc]} \BibitemShut
  {NoStop}%
\bibitem [{\citenamefont {Pathak}\ \emph {et~al.}(2022)\citenamefont {Pathak},
  \citenamefont {Reza},\ and\ \citenamefont {Sengupta}}]{Pathak:2022ktt}%
  \BibitemOpen
  \bibfield  {author} {\bibinfo {author} {\bibfnamefont {L.}~\bibnamefont
  {Pathak}}, \bibinfo {author} {\bibfnamefont {A.}~\bibnamefont {Reza}},\ and\
  \bibinfo {author} {\bibfnamefont {A.~S.}\ \bibnamefont {Sengupta}},\
  }\bibfield  {title} {\bibinfo {title} {{Rapid reconstruction of compact
  binary sources using meshfree approximation}},\ }\href@noop {} {\bibfield
  {journal} {\bibinfo  {journal} {arXiv}\ } (\bibinfo {year} {2022})},\ \Eprint
  {https://arxiv.org/abs/2210.02706} {arXiv:2210.02706 [gr-qc]} \BibitemShut
  {NoStop}%
\bibitem [{\citenamefont {{Sidery}}\ \emph {et~al.}(2014)\citenamefont
  {{Sidery}}, \citenamefont {{Aylott}}, \citenamefont {{Christensen}} \emph
  {et~al.}}]{2014PhRvD..89h4060S}%
  \BibitemOpen
  \bibfield  {author} {\bibinfo {author} {\bibfnamefont {T.}~\bibnamefont
  {{Sidery}}}, \bibinfo {author} {\bibfnamefont {B.}~\bibnamefont {{Aylott}}},
  \bibinfo {author} {\bibfnamefont {N.}~\bibnamefont {{Christensen}}}, \emph
  {et~al.},\ }\bibfield  {title} {\bibinfo {title} {{Reconstructing the sky
  location of gravitational-wave detected compact binary systems: Methodology
  for testing and comparison}},\ }\href
  {https://doi.org/10.1103/PhysRevD.89.084060} {\bibfield  {journal} {\bibinfo
  {journal} {\prd}\ }\textbf {\bibinfo {volume} {89}},\ \bibinfo {eid} {084060}
  (\bibinfo {year} {2014})},\ \Eprint {https://arxiv.org/abs/1312.6013}
  {arXiv:1312.6013 [astro-ph.IM]} \BibitemShut {NoStop}%
\bibitem [{\citenamefont {{Aasi}}\ \emph {et~al.}(2015)\citenamefont {{Aasi}},
  \citenamefont {{Abbott}}, \citenamefont {{Abbott}}, \citenamefont {{Abbott}}
  \emph {et~al.}}]{2015CQGra..32g4001L}%
  \BibitemOpen
  \bibfield  {author} {\bibinfo {author} {\bibfnamefont {J.}~\bibnamefont
  {{Aasi}}}, \bibinfo {author} {\bibfnamefont {B.~P.}\ \bibnamefont
  {{Abbott}}}, \bibinfo {author} {\bibfnamefont {R.}~\bibnamefont {{Abbott}}},
  \bibinfo {author} {\bibfnamefont {T.}~\bibnamefont {{Abbott}}}, \emph
  {et~al.},\ }\bibfield  {title} {\bibinfo {title} {{Advanced LIGO}},\ }\href
  {https://doi.org/10.1088/0264-9381/32/7/074001} {\bibfield  {journal}
  {\bibinfo  {journal} {Classical and Quantum Gravity}\ }\textbf {\bibinfo
  {volume} {32}},\ \bibinfo {eid} {074001} (\bibinfo {year} {2015})},\ \Eprint
  {https://arxiv.org/abs/1411.4547} {arXiv:1411.4547 [gr-qc]} \BibitemShut
  {NoStop}%
\bibitem [{\citenamefont {{Acernese}}\ \emph {et~al.}(2015)\citenamefont
  {{Acernese}}, \citenamefont {{Agathos}}, \citenamefont {{Agatsuma}} \emph
  {et~al.}}]{2015CQGra..32b4001A}%
  \BibitemOpen
  \bibfield  {author} {\bibinfo {author} {\bibfnamefont {F.}~\bibnamefont
  {{Acernese}}}, \bibinfo {author} {\bibfnamefont {M.}~\bibnamefont
  {{Agathos}}}, \bibinfo {author} {\bibfnamefont {K.}~\bibnamefont
  {{Agatsuma}}}, \emph {et~al.},\ }\bibfield  {title} {\bibinfo {title}
  {{Advanced Virgo: a second-generation interferometric gravitational wave
  detector}},\ }\href {https://doi.org/10.1088/0264-9381/32/2/024001}
  {\bibfield  {journal} {\bibinfo  {journal} {Classical and Quantum Gravity}\
  }\textbf {\bibinfo {volume} {32}},\ \bibinfo {eid} {024001} (\bibinfo {year}
  {2015})},\ \Eprint {https://arxiv.org/abs/1408.3978} {arXiv:1408.3978
  [gr-qc]} \BibitemShut {NoStop}%
\bibitem [{\citenamefont {{Cornish}}(2010)}]{2010arXiv1007.4820C}%
  \BibitemOpen
  \bibfield  {author} {\bibinfo {author} {\bibfnamefont {N.~J.}\ \bibnamefont
  {{Cornish}}},\ }\bibfield  {title} {\bibinfo {title} {{Fast Fisher Matrices
  and Lazy Likelihoods}},\ }\href {https://doi.org/10.48550/arXiv.1007.4820}
  {\bibfield  {journal} {\bibinfo  {journal} {arXiv e-prints}\ ,\ \bibinfo
  {eid} {arXiv:1007.4820}} (\bibinfo {year} {2010})},\ \Eprint
  {https://arxiv.org/abs/1007.4820} {arXiv:1007.4820 [gr-qc]} \BibitemShut
  {NoStop}%
\bibitem [{\citenamefont {{Wong}}\ \emph {et~al.}(2023)\citenamefont {{Wong}},
  \citenamefont {{Isi}},\ and\ \citenamefont
  {{Edwards}}}]{2023arXiv230205333W}%
  \BibitemOpen
  \bibfield  {author} {\bibinfo {author} {\bibfnamefont {K.~W.~K.}\
  \bibnamefont {{Wong}}}, \bibinfo {author} {\bibfnamefont {M.}~\bibnamefont
  {{Isi}}},\ and\ \bibinfo {author} {\bibfnamefont {T.~D.~P.}\ \bibnamefont
  {{Edwards}}},\ }\bibfield  {title} {\bibinfo {title} {{Fast gravitational
  wave parameter estimation without compromises}},\ }\href
  {https://doi.org/10.48550/arXiv.2302.05333} {\bibfield  {journal} {\bibinfo
  {journal} {arXiv e-prints}\ ,\ \bibinfo {eid} {arXiv:2302.05333}} (\bibinfo
  {year} {2023})},\ \Eprint {https://arxiv.org/abs/2302.05333}
  {arXiv:2302.05333 [astro-ph.IM]} \BibitemShut {NoStop}%
\bibitem [{\citenamefont {Green}\ \emph {et~al.}(2021)\citenamefont {Green},
  \citenamefont {Hoy}, \citenamefont {Fairhurst}, \citenamefont {Hannam},
  \citenamefont {Pannarale},\ and\ \citenamefont {Thomas}}]{Green:2020ptm}%
  \BibitemOpen
  \bibfield  {author} {\bibinfo {author} {\bibfnamefont {R.}~\bibnamefont
  {Green}}, \bibinfo {author} {\bibfnamefont {C.}~\bibnamefont {Hoy}}, \bibinfo
  {author} {\bibfnamefont {S.}~\bibnamefont {Fairhurst}}, \bibinfo {author}
  {\bibfnamefont {M.}~\bibnamefont {Hannam}}, \bibinfo {author} {\bibfnamefont
  {F.}~\bibnamefont {Pannarale}},\ and\ \bibinfo {author} {\bibfnamefont
  {C.}~\bibnamefont {Thomas}},\ }\bibfield  {title} {\bibinfo {title}
  {{Identifying when Precession can be Measured in Gravitational Waveforms}},\
  }\href {https://doi.org/10.1103/PhysRevD.103.124023} {\bibfield  {journal}
  {\bibinfo  {journal} {Phys. Rev. D}\ }\textbf {\bibinfo {volume} {103}},\
  \bibinfo {pages} {124023} (\bibinfo {year} {2021})},\ \Eprint
  {https://arxiv.org/abs/2010.04131} {arXiv:2010.04131 [gr-qc]} \BibitemShut
  {NoStop}%
\bibitem [{\citenamefont {{Schmidt}}\ \emph {et~al.}(2012)\citenamefont
  {{Schmidt}}, \citenamefont {{Hannam}},\ and\ \citenamefont
  {{Husa}}}]{2012PhRvD..86j4063S}%
  \BibitemOpen
  \bibfield  {author} {\bibinfo {author} {\bibfnamefont {P.}~\bibnamefont
  {{Schmidt}}}, \bibinfo {author} {\bibfnamefont {M.}~\bibnamefont
  {{Hannam}}},\ and\ \bibinfo {author} {\bibfnamefont {S.}~\bibnamefont
  {{Husa}}},\ }\bibfield  {title} {\bibinfo {title} {{Towards models of
  gravitational waveforms from generic binaries: A simple approximate mapping
  between precessing and nonprecessing inspiral signals}},\ }\href
  {https://doi.org/10.1103/PhysRevD.86.104063} {\bibfield  {journal} {\bibinfo
  {journal} {\prd}\ }\textbf {\bibinfo {volume} {86}},\ \bibinfo {eid} {104063}
  (\bibinfo {year} {2012})},\ \Eprint {https://arxiv.org/abs/1207.3088}
  {arXiv:1207.3088 [gr-qc]} \BibitemShut {NoStop}%
\bibitem [{\citenamefont {{Fairhurst}}\ \emph {et~al.}(2019)\citenamefont
  {{Fairhurst}}, \citenamefont {{Green}}, \citenamefont {{Hoy}}, \citenamefont
  {{Hannam}},\ and\ \citenamefont {{Muir}}}]{2019arXiv190805707F}%
  \BibitemOpen
  \bibfield  {author} {\bibinfo {author} {\bibfnamefont {S.}~\bibnamefont
  {{Fairhurst}}}, \bibinfo {author} {\bibfnamefont {R.}~\bibnamefont
  {{Green}}}, \bibinfo {author} {\bibfnamefont {C.}~\bibnamefont {{Hoy}}},
  \bibinfo {author} {\bibfnamefont {M.}~\bibnamefont {{Hannam}}},\ and\
  \bibinfo {author} {\bibfnamefont {A.}~\bibnamefont {{Muir}}},\ }\bibfield
  {title} {\bibinfo {title} {{The two-harmonic approximation for gravitational
  waveforms from precessing binaries}},\ }\href@noop {} {\bibfield  {journal}
  {\bibinfo  {journal} {arXiv e-prints}\ ,\ \bibinfo {eid} {arXiv:1908.05707}}
  (\bibinfo {year} {2019})},\ \Eprint {https://arxiv.org/abs/1908.05707}
  {arXiv:1908.05707 [gr-qc]} \BibitemShut {NoStop}%
\bibitem [{\citenamefont {Hoy}\ \emph {et~al.}(2022)\citenamefont {Hoy},
  \citenamefont {Mills},\ and\ \citenamefont {Fairhurst}}]{Hoy:2021dqg}%
  \BibitemOpen
  \bibfield  {author} {\bibinfo {author} {\bibfnamefont {C.}~\bibnamefont
  {Hoy}}, \bibinfo {author} {\bibfnamefont {C.}~\bibnamefont {Mills}},\ and\
  \bibinfo {author} {\bibfnamefont {S.}~\bibnamefont {Fairhurst}},\ }\bibfield
  {title} {\bibinfo {title} {{Evidence for subdominant multipole moments and
  precession in merging black-hole-binaries from GWTC-2.1}},\ }\href
  {https://doi.org/10.1103/PhysRevD.106.023019} {\bibfield  {journal} {\bibinfo
   {journal} {Phys. Rev. D}\ }\textbf {\bibinfo {volume} {106}},\ \bibinfo
  {pages} {023019} (\bibinfo {year} {2022})},\ \Eprint
  {https://arxiv.org/abs/2111.10455} {arXiv:2111.10455 [gr-qc]} \BibitemShut
  {NoStop}%
\bibitem [{\citenamefont {{Tiwari}}\ \emph {et~al.}(2018)\citenamefont
  {{Tiwari}}, \citenamefont {{Fairhurst}},\ and\ \citenamefont
  {{Hannam}}}]{2018ApJ...868..140T}%
  \BibitemOpen
  \bibfield  {author} {\bibinfo {author} {\bibfnamefont {V.}~\bibnamefont
  {{Tiwari}}}, \bibinfo {author} {\bibfnamefont {S.}~\bibnamefont
  {{Fairhurst}}},\ and\ \bibinfo {author} {\bibfnamefont {M.}~\bibnamefont
  {{Hannam}}},\ }\bibfield  {title} {\bibinfo {title} {{Constraining Black Hole
  Spins with Gravitational-wave Observations}},\ }\href
  {https://doi.org/10.3847/1538-4357/aae8df} {\bibfield  {journal} {\bibinfo
  {journal} {\apj}\ }\textbf {\bibinfo {volume} {868}},\ \bibinfo {eid} {140}
  (\bibinfo {year} {2018})},\ \Eprint {https://arxiv.org/abs/1809.01401}
  {arXiv:1809.01401 [gr-qc]} \BibitemShut {NoStop}%
\bibitem [{\citenamefont {Bender}\ \emph {et~al.}(1998)\citenamefont {Bender},
  \citenamefont {Brillet}, \citenamefont {Ciufolini}, \citenamefont {Cruise},
  \citenamefont {Cutler}, \citenamefont {Danzmann}, \citenamefont {Folkner},
  \citenamefont {Hough}, \citenamefont {McNamara}, \citenamefont {Peterseim}
  \emph {et~al.}}]{bender1998lisa}%
  \BibitemOpen
  \bibfield  {author} {\bibinfo {author} {\bibfnamefont {P.}~\bibnamefont
  {Bender}}, \bibinfo {author} {\bibfnamefont {A.}~\bibnamefont {Brillet}},
  \bibinfo {author} {\bibfnamefont {I.}~\bibnamefont {Ciufolini}}, \bibinfo
  {author} {\bibfnamefont {A.}~\bibnamefont {Cruise}}, \bibinfo {author}
  {\bibfnamefont {C.}~\bibnamefont {Cutler}}, \bibinfo {author} {\bibfnamefont
  {K.}~\bibnamefont {Danzmann}}, \bibinfo {author} {\bibfnamefont
  {W.}~\bibnamefont {Folkner}}, \bibinfo {author} {\bibfnamefont
  {J.}~\bibnamefont {Hough}}, \bibinfo {author} {\bibfnamefont
  {P.}~\bibnamefont {McNamara}}, \bibinfo {author} {\bibfnamefont
  {M.}~\bibnamefont {Peterseim}}, \emph {et~al.},\ }\bibfield  {title}
  {\bibinfo {title} {{LISA. Laser Interferometer Space Antenna for the
  detection and observation of gravitational waves. An international project in
  the field of Fundamental Physics in Space}},\ }\href@noop {} {\bibfield
  {journal} {\bibinfo  {journal} {Max-Planck-Institut f{\"u}r Quantenoptik}\ }
  (\bibinfo {year} {1998})}\BibitemShut {NoStop}%
\bibitem [{\citenamefont {Punturo}\ \emph {et~al.}(2010)\citenamefont {Punturo}
  \emph {et~al.}}]{Punturo:2010zz}%
  \BibitemOpen
  \bibfield  {author} {\bibinfo {author} {\bibfnamefont {M.}~\bibnamefont
  {Punturo}} \emph {et~al.},\ }\bibfield  {title} {\bibinfo {title} {{The
  Einstein Telescope: A third-generation gravitational wave observatory}},\
  }\href {https://doi.org/10.1088/0264-9381/27/19/194002} {\bibfield  {journal}
  {\bibinfo  {journal} {Class. Quant. Grav.}\ }\textbf {\bibinfo {volume}
  {27}},\ \bibinfo {pages} {194002} (\bibinfo {year} {2010})}\BibitemShut
  {NoStop}%
\bibitem [{\citenamefont {Reitze}\ \emph {et~al.}(2019)\citenamefont {Reitze},
  \citenamefont {Adhikari}, \citenamefont {Ballmer}, \citenamefont {Barish},
  \citenamefont {Barsotti}, \citenamefont {Billingsley}, \citenamefont {Brown},
  \citenamefont {Chen}, \citenamefont {Coyne}, \citenamefont {Eisenstein} \emph
  {et~al.}}]{reitze2019cosmic}%
  \BibitemOpen
  \bibfield  {author} {\bibinfo {author} {\bibfnamefont {D.}~\bibnamefont
  {Reitze}}, \bibinfo {author} {\bibfnamefont {R.~X.}\ \bibnamefont
  {Adhikari}}, \bibinfo {author} {\bibfnamefont {S.}~\bibnamefont {Ballmer}},
  \bibinfo {author} {\bibfnamefont {B.}~\bibnamefont {Barish}}, \bibinfo
  {author} {\bibfnamefont {L.}~\bibnamefont {Barsotti}}, \bibinfo {author}
  {\bibfnamefont {G.}~\bibnamefont {Billingsley}}, \bibinfo {author}
  {\bibfnamefont {D.~A.}\ \bibnamefont {Brown}}, \bibinfo {author}
  {\bibfnamefont {Y.}~\bibnamefont {Chen}}, \bibinfo {author} {\bibfnamefont
  {D.}~\bibnamefont {Coyne}}, \bibinfo {author} {\bibfnamefont
  {R.}~\bibnamefont {Eisenstein}}, \emph {et~al.},\ }\bibfield  {title}
  {\bibinfo {title} {Cosmic explorer: the us contribution to gravitational-wave
  astronomy beyond ligo},\ }\href@noop {} {\bibfield  {journal} {\bibinfo
  {journal} {arXiv preprint arXiv:1907.04833}\ } (\bibinfo {year}
  {2019})}\BibitemShut {NoStop}%
\bibitem [{\citenamefont {Harris}\ \emph {et~al.}(2020)\citenamefont {Harris},
  \citenamefont {Millman}, \citenamefont {van~der Walt}, \citenamefont
  {Gommers}, \citenamefont {Virtanen}, \citenamefont {Cournapeau},
  \citenamefont {Wieser}, \citenamefont {Taylor}, \citenamefont {Berg},
  \citenamefont {Smith}, \citenamefont {Kern}, \citenamefont {Picus},
  \citenamefont {Hoyer}, \citenamefont {van Kerkwijk}, \citenamefont {Brett},
  \citenamefont {Haldane}, \citenamefont {del R{\'{i}}o}, \citenamefont
  {Wiebe}, \citenamefont {Peterson}, \citenamefont {G{\'{e}}rard-Marchant},
  \citenamefont {Sheppard}, \citenamefont {Reddy}, \citenamefont {Weckesser},
  \citenamefont {Abbasi}, \citenamefont {Gohlke},\ and\ \citenamefont
  {Oliphant}}]{harris2020array}%
  \BibitemOpen
  \bibfield  {author} {\bibinfo {author} {\bibfnamefont {C.~R.}\ \bibnamefont
  {Harris}}, \bibinfo {author} {\bibfnamefont {K.~J.}\ \bibnamefont {Millman}},
  \bibinfo {author} {\bibfnamefont {S.~J.}\ \bibnamefont {van~der Walt}},
  \bibinfo {author} {\bibfnamefont {R.}~\bibnamefont {Gommers}}, \bibinfo
  {author} {\bibfnamefont {P.}~\bibnamefont {Virtanen}}, \bibinfo {author}
  {\bibfnamefont {D.}~\bibnamefont {Cournapeau}}, \bibinfo {author}
  {\bibfnamefont {E.}~\bibnamefont {Wieser}}, \bibinfo {author} {\bibfnamefont
  {J.}~\bibnamefont {Taylor}}, \bibinfo {author} {\bibfnamefont
  {S.}~\bibnamefont {Berg}}, \bibinfo {author} {\bibfnamefont {N.~J.}\
  \bibnamefont {Smith}}, \bibinfo {author} {\bibfnamefont {R.}~\bibnamefont
  {Kern}}, \bibinfo {author} {\bibfnamefont {M.}~\bibnamefont {Picus}},
  \bibinfo {author} {\bibfnamefont {S.}~\bibnamefont {Hoyer}}, \bibinfo
  {author} {\bibfnamefont {M.~H.}\ \bibnamefont {van Kerkwijk}}, \bibinfo
  {author} {\bibfnamefont {M.}~\bibnamefont {Brett}}, \bibinfo {author}
  {\bibfnamefont {A.}~\bibnamefont {Haldane}}, \bibinfo {author} {\bibfnamefont
  {J.~F.}\ \bibnamefont {del R{\'{i}}o}}, \bibinfo {author} {\bibfnamefont
  {M.}~\bibnamefont {Wiebe}}, \bibinfo {author} {\bibfnamefont
  {P.}~\bibnamefont {Peterson}}, \bibinfo {author} {\bibfnamefont
  {P.}~\bibnamefont {G{\'{e}}rard-Marchant}}, \bibinfo {author} {\bibfnamefont
  {K.}~\bibnamefont {Sheppard}}, \bibinfo {author} {\bibfnamefont
  {T.}~\bibnamefont {Reddy}}, \bibinfo {author} {\bibfnamefont
  {W.}~\bibnamefont {Weckesser}}, \bibinfo {author} {\bibfnamefont
  {H.}~\bibnamefont {Abbasi}}, \bibinfo {author} {\bibfnamefont
  {C.}~\bibnamefont {Gohlke}},\ and\ \bibinfo {author} {\bibfnamefont {T.~E.}\
  \bibnamefont {Oliphant}},\ }\bibfield  {title} {\bibinfo {title} {Array
  programming with {NumPy}},\ }\href
  {https://doi.org/10.1038/s41586-020-2649-2} {\bibfield  {journal} {\bibinfo
  {journal} {Nature}\ }\textbf {\bibinfo {volume} {585}},\ \bibinfo {pages}
  {357} (\bibinfo {year} {2020})}\BibitemShut {NoStop}%
\bibitem [{\citenamefont {Virtanen}\ \emph {et~al.}(2020)\citenamefont
  {Virtanen}, \citenamefont {Gommers}, \citenamefont {Oliphant}, \citenamefont
  {Haberland}, \citenamefont {Reddy}, \citenamefont {Cournapeau}, \citenamefont
  {Burovski}, \citenamefont {Peterson}, \citenamefont {Weckesser},
  \citenamefont {Bright}, \citenamefont {{van der Walt}}, \citenamefont
  {Brett}, \citenamefont {Wilson}, \citenamefont {Millman}, \citenamefont
  {Mayorov}, \citenamefont {Nelson}, \citenamefont {Jones}, \citenamefont
  {Kern}, \citenamefont {Larson}, \citenamefont {Carey}, \citenamefont {Polat},
  \citenamefont {Feng}, \citenamefont {Moore}, \citenamefont {{VanderPlas}},
  \citenamefont {Laxalde}, \citenamefont {Perktold}, \citenamefont {Cimrman},
  \citenamefont {Henriksen}, \citenamefont {Quintero}, \citenamefont {Harris},
  \citenamefont {Archibald}, \citenamefont {Ribeiro}, \citenamefont
  {Pedregosa}, \citenamefont {{van Mulbregt}},\ and\ \citenamefont {{SciPy 1.0
  Contributors}}}]{2020SciPy-NMeth}%
  \BibitemOpen
  \bibfield  {author} {\bibinfo {author} {\bibfnamefont {P.}~\bibnamefont
  {Virtanen}}, \bibinfo {author} {\bibfnamefont {R.}~\bibnamefont {Gommers}},
  \bibinfo {author} {\bibfnamefont {T.~E.}\ \bibnamefont {Oliphant}}, \bibinfo
  {author} {\bibfnamefont {M.}~\bibnamefont {Haberland}}, \bibinfo {author}
  {\bibfnamefont {T.}~\bibnamefont {Reddy}}, \bibinfo {author} {\bibfnamefont
  {D.}~\bibnamefont {Cournapeau}}, \bibinfo {author} {\bibfnamefont
  {E.}~\bibnamefont {Burovski}}, \bibinfo {author} {\bibfnamefont
  {P.}~\bibnamefont {Peterson}}, \bibinfo {author} {\bibfnamefont
  {W.}~\bibnamefont {Weckesser}}, \bibinfo {author} {\bibfnamefont
  {J.}~\bibnamefont {Bright}}, \bibinfo {author} {\bibfnamefont {S.~J.}\
  \bibnamefont {{van der Walt}}}, \bibinfo {author} {\bibfnamefont
  {M.}~\bibnamefont {Brett}}, \bibinfo {author} {\bibfnamefont
  {J.}~\bibnamefont {Wilson}}, \bibinfo {author} {\bibfnamefont {K.~J.}\
  \bibnamefont {Millman}}, \bibinfo {author} {\bibfnamefont {N.}~\bibnamefont
  {Mayorov}}, \bibinfo {author} {\bibfnamefont {A.~R.~J.}\ \bibnamefont
  {Nelson}}, \bibinfo {author} {\bibfnamefont {E.}~\bibnamefont {Jones}},
  \bibinfo {author} {\bibfnamefont {R.}~\bibnamefont {Kern}}, \bibinfo {author}
  {\bibfnamefont {E.}~\bibnamefont {Larson}}, \bibinfo {author} {\bibfnamefont
  {C.~J.}\ \bibnamefont {Carey}}, \bibinfo {author} {\bibfnamefont
  {{\.I}.}~\bibnamefont {Polat}}, \bibinfo {author} {\bibfnamefont
  {Y.}~\bibnamefont {Feng}}, \bibinfo {author} {\bibfnamefont {E.~W.}\
  \bibnamefont {Moore}}, \bibinfo {author} {\bibfnamefont {J.}~\bibnamefont
  {{VanderPlas}}}, \bibinfo {author} {\bibfnamefont {D.}~\bibnamefont
  {Laxalde}}, \bibinfo {author} {\bibfnamefont {J.}~\bibnamefont {Perktold}},
  \bibinfo {author} {\bibfnamefont {R.}~\bibnamefont {Cimrman}}, \bibinfo
  {author} {\bibfnamefont {I.}~\bibnamefont {Henriksen}}, \bibinfo {author}
  {\bibfnamefont {E.~A.}\ \bibnamefont {Quintero}}, \bibinfo {author}
  {\bibfnamefont {C.~R.}\ \bibnamefont {Harris}}, \bibinfo {author}
  {\bibfnamefont {A.~M.}\ \bibnamefont {Archibald}}, \bibinfo {author}
  {\bibfnamefont {A.~H.}\ \bibnamefont {Ribeiro}}, \bibinfo {author}
  {\bibfnamefont {F.}~\bibnamefont {Pedregosa}}, \bibinfo {author}
  {\bibfnamefont {P.}~\bibnamefont {{van Mulbregt}}},\ and\ \bibinfo {author}
  {\bibnamefont {{SciPy 1.0 Contributors}}},\ }\bibfield  {title} {\bibinfo
  {title} {{{SciPy} 1.0: Fundamental Algorithms for Scientific Computing in
  Python}},\ }\href {https://doi.org/10.1038/s41592-019-0686-2} {\bibfield
  {journal} {\bibinfo  {journal} {Nature Methods}\ }\textbf {\bibinfo {volume}
  {17}},\ \bibinfo {pages} {261} (\bibinfo {year} {2020})}\BibitemShut
  {NoStop}%
\bibitem [{\citenamefont {Nitz}\ \emph {et~al.}(2022)\citenamefont {Nitz} \emph
  {et~al.}}]{PyCBCSoftware}%
  \BibitemOpen
  \bibfield  {author} {\bibinfo {author} {\bibfnamefont {A.}~\bibnamefont
  {Nitz}} \emph {et~al.},\ }\href {https://doi.org/10.5281/zenodo.6912865}
  {\bibinfo {title} {gwastro/pycbc: v2.0.5 release of pycbc}} (\bibinfo {year}
  {2022})\BibitemShut {NoStop}%
\bibitem [{\citenamefont {Hunter}(2007)}]{Hunter:2007}%
  \BibitemOpen
  \bibfield  {author} {\bibinfo {author} {\bibfnamefont {J.~D.}\ \bibnamefont
  {Hunter}},\ }\bibfield  {title} {\bibinfo {title} {Matplotlib: A 2d graphics
  environment},\ }\href {https://doi.org/10.1109/MCSE.2007.55} {\bibfield
  {journal} {\bibinfo  {journal} {Computing in Science \& Engineering}\
  }\textbf {\bibinfo {volume} {9}},\ \bibinfo {pages} {90} (\bibinfo {year}
  {2007})}\BibitemShut {NoStop}%
\bibitem [{\citenamefont {Waskom}(2021)}]{Waskom2021}%
  \BibitemOpen
  \bibfield  {author} {\bibinfo {author} {\bibfnamefont {M.~L.}\ \bibnamefont
  {Waskom}},\ }\bibfield  {title} {\bibinfo {title} {seaborn: statistical data
  visualization},\ }\href {https://doi.org/10.21105/joss.03021} {\bibfield
  {journal} {\bibinfo  {journal} {Journal of Open Source Software}\ }\textbf
  {\bibinfo {volume} {6}},\ \bibinfo {pages} {3021} (\bibinfo {year}
  {2021})}\BibitemShut {NoStop}%
\bibitem [{\citenamefont {Held}\ and\ \citenamefont
  {Bov{\'e}}(2020)}]{held2020likelihood}%
  \BibitemOpen
  \bibfield  {author} {\bibinfo {author} {\bibfnamefont {L.}~\bibnamefont
  {Held}}\ and\ \bibinfo {author} {\bibfnamefont {D.}~\bibnamefont
  {Bov{\'e}}},\ }\href {https://books.google.co.uk/books?id=EVHaDwAAQBAJ}
  {\emph {\bibinfo {title} {Likelihood and Bayesian Inference: With
  Applications in Biology and Medicine}}},\ Statistics for Biology and Health\
  (\bibinfo  {publisher} {Springer Berlin Heidelberg},\ \bibinfo {year}
  {2020})\BibitemShut {NoStop}%
\bibitem [{\citenamefont {{Tiwari}}(2018)}]{2018CQGra..35n5009T}%
  \BibitemOpen
  \bibfield  {author} {\bibinfo {author} {\bibfnamefont {V.}~\bibnamefont
  {{Tiwari}}},\ }\bibfield  {title} {\bibinfo {title} {{Estimation of the
  sensitive volume for gravitational-wave source populations using weighted
  Monte Carlo integration}},\ }\href {https://doi.org/10.1088/1361-6382/aac89d}
  {\bibfield  {journal} {\bibinfo  {journal} {Classical and Quantum Gravity}\
  }\textbf {\bibinfo {volume} {35}},\ \bibinfo {eid} {145009} (\bibinfo {year}
  {2018})},\ \Eprint {https://arxiv.org/abs/1712.00482} {arXiv:1712.00482
  [astro-ph.HE]} \BibitemShut {NoStop}%
\end{thebibliography}%

\end{document}